\documentclass[12pt]{article}
\pdfoutput=1
\usepackage{graphicx,amssymb,amsmath}
\usepackage{epsfig}
\usepackage{color,graphicx}
\usepackage{cite}
\usepackage{amssymb,amsmath}
\usepackage{multirow}
\usepackage{hyperref}
\usepackage[top=3cm, bottom=3cm, left=2cm, right=2cm]{geometry}	

\newcommand\fverb{\setbox\pippobox=\hbox\bgroup\verb}
\newcommand\fverbdo{\egroup\medskip\noindent%
                              \fbox{\unhbox\pippobox}\ }
\newcommand\fverbit{\egroup\item[\fbox{\unhbox\pippobox}]}
\newbox\pippobox

\newcommand{\beq} {\begin{equation}}
\newcommand{\eeq} {\end{equation}}
\newcommand{\beqa} {\begin{eqnarray}}
\newcommand{\eeqa} {\end{eqnarray}}

\newcommand{\rc}{\nonumber\\}
\newcommand{\bear}{\begin{eqnarray}}
\newcommand{\eear}{\end{eqnarray}}

\newcommand{\ie}{{\it i.e.}}

\def\to{\rightarrow}

\newcommand{\Tr}{\mbox{Tr}}    

\newcommand{\be}{\begin{equation}}
\newcommand{\ee}{\end{equation}}
\newcommand{\bea}{\begin{eqnarray}}
\newcommand{\eea}{\end{eqnarray}}

\def\om{\omega}

\begin{document}
 
\begin{flushright}
HIP-2019-21/TH
\end{flushright}

\begin{center}

\centering{\Large {\bf Holographic fundamental matter in  multilayered media}}

\vspace{8mm}

\renewcommand\thefootnote{\mbox{$\fnsymbol{footnote}$}}

Ulf Gran,${}^{1}$\footnote{ulf.gran@chalmers.se}
Niko Jokela,${}^{2,3}$\footnote{niko.jokela@helsinki.fi}
Daniele Musso,${}^{4,5}$\footnote{daniele.musso@usc.es} \\
Alfonso V. Ramallo,${}^{4,5}$\footnote{alfonso@fpaxp1.usc.es} and
Marcus Torns\"o${}^1$\footnote{marcus.tornso@chalmers.se}

\vspace{4mm}
${}^1${\small \sl Department of Physics, Division for Theoretical Physics} \\
{\small \sl Chalmers University of Technology} \\
{\small \sl SE-412 96 G\"oteborg, Sweden} 

\vspace{2mm}
${}^2${\small \sl Department of Physics} and ${}^3${\small \sl Helsinki Institute of Physics} \\
{\small \sl P.O.Box 64} \\
{\small \sl FIN-00014 University of Helsinki, Finland} 

\vspace{2mm}
${}^4${\small \sl Departamento de F\'isica de Part\'iculas,  Universidade de Santiago de Compostela and} \\
${}^5${\small \sl Instituto Galego de F\'isica de Altas Enerx\'ias (IGFAE), Santiago de Compostela, Spain} 

\end{center}

\vspace{8mm}
\numberwithin{equation}{section}
\setcounter{footnote}{0}
\renewcommand\thefootnote{\mbox{\arabic{footnote}}}

\begin{abstract}
\noindent 
We describe a strongly coupled layered system in 3+1 dimensions by means of a top-down D-brane construction. Adjoint matter is encoded in a large-$N_c$ stack of D3-branes, while fundamental matter is confined to $(2+1)$-dimensional defects introduced by a large-$N_f$ stack of smeared D5-branes. To the anisotropic Lifshitz-like background geometry, we add a single flavor D7-brane treated in the probe limit. Such bulk setup corresponds to a partially quenched approximation for the dual field theory.  The holographic model sheds light on the anisotropic physics induced by the layered structure, allowing one to disentangle flavor physics along and orthogonal to the layers as well as identifying distinct scaling laws for various dynamical quantities. We study the thermodynamics and the fluctuation spectrum with varying valence quark mass or baryon chemical potential. We also focus on the density wave propagation in both the hydrodynamic and collisionless regimes where analytic methods complement the numerics, while the latter provides the only resource to address the intermediate transition regime. 
\end{abstract}

\newpage
\tableofcontents
\newpage


\section{Introduction}\label{sec:introduction}

The physics of $(2+1)$-dimensional layers is at the core of two active research fronts in condensed matter: 
graphene and layered hetero-structures. 
The most notable examples of layered systems are the copper oxides and high-$T_c$ superconductors in general. 
Although the attention is usually focused on the physics along the layers, relevant information is also associated with the dynamics in the orthogonal directions \cite{Hepting2018}. 
This motivates the quest for theoretical models which can simultaneously capture the whole in-/off-plane behavior of multilayered systems.

The layers are in general expected to induce characteristic anisotropy both to the equilibrium and out-of-equilibrium properties of the system. 
However, it should be borne in mind that the anisotropy could also be regarded as an emergent effect due to the coarse-graining of the microscopic scales dominated by strongly-coupled fundamental degrees of freedom. 
Thus, to genuinely describe layered hetero-structures, the construction of a bottom-up model which just reproduces the low-energy anisotropic physics may be unsatisfactory.  

The description of strongly-coupled dynamics does not only require an alternative to perturbative approaches,
but it can even radically affect fundamental schematizations on which intuition and physical models are founded. 
Most notably, the concept of quasi-particles is typically unsuited to account for the low-energy response of a strongly-correlated system, especially in a dense medium. 
In other words, the collective dynamics has to be directly addressed without any pre-established paradigm based on long-lived modes. 
This constitutes a fundamental motivation to resort to holographic descriptions \cite{Maldacena:1997re,Ramallo:2013bua} that, through the study of the small fluctuations of the gravity dual, provide explicit information on the low-energy collective behavior of the strongly-coupled boundary field theory. 
Layered systems are no exception. 
Layers introduce two qualitatively different ``sectors'' associated to the in- and off-plane physics, respectively. 
Accessing direct information about the collective low-energy response \cite{Gushterov:2018spg} is particularly important for layered systems. 
Recent experiments on copper oxides present puzzling outcomes, for instance the characterization of the modes emerging from the low-energy transport properties contrasts the featureless spectral density-density response \cite{Mitrano5392}.

In this paper we follow a top-down strategy, where we can, at least in principle, keep track of microscopic degrees of freedom. To describe the strongly-coupled physics of fundamental matter, we resort to AdS/CFT and consider a top-down intersecting D-brane construction.\footnote{For previous work on anisotropic systems via holography, see, {\emph{e.g.}}, 
\cite{Azeyanagi:2009pr,Mateos:2011tv,Kiritsis:2011zq,Ammon:2012qs,Cheng:2014qia,Jain:2014vka,Roychowdhury:2015cva,Banks:2015aca,Itsios:2018hff,Giataganas:2017koz,Rahimi:2018ica,Gursoy:2018ydr}.} The most natural intersecting D-brane configuration is the D3-D5-brane construction of \cite{DeWolfe:2001pq,Erdmenger:2002ex}. In this setup the layer associated with directions along the $(2+1)$-dimensional D5-brane worldvolume introduces a co-dimension one defect in the ambient $(3+1)$-dimensional ${\cal{N}}=4$ super Yang-Mills theory. Previous studies of the D3-D5 systems where the flavor branes are considered in the probe approximation include \cite{Arean:2006pk,Gomis:2006cu,Filev:2009xp,Jensen:2010ga,Evans:2010hi,Kristjansen:2012ny,Kristjansen:2013hma,Evans:2014mva}.

In more detail, the large-$N$ stack of space-filling D3-branes introduces adjoint matter living in $3+1$ dimensions and a stack of D5-branes, partially wrapped along the internal manifold, introduce fundamental hypermultiplets on a ${\mathbb{R}}^{1,2}\subset {\mathbb{R}}^{1,3}$ defect. However, we do not attempt to solve this localized configuration, but instead we introduce the defects through continuously smearing them over the orthogonal directions \cite{Nunez:2010sf,Conde:2016hbg,Penin:2017lqt}. In addition to computational advantages, this procedure is also a physically appealing approximation. The smearing allows one to disentangle two consequences due to the layers: the breaking of rotations and that of orthogonal translations. Introducing the smeared distribution of layers, the interlayer separation is formally vanishing. This procedure then does not break orthogonal translations and leads to genuinely homogeneous but anisotropic configuration. In the dual ten-dimensional gravity side, the anisotropy is then encoded, {\emph{e.g.}}, in the resulting metric featuring a Lifshitz-like scaling and where the spatial direction orthogonal to the layers is singled out.

In this paper, we furthermore introduce flavor D7-branes \cite{Karch:2002sh} in the anisotropic D3-D5 background. 
We treat these D7-branes in the probe approximation, which on the dual field theory side corresponds to the partially quenched approximation, where the defect ``sea quark'' degrees of freedom stemming from D5-branes are unquenched while the ``valence quarks'' (D3-D7 strings) are quenched. We allow finite bare valence quark masses as well as a non-vanishing baryonic chemical potential for them, though we do not consider them simultaneously. 

In addition to studying the thermodynamic phase diagram of the model we also discuss the excitation spectrum. In particular, we study the various propagating (sound) and purely dissipating (diffusion) regimes \cite{Grozdanov:2018fic} of the longitudinal modes, both along and perpendicular to the defects. We especially focus on the peculiarities of the off-plane sector and comment on its analogies to Lifshitz models \cite{Hoyos:2013qna,Hoyos:2013eza}. As for future directions, the characterization of the dynamical polarizability of the layered medium is a necessary step in view of considering 
semi-holographic generalizations \cite{Faulkner:2010tq}. Assigning a dynamical character to the boundary conditions of the bulk gauge field \cite{Witten:2003ya,Burgess:2000kj,Burgess:2006fw,Jokela:2013hta,Jokela:2014wsa,Ihl:2016sop,Jokela:2017fwa}, one can address physically relevant questions, for instance: the propagation of light-waves through the layered medium \cite{Amariti:2010jw,Amariti:2010hw,Forcella:2014dwa} and the introduction of long-range (Coulomb-like) interactions relevant for plasmonic physics \cite{Aronsson:2017dgf,Aronsson:2018yhi,Gran:2018vdn,Gran:2018jnt,Mauri:2018pzq,Krikun:2018agd,Romero-Bermudez:2019lzz,Baggioli:2019aqf}.

The paper is structured as follows. In Section~\ref{sec:setup} we describe the D3-D5-brane setup and briefly review the zero-temperature background and the corresponding black hole solutions. We also introduce a probe D7-brane and study its embedding. In Section~\ref{tempe} we start analyzing the thermodynamics of the background at vanishing chemical potential. We first recall the supersymmetric solution at zero temperature and show that it can be treated fully analytically. We then turn to a finite temperature analysis and discuss the two competing phases and the associated meson melting phase transition. The two phases correspond to Minkowski and black-hole embeddings of the D7-brane. In Section~\ref{non_zero_density} we start analyzing the system at non-zero chemical potential. This allows us to approach questions that are relevant for the condensed matter systems. The Minkowski embeddings describe insulators while black hole embeddings are associated with metallic behavior. We will, however, mainly focus on the latter in this paper. We are especially interested in the fluctuation spectrum, the propagation of modes and their dissipation. In particular, we study the longitudinal zero sound both along and orthogonal to the layers. Adopting the two alternative real-momentum and real-frequency  approaches, we characterize the response of the system. Section~\ref{sec:discussion} contains our discussion and outlines directions for future research. The appendices contain several technical details omitted in the body of the paper.

\section{Setup}\label{sec:setup}

In this section we review the background generated by intersecting D3- and D5-branes \cite{DeWolfe:2001pq,Erdmenger:2002ex,Conde:2016hbg} to which we then add an extra D7-brane. 
The flat space and low-energy configuration of the set of all the branes can be summarized in the following table:
\beq
\label{D3D5intersection}
\begin{array}{cccccccccccl}
 & 0&1&2&3& 4& 5&6 &7&8&9 &  \\
D3: & \times & \times &\times &\times &\_ &\_ & \_&\_ &\_ &\_ &      \\
D5: & \times &\times&\times&\_&\times&\times&\times&\_&\_&\_ &\\
D7: & \times &\times&\times&\times&\times&\times&\times&\times&\_&\_ &
\end{array}
\eeq
The study of the complete dynamical system is rather involved, we therefore adopt an approximation in which the D7-brane is considered a probe in the geometry generated by the set of intersecting D3- and D5-branes. 
On the field theory side this corresponds to a partially quenched approximation where the fundamentals dual to strings attached to the D7-brane are not dynamical. 

The geometry following from stacks of (coincident) D3- and (delocalized) D5-branes was obtained at zero temperature in \cite{Conde:2016hbg}, with subsequent generalization to a black hole geometry in \cite{Penin:2017lqt}.  The solutions to the field equations were obtained by the smearing approximation \cite{Nunez:2010sf}, in which the D5-branes are distributed homogeneously along the internal directions orthogonal to the D3's, as well as along one of the directions parallel to the D3's. The D5-branes can then be viewed as multiple parallel layers that create a  codimension-one defect in the (3+1)-dimensional theory living on the worldvolume of the stack of D3-branes. This theory is very rich and has lots of interesting features \cite{Jokela:2019tsb}. 
The present analysis focuses on the extra D7-brane probing the multilayer geometry and introducing new flavor degrees of freedom in the fundamental representation of the gauge group (quarks). 
Similar bulk constructions \cite{Conde:2011sw,Jokela:2012dw,Bea:2013jxa,Jokela:2013qya,Bea:2014yda,Bea:2017iqt} but with a completely (2+1)-dimensional field theory dual are connected with ABJ(M) Chern-Simons matter theories \cite{Aharony:2008ug,Aharony:2008gk} coupled with fundamentals \cite{Hohenegger:2009as,Gaiotto:2009tk}.

Let us now review in detail the D3-D5 geometry of \cite{Conde:2016hbg,Penin:2017lqt}. The ten-dimensional Einstein frame metric at finite temperature can be written as:
\beq
ds^2_{10}\,=\,ds^2_5\,+\,d\hat s_5^2\,\,,
\label{10d_flavored_metric}
\eeq
where $ds^2_5$ is the following five-dimensional metric:
\beq
ds^2_5\,=\,h^{-{1\over 2}}\,\Big[-B\,(dx^0)^2\,+\,(dx^1)^2\,+\,(dx^2)^2\,+\,
e^{-2\phi}\,\,
(dx^3)^2\,\Big]\,+\,
\,{h^{{1\over 2}} \over B}\,dr^2\ ,
\label{black-anisotropic}
\eeq
where $\phi$ is the dilaton. The anisotropy in the spatial directions in (\ref{black-anisotropic})  is due to the fact that the D5-branes are extended along $x^1\,x^2$ and smeared along $x^3$.  The  warp factor $h$ and the emblackening factor $B$  in (\ref{black-anisotropic}) are given by:
\beq
h\,=\,{R^4\over r^4}\,\,,
\qquad\qquad
B(r)\,=\,1\,-\,\Big({r_h\over r}\Big)^{{10\over 3}}\,\,.
\label{warp_blackening}
\eeq
The radius of curvature $R$ is determined by the number of colors $N_c$, \ie, the number of D3-branes:
\beq
R^4\,=\,{256\over 1215}\,Q_c\,\,,
\label{R_Qc}
\eeq
where $Q_c\sim N_c$ (see below). The dilaton field $\phi$ for this geometry is non-trivial and determines the spatial anisotropy:
\beq
e^{{3\phi\over 2}}\,=\,{3\over 4\,Q_f}\,r\,\,,
\label{dilaton}
\eeq
where $Q_f\sim N_f$, with $N_f$ being the number of D5-branes per unit length in the $x^3$ direction.  The precise relations between $Q_c, Q_f$, and $N_c, N_f$ are
\beq
Q_c\,=\,16\pi\,N_c\,\,,
\qquad\qquad
Q_f\,=\,{4\,\pi\,N_f\over 9\,\sqrt{3}}\,\,,
\label{Qf_Qc_Nf_N_c}
\eeq
where we are working in units in which $g_s=\alpha'=1$. The metric $d\hat s_5^2$  of the internal part is \cite{Conde:2016hbg}:
\beq
d\hat s_5^2\,=\,\bar R^2\,\Big[
d\chi^2+{\cos^2\chi\over 4}((\omega^1)^2+(\omega^2)^2)+
{\cos^2\chi \sin^2\chi \over 4} (\omega^3)^2
+{1\over b}\,\big(d\tau+{1\over 2}\,\cos^2\chi\, \omega^3\big)^2\Big]\,\,,
\label{compact_flavored_metric}
\eeq
where
\beq
b\,=\,{8\over 9}\ ,  \quad\qquad \bar R^2\,=\,{r^2\,h^{{1\over 2}}\over b}\,=\,{R^2\over b}\,=\,{4\over 15}\,Q_c\ .
\eeq
In (\ref{compact_flavored_metric}) $\chi$ and $\tau$ are angular coordinates taking values in the range 
$0\le \chi\le \pi/2$ and $0\le\tau\le 2\pi$, whereas $\omega^1$, $\omega^2$, and $\omega^3$ are left-invariant 
$SU(2)$ one-forms (see (\ref{SU2_oneforms}) in Appendix \ref{kappa} for their explicit representation in terms of three angles). The complete type IIB supergravity background contains Ramond-Ramond five- and three-forms, whose explicit expression is not needed in this work (see \cite{Conde:2016hbg,Penin:2017lqt}). 
The different fields satisfy the equations of motion of ten-dimensional type IIB supergravity with D5-brane sources. 
When $r_h=0$  ($B=1$) our solution is supersymmetric and invariant under a set of Lifshitz-like anisotropic 
scale transformations in which the $x^3$ coordinate transforms with an anomalous exponent $z=3$ \cite{Conde:2016hbg}. 

Let us recall that the temperature $T$ of a black hole is given in terms of the $tt$ and $rr$ components of the metric by the general formula, leading to:
\beq
T={1\over 2\pi}\Big[\,{1\over \sqrt{g_{rr}}}\,\,{d\over dr}\Big(\,\sqrt{\,-g_{tt}}\,\Big)\Big]_{r=r_h} = {5\,r_h\over 6\pi\, R^2}\ . \label{T_rh_R}
\eeq
Using (\ref{R_Qc}) we can recast this relation in terms of $Q_c$ as:
\beq
r_h\,=\,{2^5\,\pi\over 3^{{3\over 2}}\,5^{{3\over 2}}}\,
Q_c^{{1\over 2}}\,T\,\,.
\label{rh_T_Q_c}
\eeq

Let us now add a D7-brane probe to the D3-D5 geometry. 
According to the array (\ref{D3D5intersection}), we extend the D7-brane along $(x^0, x^1, x^2, x^3, r)$ and the three angular directions of the three-sphere corresponding to the one-forms $\omega^1$, $\omega^2$, and $\omega^3$. In addition we can consistently set $\tau={\rm constant}$ and consider the embedding
\beq
\chi\,=\,\chi(r)\,\,.
\eeq
Since we want to add charge in the fundamental representation in the dual theory, we consider a D7-brane with a potential one-form  $A$ on the worldvolume:
\beq
A\,=\,A_t(r)\,dt\,\,.
\label{A_t_charge}
\eeq
Therefore, in order to fix completely the configuration, we have to determine the functions $\chi(r)$ and $A_t(r)$  by solving the equations of motion derived from the DBI action:
\beq
S_{DBI}\,=\,-T_7\,\int d^8\xi\,e^{\phi}\,\sqrt{-\det (g_8+e^{-{\phi\over 2}}\,F)}\,\,,
\eeq
where  $F=dA$ and the factors of $e^{\phi}$ result from working in the Einstein frame.\footnote{Our background contains an RR seven-form $F_7\sim e^{\phi}\,*F_3$ which could contribute to the Wess-Zumino part of the action of the probe D7-brane through a term of the form $\int \hat F_7\wedge A$, where the hat denotes the pullback to the worldvolume.  It turns out, however, that $\hat F_7$ vanishes for our embedding ansatz and therefore this Wess-Zumino term does not contribute to the action of the probe.} The induced metric on the worldvolume for our ansatz is:
\bear
&&ds^2_8\,=\,h^{-{1\over 2}}\,\big[-B(dx^0)^2+(dx^1)^2+(dx^2)^2\,+\,e^{-2\phi}\,(dx^3)^2\big]+
h^{{1\over 2}}\,\Big[{1\over B}+{r^2\over b}\,\chi'^2\Big]dr^2\rc\rc
&&\qquad\qquad\qquad
+{r^2\,h^{{1\over 2}}\over 4\, b}\,\cos^2\chi\,\Big[(\omega^1)^2+(\omega^2)^2\,+\,
\big(1+{1-b\over b}\,\cos^2\chi\big)(\omega^3)^2\,\Big]\,\,,
\label{induced_metric_chi}
\eear
where $\chi'=d\chi/dr$.  The integrand of the DBI reads
\bear
&&e^{\phi}\,\sqrt{-\det (g_8+e^{-{\phi\over 2}}\,F)}\,=\,
{\sin\theta\over 8\,b^{{3\over 2}}}\,r^3\,\cos^3\chi\,
\sqrt{1+{1-b\over b}\,\cos^2\chi}\rc\rc
&&\qquad\qquad\qquad\qquad
\times
\sqrt{1+{r^2\,B\over b}\,\chi'^2\,-\,e^{-\phi}\,A_t'^{\,2}}\,\,,
\label{DBI_det_general}
\eear
where $\theta$ is the polar angle used to represent the one-forms $\omega^i$ (\ref{SU2_oneforms}). The Lagrangian then is:
\beq
L\propto r^3\,\cos^3\chi\,
\sqrt{1+{1-b\over b}\,\cos^2\chi}\,
\sqrt{1+{r^2B\over b}\,\chi'^2-\,e^{-\phi}\,A_t'^{\,2}}\ ,
\label{L_DBI_general}
\eeq
from where one can derive the equations of motion for the embedding and the gauge potential. The equation of motion for $A_t$ can be integrated once, giving
\beq
{r^3\,\cos^3\chi\,\sqrt{\cos^2\chi+b\,\sin^2\chi}\,\,e^{-\phi}\,A_t'
\over \sqrt{1+{r^2B\over b}\,\chi'^2-\,e^{-\phi}\,A_t'^{\,2}}}\,=\,d\,\,,
\label{At_eq}
\eeq
where $d$ is a constant of integration which is proportional to the charge density. It is rather straightforward to solve (\ref{At_eq}) for $A_t'$, namely:
\beq
A_t'\,=\,{e^{\phi}\,\sqrt{1+{r^2B\over b}\,\chi'^2}
\over \sqrt{d^2\,e^{\phi}+r^6\,\cos^6\chi\,\big(\cos^2\chi+b\,\sin^2\chi\big)
}}\,\,d\,\,.
\label{At_d}
\eeq
We can now write the equation of motion for the embedding function $\chi(r)$ and use (\ref{At_d}) to eliminate the worldvolume gauge field in favor of the density $d$:
\bear
&&
\partial_r\,\Bigg[
{r^2\,B\,\sqrt{d^2\,e^{\phi}+r^6\,\cos^6\chi\,\big(\cos^2\chi+b\sin^2\chi\big)}
\over \sqrt{1+{r^2B\over b}\,\chi'^2}}\,\,\chi\,'\Bigg]\qquad\qquad
\rc\rc
&&
+r^6\cos^5\chi\,\sin\chi{b\big(3\,b+4(1-b)\,\cos^2\chi\big)\over 
\sqrt{d^2\,e^{\phi}+r^6\,\cos^6\chi\,\big(\cos^2\chi+b\,\sin^2\chi\big)}}
\sqrt{1+{r^2B\over b}\,\chi'^2}\,=\,0\,\,.
\qquad\qquad
\label{eom_chi_general}
\eear
Notice that (\ref{eom_chi_general}) is trivially satisfied by taking $\chi=0$. This solution corresponds to the so-called massless embedding which we consider in Section~\ref{non_zero_density}. 

Let us now discuss the asymptotic UV behavior of the solutions to (\ref{eom_chi_general}), {\emph{i.e.}}, at large $r$. In this regime we can safely assume that $\chi$ is small, giving us the following differential equation
\beq
\partial_r\,(r^5\,\chi ')\,+\,b(4-b)\,r^3\,\chi\,=\,0\ .
\eeq
An ansatz $\chi\sim r^{\alpha}$ leads to an algebraic equation to be solved for the exponent $\alpha$:
\beq
\alpha(\alpha+4)+b(4-b)\,=\,0\ ,
\eeq
which has two solutions
\be
\alpha\,=\,-b \ , \ -4+b = \left\{ \begin{array}{ll} 
 -{8\over 9} & \\
 -{28\over 9}& \ .\end{array} \right. 
\ee
Therefore, deep in the UV, the angle $\chi$ behaves as:
\beq
\chi\sim {\cal{A}}\ \Big({1\over r^b}+\ldots\Big) + {\cal{B}}\ \Big({1\over r^{4-b}}+\ldots\Big)\ .
\label{UV_chi}
\eeq
The first term corresponds to the leading UV behavior. According to the holographic dictionary, its coefficient ${\cal{A}}$ is related to the mass $m_q$ of the quarks 
introduced by the D7-brane. Moreover, the coefficient ${\cal{B}}$ of the subleading solution determines the vacuum expectation value of the quark-antiquark bilinear operator ${\cal O}_m=\bar\psi\,\psi+\cdots$ (see below). 

In the sections that follow, we analyze different particular solutions of (\ref{eom_chi_general}).  
We begin by analyzing in Section~\ref{SUSY} the case with $T=d=0$, for which we find a simple analytic solution. 
In Appendix~\ref{kappa} we further show that this embedding is supersymmetric.

\section{Vanishing density}\label{tempe}

\subsection{SUSY configuration}\label{SUSY}

Let us consider the D3-D7 background at zero temperature and suppose that the gauge potentials on the worldvolume of the probe D7-brane are vanishing. In this $d=0$, $B=1$ case the embedding equation (\ref{eom_chi_general}) can be solved analytically. Indeed, one can directly verify that $\chi=\chi(r)$ given by:
\beq
\sin\chi(r)\,=\,{m_r\over r^{b}}\,\,,
\label{sin_chi_BPS}
\eeq
where $m_r$ is a constant proportional to the mass (discussed at length below), is a solution of (\ref{eom_chi_general}).  Notice that (\ref{sin_chi_BPS}) behaves in the UV  as in the general equation (\ref{UV_chi}) with ${\cal{B}}=0$. This means that the field theory dual to this solution has vanishing quark-antiquark condensate $\langle {\cal O}_m\rangle$ and that the constant $m_r$ in (\ref{sin_chi_BPS}) is proportional to the quark mass. Interestingly, one can verify that (\ref{sin_chi_BPS}) is the general solution of a first-order differential equation, which in turn can be obtained as the saturation condition  of a BPS bound for the action. In order to prove this statement, let us write the Lagrangian density of this $d=0$, $B=1$ case as:
\beq
{\cal{L}}_0\,=\,r^3\,\cos^3\chi\,
\sqrt{\sin^2\chi+{\cos^2\chi\over b}}\,
\sqrt{1+{r^2\over b}\,\chi'^2}\,\,,
\label{L_DBI_zeroT}
\eeq
where we have omitted the irrelevant prefactor.  Let us now write the Lagrangian (\ref{L_DBI_zeroT}) for a general function $\chi(r)$  as  the following square root of a sum of squares:
\beq
{\cal{L}}_0\,=\,\sqrt{{\cal Z}^2\,+\,\Lambda^2}\,\,,
\eeq
where ${\cal Z}$ and $\Lambda$ are given by:
\bear
&&{\cal Z}\,=\,{1\over 4\sqrt{b}}\,\partial_r\,
\Big[r^4\,\cos^4\chi\Big]\,,\,\,\rc\rc 
&&\Lambda\,=\,r^3\,\cos^4\chi\,\Big[
{r\chi'\over b}\,+\,\tan\chi\Big]\,\,.
\eear
Clearly, ${\cal{L}}_0$ obeys the bound
\beq
L\,\ge\,{\cal Z}\,\,,
\eeq
which, taking into account that ${\cal Z}$  is a total radial derivative,  implies the following bound for the action $S_0=\int dr {\cal{L}}_0$:
\beq
S_0\,\ge\,{1\over 4\sqrt{b}}\,r^4\,\cos^4\chi\Big|_{{\rm boundary}}\,\,.
\label{BPS_bound}
\eeq
This lower bound only depends on the boundary values of the fields and is saturated  when $\Lambda=0$, \ie\, when the following first-order equation holds:
\beq
{r\chi'\over b}\,+\,\tan\chi\,=\,0\,\,,
\label{BPS_ODE}
\eeq
whose integration yields precisely (\ref{sin_chi_BPS}).  Notice that the previous argument shows that (\ref{sin_chi_BPS}) minimizes $S_0$ for fixed boundary conditions at the UV. Thus, it solves the variational problem 
and, therefore, it must fulfill the Euler-Lagrange equations derived from ${\cal{L}}_0$.  Actually, our derivation of (\ref{BPS_ODE}) suggests that (\ref{sin_chi_BPS}) is a supersymmetric worldvolume soliton.  We check this fact directly in Appendix~\ref{kappa} by using the kappa symmetry of the D7-brane probe. 

In this paper we focus mostly on non-supersymmetric configurations. However, the supersymmetric configuration (\ref{sin_chi_BPS}) is quite useful in these studies since it allows us to regulate the thermodynamic functions of the probe at non-zero temperature. For this purpose, it is interesting to notice that the on-shell action for the BPS solution diverges as:
\beq
S_{0}^{on-shell}\sim r_{max}^4-2m_r^2\,r_{max}^{4-2b}\,+\,m_r^4\,r_{max}^{4-4b}\,\,,
\label{on-shell_action_zeroT}
\eeq
as can be easily derived by evaluating the right-hand side of (\ref{BPS_bound}) on the solution (\ref{sin_chi_BPS}) for $r=r_{max}\to \infty$.

\subsection{Finite temperature}

In the rest of this section we restrict ourselves to the case in which the charge density $d$ still vanishes while keeping the temperature $T$ finite. We want to study the thermodynamics of the probe D7-brane. For that purpose it is quite convenient to introduce  a new isotropic radial variable $u$, similar to the one defined in \cite{Mateos:2007vn,Jokela:2012dw}. We define $u$ by means of the following differential relation with $r$:
\beq
{dr\over r\sqrt{B(r)}}\,=\,{1\over b}\,{du\over u}\,\,.
\label{r_u_diff}
\eeq
 This equation can be readily integrated as:
 \beq
 u^{{5\over 3b}}\,=\,\Big({r\over r_h}\Big)^{{5\over 3}}\,+\,
 \sqrt{\Big({r\over r_h}\Big)^{{10\over 3}}\,-\,1 }\,\,.
 \label{u_r}
 \eeq
In this new variable the horizon is located at $u=1$.  Notice also that asymptotically in the UV:
\beq
 u\approx 2^{{3b\over 5}}\,\Big({r\over r_h}\Big)^b\,, \qquad\qquad  
  \ r\to\infty .
\eeq
 Eq. (\ref{u_r})  can be easily inverted as:
 \beq
 \Big({r\over r_h}\Big)^{{5\over 3}}\,=\,{1\over 2}\,
 \Big[u^{{5\over 3b}}\,+\,u^{-{5\over 3b}}\Big]\,=\,
 {1\over 2}\,u^{{5\over 3b}}\,\tilde f(u)\,\,,
 \label{r_u}
 \eeq
where $\tilde f(u)$ is defined as:
\beq
\tilde f(u)\,=\,1\,+\,u^{-{10\over 3b}}\,=\,1+u^{-{15\over 4}}\,\,.
\label{tilde_f}
\eeq
 Let also define $f(u)$ as:
 \beq
  f(u)\,=\,1\,-\,u^{-{10\over 3b}}\,=\,1-u^{-{15\over 4}}\,\,.
 \label{no_tilde_f}
\eeq
The emblackening factor $B$ in terms of $f$ and $\tilde f$ can be written as:
 \beq
 \sqrt{B}\,=\,{f\over \tilde f}\,\,.
 \eeq
Let us now use these results to write the 10d metric in the variable $u$. The internal metric $d\hat s_5^2$ is the same as in (\ref{compact_flavored_metric}). The non-compact part $d s_5^2$ takes the form:
 \beq
 ds_5^2\,=\,{r_h^2\over 2^{{6\over 5}}\,R^2}\,u^{{2\over b}}\,\tilde f^{{6\over 5}}\,
 \Big[-{f^2\over \tilde f^2}(dx^0)^2\,+\,(dx^1)^2\,+\,(dx^2)^2\,+\,e^{-2\phi}\,(dx^3)^2\Big]\,+\,
 {R^2\over b^2}\,{du^2\over u^2}\,\,,
 \eeq
with the dilaton $\phi$ given by:
 \beq
 e^{\phi}\,=\,\Big({3\over 4 Q_f}\Big)^{{2\over 3}}\,{r_h^{{2\over 3}}\over 2^{{2\over 5}}}\,
 u^{{2\over 3b}}\,\tilde f^{{2\over 5}}\,\,.
 \eeq
Let us  consider now an embedding ansatz for the D7-brane in which $\tau$ is constant and  $\chi=\chi(u)$. The induced metric is:
\bea
  ds_8^2 &  = & {r_h^2\over 2^{{6\over 5}}R^2}u^{{2\over b}}\tilde f^{{6\over 5}} \Big[-{f^2\over \tilde f^2}(dx^0)^2+(dx^1)^2+(dx^2)^2+e^{-2\phi}(dx^3)^2\Big]+ {R^2\over b^2u^2}\big[1+b u^2\dot \chi^2\big] du^2 \nonumber \\
  & & +{R^2\over 4\, b}\,\cos^2\chi\,\Big[(\omega^1)^2+(\omega^2)^2+\big(1+{1-b\over b}\,\cos^2\chi\big)(\omega^3)^2\,\Big]\ ,
\eea
where $\dot\chi=d\chi/du$.  Since we are considering the case in which the charge density vanishes, the worldvolume gauge field is also zero. Therefore,  the Lagrangian for the probe in the new radial variable is:
 \beq
 {\cal{L}}\sim u^{{4\over b}\,-\,1}\,\tilde f^{{7\over 5}}\,f\,
 \cos^3\chi\,
\sqrt{1+{1-b\over b}\,\cos^2\chi}\,
\sqrt{1+b\, u^2\dot \chi^2 }\,\,,
\eeq
from which the equation of motion can be readily obtained,
\bear
&&{\partial\over \partial u}\,\,
\Bigg[u^{{4\over b}+1}\,\tilde f^{{7\over 5}}\,f \,\cos^3\chi\,
{\sqrt{1+{1-b\over b}\,\cos^2\chi}\over \sqrt{1+b\, u^2\dot \chi^2 }}\,\dot\chi\Bigg] \rc\rc\rc
&&+{1\over b^2}\,u^{{4\over b}-1} \tilde f^{{7\over 5}}\,f \,\cos^2\chi\,\sin\chi\,
{3\,b+4(1-b)\,\cos^2\chi\over 
\sqrt{1+{1-b\over b}\,\cos^2\chi}}\,
\sqrt{1+b\, u^2\dot \chi^2 }\,=\,0\,\,.\qquad\qquad
\label{eom_chi_u}
\eear
Let us now study the UV behavior of the solutions of (\ref{eom_chi_u}). By expanding $\chi$ in power series around 
$u=\infty$ we can solve (\ref{eom_chi_u}) asymptotically. We get:
\beq
\chi\,=\,{m\over u}\,+\,{m^3\over 6\,u^3}\,+\, {c\over u^{{4\over b}-1}}+\ldots \ ,
\label{chi_UV}
\eeq
where $m$ and $c$ are  related to the quark  mass and condensate, respectively. The precise relation is worked out in detail in Appendix \ref{sec:dictionary}. It turns out that the bare quark mass $m_q$ can be written in terms of the parameter $m$ as
\beq
 m_q\,\propto 
 \Bigg({Q_c^2\over Q_f}\Bigg)^{{1\over 3}}\,T^{{4\over 3}}\,m^{{3\over 2}}\,\,, \label{m_q_dict}
\eeq
where, in units in which $\alpha'=g_s=1$,  the proportionality constant is a pure number (\ref{mass_dictionary}). It follows from (\ref{m_q_dict}) that, for fixed values of the physical parameters  $Q_c$, $Q_f$, and $m_q$, we have $m\sim T^{-b}$ and thus small (large) $m$ corresponds to large (small) temperature. Similarly, we have the following relation between the quark-antiquark condensate $\langle {\cal O}_m\rangle$ and the parameter $c$:
\beq
\langle {\cal O}_m\rangle \propto  Q_c^{{14\over 9}}\,Q_f^{{2\over 9}}\,m_q^{-{1\over 3}}\,
T^{{28\over 9}}\,c\,\,,
\label{cond_dict}
\eeq
where the coefficient $c$ is a pure number in our units (\ref{Om_c}). 
 
As in \cite{Mateos:2007vn}, given the UV limit (\ref{chi_UV}), there are two types of embeddings depending on how the brane behaves at the IR.  At low temperature (or large mass parameter $m$) the D7-brane probe closes off outside the horizon and we have a so-called Minkowski embedding. On the other hand, if the temperature is large enough (small $m$) the probe terminates at the horizon and we have a so-called black hole embedding. For some intermediate value of $m$ there is a phase transition between these two types of configurations. Due to the different boundary conditions they satisfy at the IR it is quite convenient to choose different variables to analyze these two types of embeddings. We do it separately in the two subsections that follow. In Appendix \ref{CE} we study the critical embeddings close to the transition, while in Appendix \ref{TS} we analyze the thermal screening of the quark-antiquark potential.

\subsubsection{Black hole embeddings}\label{sec:BHembeddings}

In order to study the black hole embeddings of the probe, let us introduce the variable $\eta$, defined as:
\beq
 \eta\,=\,\sin\chi\,\,.\label{eta-def}
\eeq
We parametrize our embeddings by a function $\eta=\eta(u)$. Since 
\beq
 \dot\chi\,=\,{\dot\eta\over \sqrt{1-\eta^2}}\,\,,
\eeq
the induced metric on the worldvolume of the D7-brane for this parametrization  becomes: 
\bea
 ds_8^2 & = & {r_h^2\over 2^{{6\over 5}}\,R^2}\,u^{{2\over b}}\,\tilde f^{{6\over 5}}\, \Big[-{f^2\over \tilde f^2}(dx^0)^2\,+\,(dx^1)^2\,+\,(dx^2)^2\,+\,e^{-2\phi}\,(dx^3)^2\Big] \nonumber \\
 && +{R^2\over b^2\,u^2\,(1-\eta^2)}\,\Big[1-\eta^2+b\,u^2\,\dot\eta^2\Big]\,du^2 \nonumber \\
&& +{R^2\over 4b}\,(1-\eta^2) \Big[(\omega^1)^2+(\omega^2)^2\,+\,{1+(b-1)\,\eta^2\over b}(\omega^3)^2\,\Big]\,\,.
\eea
It follows that, in these variables, we have for the DBI determinant:
\beq
 e^{\phi}\,\sqrt{-\det g_8}\,=\,{r_h^4\over 32\,\cdot 2^{{2\over 5}}\,b^3}\,
 u^{{4\over b}-1}\,\tilde f^{{7\over 5}}\,f \,(1-\eta^2)\,
 \sqrt{1+(b-1)\,\eta^2}\,\sqrt{1-\eta^2\,+\,b\,u^2\,\dot\eta^2}\,\,.
\eeq
Let us now obtain the reduced action of the probe, denoted by ${\cal I}_{bulk}$, defined as:
\beq
 {\cal I}_{bulk}\,=\,-\,{S\over {\cal N}\,V_4}\,\,,
\eeq
where $S$ is the action, $V_4$ is the (infinite) volume of the 4d Minkowski spacetime and ${\cal N}$ is:
 \beq
{\cal N}\,=\,{\pi^2\,r_h^4\over 2^{{7\over 5}}\,b^3}\,T_{D7}\,=\,{2\cdot 2^{{2\over 3}}\over 5^6\cdot \pi}\,Q_c^2\,T^4\,\,.
\label{calN_eta}
\eeq
The explicit form of $ {\cal I}_{bulk}$ is given by
\be
  {\cal I}_{bulk}\,=\,\int du\,   u^{{4\over b}-1}\,\tilde f^{{7\over 5}}\,f \,(1-\eta^2)\, \sqrt{1+(b-1)\,\eta^2}\,\sqrt{1-\eta^2\,+\,b\,u^2\,\dot\eta^2}\,\,.\label{I_bulk}
\ee
Taking into account (\ref{eta-def}) and that $\sin\chi\approx \chi-\chi^3/6+\ldots$, we obtain from (\ref{chi_UV}) the UV behavior of $\eta$:
\beq
\eta\,=\,{m\over u}\,+\,\,{c\over u^{{7\over 2}}}\,+\,\ldots\,\,.\label{eta_UV}
\eeq
This behavior can be directly obtained by solving the equation of motion  derived from (\ref{I_bulk}) as a power expansion around the UV.  Notice that the term $u^{-3}$ is not present in the UV expansion of $\eta$. 

\begin{figure}[ht]
\center
 \includegraphics[width=0.50\textwidth]{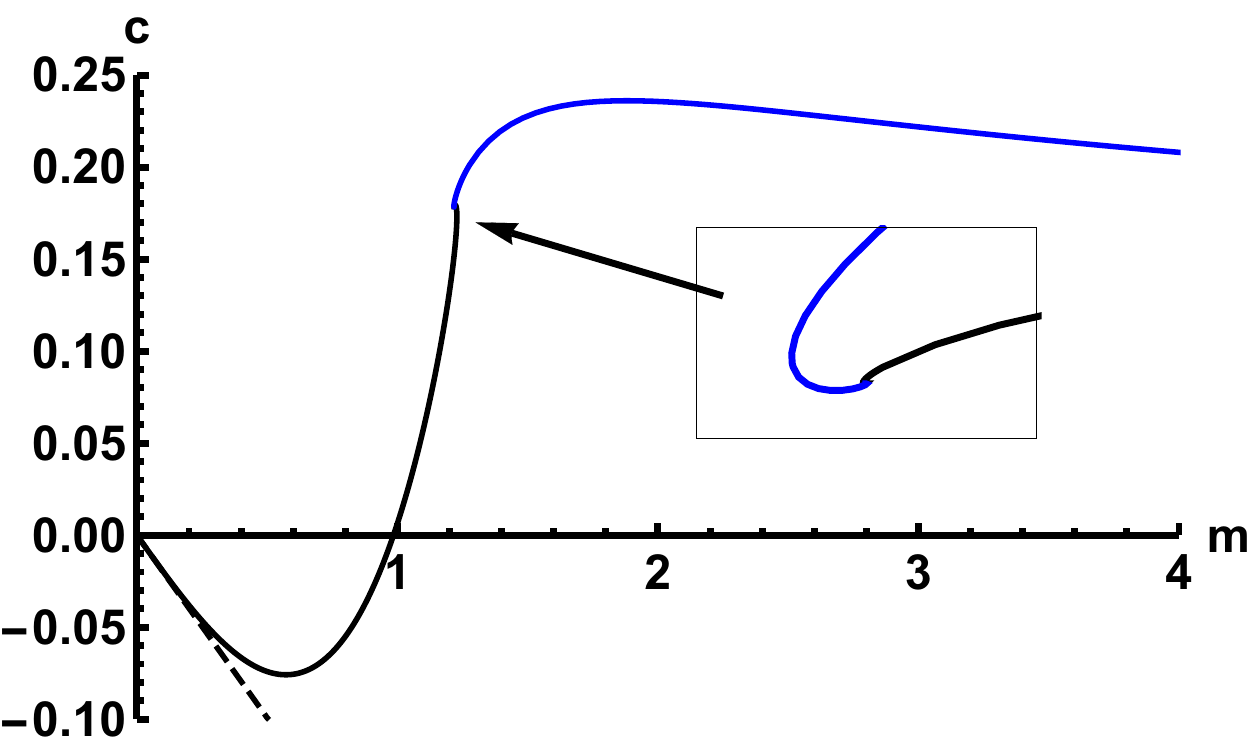}
  \caption{In this figure we plot the condensate parameter $c$ versus the mass parameter $m$. The solid curves have been found numerically both for black hole embeddings (black curve) and Minkowski embeddings (blue curve). The transition between them occurs at $m\approx 1.23$. The condensate $c$ vanishes for a black hole embedding at approximately $m\approx 0.985$.  The dashed linear curve near $m=0$ corresponds to the high temperature analytic result (\ref{c_m_highT}).  The condensate of the Minkowski embeddings with  $m\to\infty$  tends to zero as $c\propto m^{-{14\over 3}}$.  } 
  \label{fig:c_vs_m}
\end{figure}

In order to get the embedding function $\eta(u)$ in the whole range of the holographic coordinate $u$ we must solve numerically the Euler-Lagrange equation of motion derived from the action (\ref{I_bulk}). For a black hole embedding we must impose regularity conditions at the horizon $u=1$, which are easily seen to be:
\beq
\eta(u=1)\,=\,\eta_h\,\,,
\qquad\qquad
\dot\eta(u=1)\,=\,0\,\,,
\eeq
where $\eta_h$ is an IR constant which determines the UV constants $m$ and $c$. By varying $\eta_h$ we get a relation $c=c(m)$, which determines the condensate as a function of the quark mass at fixed temperature or, equivalently, the condensate as a function of the temperature at fixed $m_q$. The corresponding results are plotted in Fig.~\ref{fig:c_vs_m}. We notice that $c$ is negative for small $m$ (or large $T$), vanishes near $m\approx 1$ and becomes positive for larger values of $m$. For high temperatures (or low $m$) the embedding function $\eta$ remains small for all values of the holographic coordinate $u$ and the equation for the embedding can be linearized and solved analytically. This analysis is presented in detail in Appendix~\ref{HighT_BH_embed}.  We find that $c$ decreases linearly with $m$ ($c\approx -0.198m$, see (\ref{c_m_highT})).  This result is in good agreement with the numerical results, as shown in Fig.~\ref{fig:c_vs_m}.

We want now to address the problem of determining the thermodynamic functions of the probe. The first step in this analysis is finding the free energy density $F$ of the D7-brane. The expression of $F$ can be obtained from the Euclidean on-shell action of the brane, which is divergent and should be regulated appropriately by adding a suitable boundary term. Indeed,  plugging the UV behavior ({\ref{eta_UV}) into the right-hand side of (\ref{I_bulk}}),  we find that $ {\cal I}_{bulk}$  diverges in the UV as:
\bea
 {\cal I}_{bulk}\Big|_{div} & = & \int^{u_{max}}du\Big[u^{{7\over 2}}\,+\,{2\over 5}\,u^{-{1\over 4}}\,-\,{10\over 9}\,m^2\,u^{{3\over 2}}\,+\,{m^4\over 9}\,u^{-{1\over 2}}\Big]\nonumber \\
& = & {2\over 9}\,u_{max}^{{9\over 2}}\,+\,{8\over 15}\,u_{max}^{{3\over 4}}\,-\,{4\,m^2\over 9}\,u_{max}^{{5\over 2}}\,+\,{2\over 9}\,m^4\,u_{max}^{{1\over 2}}\,\,.\label{I_bulk_div}
\eea
It is quite instructive to compare (\ref{I_bulk_div}) to the behavior (\ref{on-shell_action_zeroT}) for the SUSY embedding at zero temperature.  The three terms in  (\ref{on-shell_action_zeroT})  correspond to the first, third, and fourth terms in  (\ref{I_bulk_div}) (take into account that $r_{max}\sim u_{max}^{9/8}$). The second term in 
(\ref{I_bulk_div})  is independent  of the embedding and is due to the finite temperature. 

To construct a boundary action that cancels the divergences of  ${\cal I}_{bulk}$ we first consider the boundary metric at $u=u_{max}$
\beq
 ds^2(\gamma)\,\sim\,u_{max}^{{2\over b}}\,\tilde f^{{6\over 5}}(u_{max})\,
 \Big[-{f^2(u_{max})\over \tilde f^2(u_{max})}(dx^0)^2\,+\,(dx^1)^2\,+\,(dx^2)^2\,+\,e^{-2\phi}\,(dx^3)^2\Big]\,\,.
\eeq
It follows that:
\beq
 e^{\phi}\,\sqrt{-\det \gamma}\,=\,u_{max}^{{4\over b}}\tilde f^{{7\over 5}}(u_{max})\,
 f(u_{max})\,\,.
\eeq
We want the renormalized action to vanish for the SUSY embeddings at $T=0$.  To fulfill this requirement we notice that the on-shell action of such embeddings depends on $\eta$ as $\cos^4\chi=(1-\eta^2)^2$ evaluated at the boundary (see (\ref{BPS_bound})). Taking this fact into account,   let us write the boundary action as:
 \beq
 {\cal I}_{bdy}\,=\,\alpha\,f^{\beta_1}\,\tilde f^{\beta_2}\, e^{\phi}\,\sqrt{-\det \gamma}\,\, (1-\eta^2)^2\,\Big|_{u=u_{max}}\,\,,\label{I_bdy_ansatz}
 \eeq
where $\alpha$, $\beta_1$, and $\beta_2$ are constants to be determined by imposing the condition that 
 $ {\cal I}_{bulk}+{\cal I}_{bdy}$ is finite as $u_{max}\to\infty$. We have included in (\ref{I_bdy_ansatz}) a prefactor containing powers of the functions $f$ and $\tilde f$ to take into account the redshift occurring at $T\not=0$ due to the emblackening factor of the metric. Actually, this term is needed  to cancel the second term on the right-hand side of (\ref{I_bulk_div}).  Plugging in the UV behavior of the embedding function $\eta$ (eq. (\ref{eta_UV})), we find
\beq
{\cal I}_{bdy}\,=\,\alpha\,\Big[u_{max}^{{9\over 2}}\,+\,\Big({2\over 5}\,-\,\beta_1+\beta_2\Big)\,
u_{max}^{{3\over 4}}\,-\,2 m^2\,u_{max}^{{5\over 2}}\,+\,m^4\,u_{max}^{{1\over 2}}\,-\,4\,m\,c\,\Big]\,\,.
\eeq
By requiring  the cancellation of the leading term of  $ {\cal I}_{bulk}+{\cal I}_{bdy}$, we fix the value of the constant $\alpha$ to be:
\beq
\alpha\,=\,-{2\over 9}\,\,.
\label{alpha_bdy_action}
\eeq
Moreover, the cancellation of the $u_{max}^{{3\over 4}}$ term requires that:
\beq
\beta_2-\beta_1\,=\,2\,\,.
\label{beta_eq}
\eeq
Using these values of the constants, we can write ${\cal I}_{bdy}$ as:
\beq
 {\cal I}_{bdy}\,=\,-{2\over 9}\,u_{max}^{{1\over 2}}\,
 (u_{max}^2-m^2)^2 -\,{8\over 15}\,u_{max}^{{3\over 4}}\,+\,{8\over 9}\,m\,c\,\,. \label{I_bdy_explicit}
\eeq
Notice also that taking  the solution $\beta_2=-\beta_1=1$ of (\ref{beta_eq}), the boundary action can be written in terms of the emblackening factor $B$ as:
 \beq
 {\cal I}_{bdy}\,=\,-{2\over 9}\,
 {e^{\phi}\over \sqrt{B}}\,\sqrt{-\det \gamma}\,\,
 (1-\eta^2)^2\,\Big|_{u=u_{max}}\,\,.
 \eeq
Let us rewrite the divergent terms on the right-hand side of (\ref{I_bdy_explicit}) as:
\bea
 & & -{2\over 9}u_{max}^{{1\over 2}} (u_{max}^2-m^2)^2 = -{1\over 9}\int_{u_{min}}^{u_{max}} {(u^2-m^2)(9u^2-m^2) \over u^{{1\over 2}}}du- {2\over 9}u_{min}^{{1\over 2}} (u_{min}^2-m^2)^2  \nonumber \\
 & & -{8\over 15}\,u_{max}^{{3\over 4}}\,=\, -{2\over 5}\,\int_{u_{min}}^{u_{max}}\,u^{-{1\over 4}}\,du\, -{8\over 15}\,u_{min}^{{3\over 4}}\ .
\eea
Using this result, we can write $ {\cal I}_{bdy}$ as the following integral:
\bea
{\cal I}_{bdy} & = & -\int_{u_{min}}^{u_{max}}{du\over \sqrt{u}} \Big[{1\over 9}\,(u^2-m^2)(9u^2-m^2)  +{2\over 5}u^{{1\over 4}}\Big] \rc\rc
 & & -{2\over 9}\,u_{min}^{{1\over 2}}\, (u_{min}^2-m^2)^2\,-\,   {8\over 15}\,u_{min}^{{3\over 4}}\,+\,{8\over 9}\,m\,c\,\,.\label{eq:BHregulator}
\eea
The free energy density of the probe is given by:
\bea
 {F\over {\cal N}} & = &  {\cal I}_{bulk}\,+\, {\cal I}_{bdy}\nonumber \\
 & = & {\cal G}(m)-{2\over 9}u_{min}^{{1\over 2}} (u_{min}^2-m^2)^2 -{8\over 15}u_{min}^{{3\over 4}}+{8\over 9}\,m\,c\,\,,   \label{F_bh}
\eea
where $ {\cal G}(m)$ is the following integral:
\bear
  &&{\cal G}(m)=\int_{u_{min}}^{\infty}{du\over \sqrt{u}}  \Big[u^4\,f\,\tilde f^{{7\over 5}}\,(1-\eta^2)\,\sqrt{1-{\eta^2\over 9}}\,  \sqrt{1-\eta^2+{8u^2\over 9}\,\dot\eta^2}\rc\rc
   &&\qquad\qquad\qquad\qquad\qquad\qquad   -\,{1\over 9}(u^2-m^2)(9u^2-m^2) \,-\,{2\over 5}\,u^{{1\over 4}}\Big]\ .
   \label{cal_G_BH_embed}
\eear
We have extended the upper limit of integration to $\infty$ since it is a convergent integral.  For black hole embeddings, $u_{\min}=1$ in (\ref{F_bh}) and (\ref{cal_G_BH_embed}).

\begin{figure}[ht]
\center
 \includegraphics[width=0.45\textwidth]{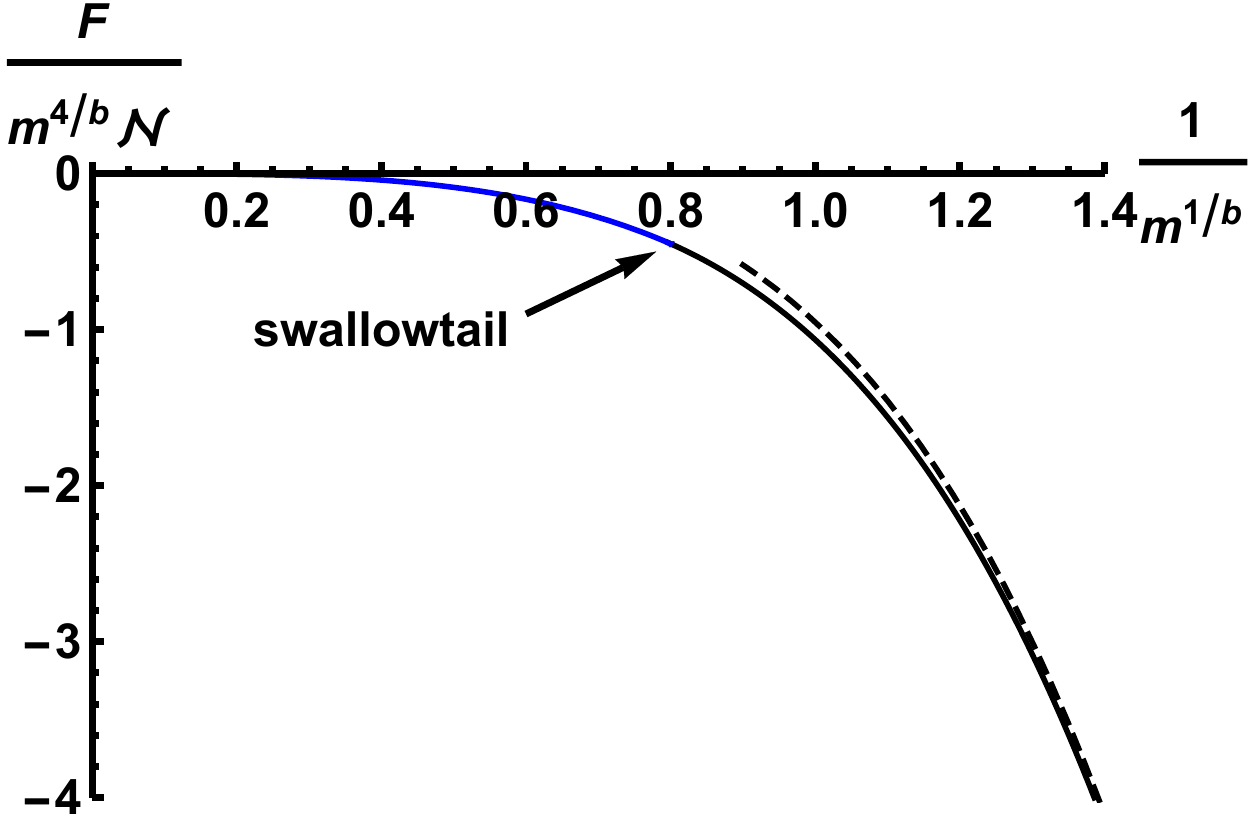}
 \qquad
 \includegraphics[width=0.45\textwidth]{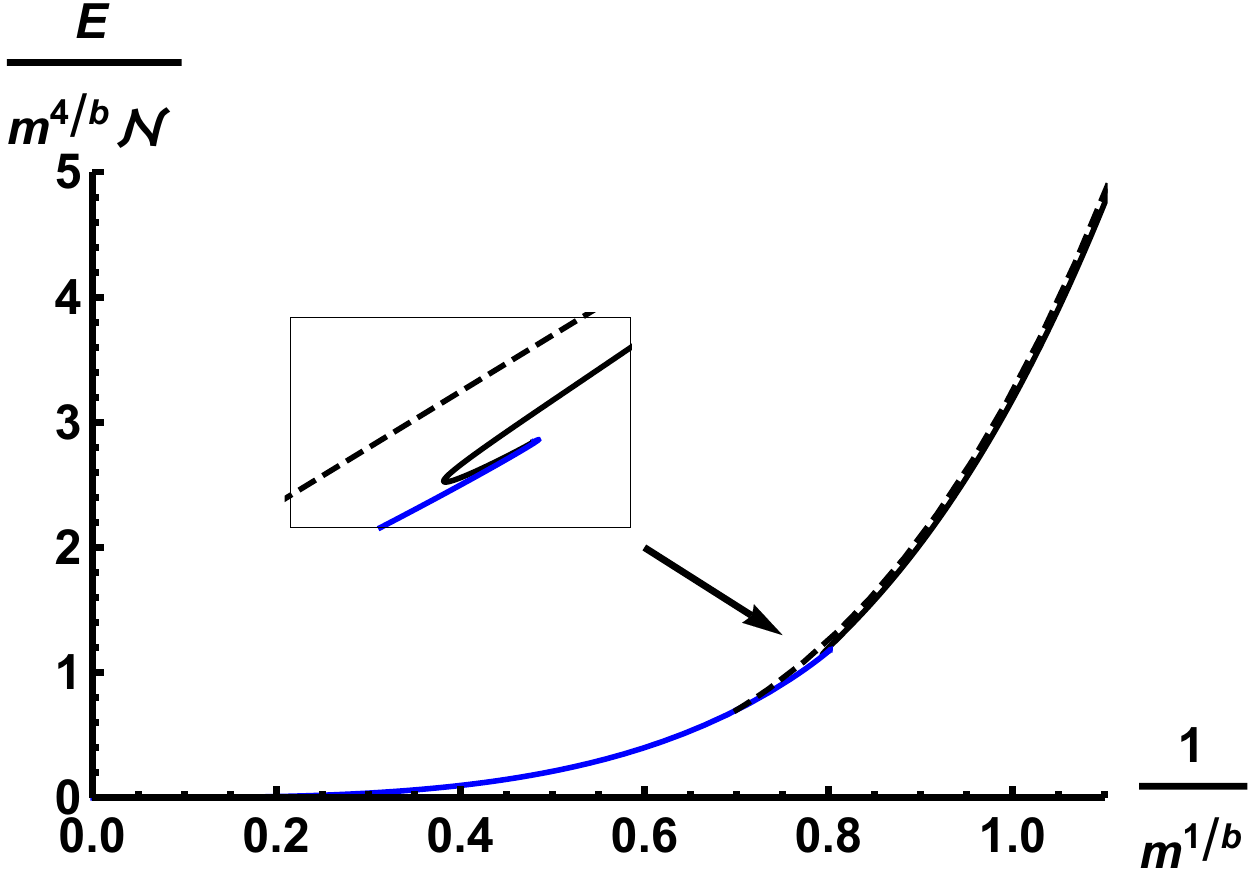}
 \\ \includegraphics[width=0.45\textwidth]{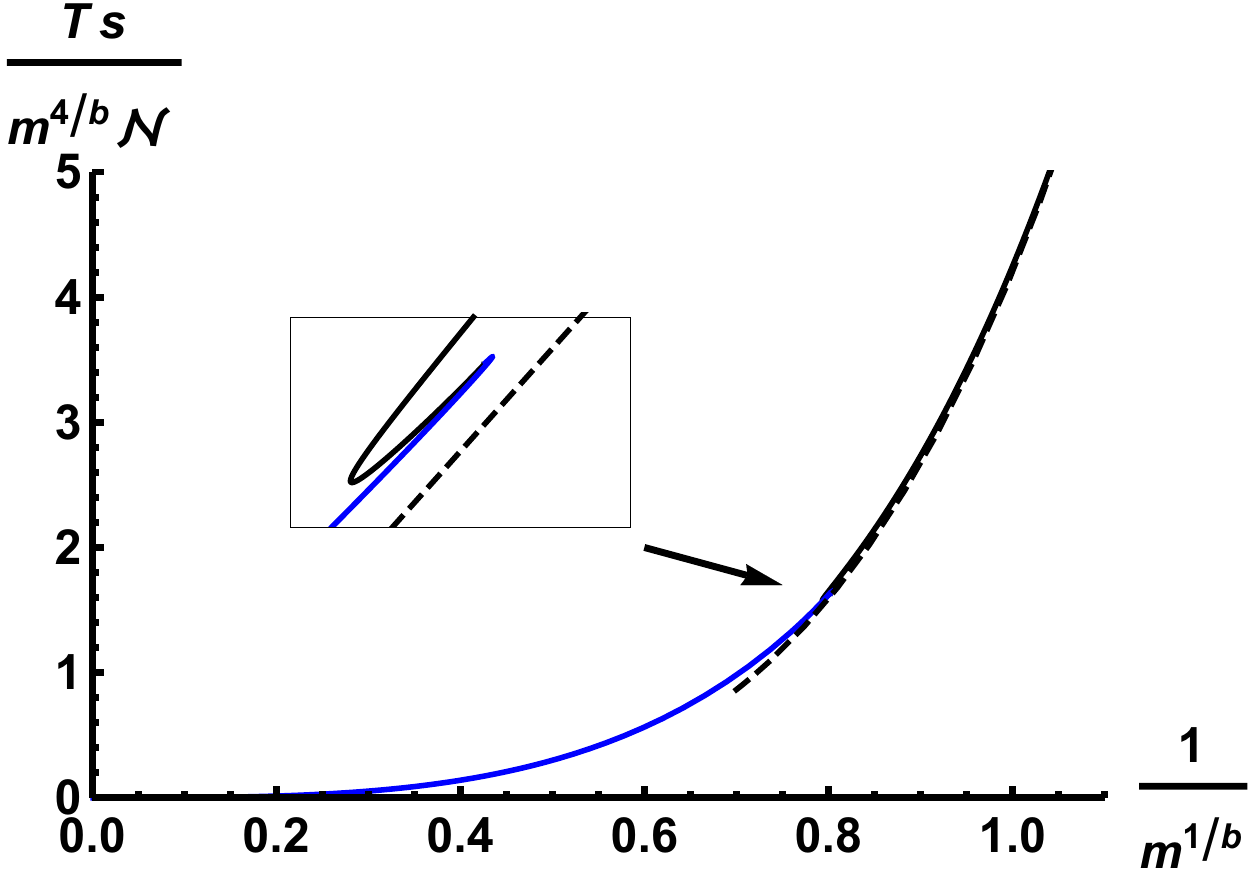}
 \caption{In these plots we represent the free energy density, internal energy density, and entropy density  as a function of $m^{-{1\over b}}\sim T$.  Notice that the normalization factor $m^{{4\over b}}\,{\cal N}$ is independent of the temperature. The continuous black (blue) curve corresponds to black hole (Minkowski) embeddings. Near the transition point the expected swallowtail behavior is obtained. The dashed curves are the result of the analytic high temperature calculation of Appendix~\ref{HighT_BH_embed}.}\label{fig:F_E_S_vsT}
\end{figure}

Once $F$ is known, we can obtain the other thermodynamic functions by using the standard relations between them. For example,  the entropy density $s$ of the probe  can be  computed as:
\beq
s\,=\,-{\partial F\over \partial T}\,=\,-{F\over {\cal N}}\,{\partial {\cal N}\over \partial T}\,-\,{\cal N}\,
{\partial\over \partial T}\Big({F\over {\cal N}}\Big)\,\,.
\eeq
Since ${\cal N}\sim T^4$, $\partial_T {\cal N}=4{\cal N}/T$ and therefore
\beq
s\,=\,-{4\,F\over  T}\,-\,{\cal N}\,{\partial\over \partial T}\Big({F\over {\cal N}}\Big)\,\,.
\eeq
Let us now calculate the second term in this expression
\beq
{\partial\over \partial T}\Big({F\over {\cal N}}\Big)\,=\,
{\partial\over \partial m}\Big({F\over {\cal N}}\Big)\,{\partial m\over \partial T}\,=\,-
{b\,m\over T}\,{\partial\over \partial m}\Big({F\over {\cal N}}\Big)\,\,,
\eeq
where we took into account that, for fixed quark mass $m_q$, the mass parameter $m$ behaves as $m\approx T^{-b}$ (\ref{mass_dictionary}). The derivative of $F/{\cal N}$ with respect to $m$ can be extracted from (\ref{mass_derivative_F}), yielding
\beq
T\,{ s\over {\cal N}}\,=\,-4\,{F\over {\cal N}}\,-\,{160\over 81}\,c\,m\,\,, \label{S_F}
\eeq
where $F/{\cal N}$ can be obtained from (\ref{F_bh}) for black hole embeddings or from (\ref{F_Min}) for Minkowski embeddings. The internal energy $E$ is given by $E=F+Ts$:
\beq
{ E\over {\cal N}}\,=\,-3\,{F\over {\cal N}}\,-\,{160\over 81}\,c\,m\,\,.
\label{E_F}
\eeq
Let us next compute the heat capacity, defined as:
\beq
c_v\,=\,{\partial E\over \partial T}\,\,.
\eeq
Proceeding as with $F$ to compute this derivative, we get:
\bea
T\,{c_v\over {\cal N}} & = &-12\,{F\over {\cal N}}\,\,-\,{160\over 729}\,\Big[1\,-\,8\,{\partial \log c\over \partial\log m}\Big]\,c\,m  \\ \label{cv_F}
& = & 3\,T\,{s\over {\cal N}}\,+\,{320\over 729}\Big[13\,+\,4\,{\partial \log c\over \partial\log m}\Big]\,c\,m \ .
\eea
From the behavior of $F$ at high temperature obtained in Appendix~\ref{HighT_BH_embed}, (eq. (\ref{high_T_F_subleading})), it is immediate to find the leading behavior of $s$, $E$, and $c_v$:
\beq
\lim_{m\to 0}\,T\,{ s\over {\cal N}}\,=\,{32\over 9}\,2^{{2\over 5}}\,\,,
\qquad\qquad
\lim_{m\to 0}\,{ E\over {\cal N}}\,=\,{8\over 3}\,2^{{2\over 5}}\,\,,
\qquad\qquad
\lim_{m\to 0}\,T\,{c_v\over {\cal N}}\,=\,{32\over 3}\,2^{{2\over 5}}\,\,.
\eeq
We have checked this high temperature  behavior numerically (see Fig.~\ref{fig:F_E_S_vsT}).

\subsubsection{Minkowski embeddings}
The $\eta=\eta(u)$ parametrization is convenient for the black hole embeddings. For Minkowski embeddings,  Cartesian-like coordinates $(\rho, P)$ are better suited. They are defined in terms of $u$ and $\chi$ as follows:
\beq
P\,=\,u\sin\chi\,\,,
\qquad\qquad
\rho\,=\,u\cos\chi\,\,.
\label{P_rho_def}
\eeq
The inverse relation is:
\beq
u\,=\,\big[\rho^2\,+\,P^2\big]^{{1\over 2}}\,\,,
\qquad\qquad
\tan\chi\,=\,{P\over \rho}\,\,.
\eeq
When using these variables, we parametrize our embeddings by a function $P=P(\rho)$, then
  \beq
   {dr\over r\,\sqrt{B}}\,=\,{1\over b}\,{du\over u}\,=\,{1\over b}\,
 {\rho+P\,P'\over \rho^2\,+\,P^2}\,d\rho\,\,,
 \qquad\qquad
 d\chi\,=\,{\rho\,P'\,-\,P\over \rho^2+P^2}\,d\rho\,\,,
 \eeq
where $P'=dP/d\rho$.  Let us now write down the induced metric on the worldvolume. Taking into account that:
\beq
 {h^{{1\over 2}}\,dr^2\over B}\,+\,h^{{1\over 2}}e^{2g}\,(d\chi)^2\,=\,
 {R^2\over b^2(\rho^2+P^2)^2}\,
 \Big[(\rho+P\,P')^2\,+\,b\,(\rho\,P'-P)^2\Big]\,d\rho^2\,\,,
\eeq
as well as:
\beq
 r\,=\,{r_h\over 2^{{3\over 5}}}\,\big[\rho^2+P^2\big]^{{1\over 2b}}\,\tilde f^{{3\over 5}}\,\,,
 \qquad\qquad
 \cos^2\chi\,=\,{\rho^2\over \rho^2+P^2}\,\,,
 \eeq
we get:
\bea
 ds_8^2 & = & {r_h^2\over 2^{{6\over 5}}\,R^2}\big[\rho^2+P^2\big]^{{1\over b}}\,\tilde f^{{6\over 5}} \Big[-{f^2\over \tilde f^2}(dx^0)^2+(dx^1)^2+(dx^2)^2+e^{-2\phi}\,(dx^3)^2\Big]\rc\rc
 & & +{R^2\over b^2(\rho^2+P^2)^2}\Big[(\rho+P\,P')^2\,+\,b\,(\rho\,P'-P)^2\Big]d\rho^2\rc\rc
 & & + {R^2\over 4b}\,{\rho^2\over \rho^2+P^2}\Big[(\omega^1)^2+(\omega^2)^2\,+\,\Big(1+{1-b\over b}\,{\rho^2\over \rho^2+P^2}\Big)(\omega^3)^2\,\Big]\,\,.
\eea
In this expression $\tilde f$ and $f$ should be understood as the functions:
\beq
\tilde f\,=\,1+(\rho^2+P^2)^{-{15\over 8}}\,\,,
\qquad\qquad
 f\,=\,1-(\rho^2+P^2)^{-{15\over 8}}\,\,.
\eeq
 It follows that the DBI Lagrangian density in these variables is:
  \beq
 {\cal L}\,\sim\,-\rho^3\,\tilde f^{{7\over 5}}\,f\,(\rho^2+P^2)^{-{3\over 4}}\,
 \Big(\rho^2+{8\over 9}\,P^2\Big)^{{1\over 2}}\,
 \Big[(\rho+P\,P')^2\,+\,{8\over 9}\,(\rho\,P'-P)^2\Big]^{{1\over 2}}\,\,.
 \eeq
 Let us now study the free energy $F$ in these variables. First of all, let us write the bulk on-shell action as:
\beq
{\cal I}_{bulk}\,=\,\int_0^{\rho_{max}}\,d\rho\,\rho^3\,
\,\tilde f^{{7\over 5}}\,f\,(\rho^2+P^2)^{-{3\over 4}}\,
 \Big(\rho^2+{8\over 9}\,P^2\Big)^{{1\over 2}}\,
 \Big[(\rho+P\,P')^2\,+\,{8\over 9}\,(\rho\,P'-P)^2\Big]^{{1\over 2}}\,\,.
 \eeq
This action diverges at the UV. To characterize this divergence, we use the UV behavior of the embedding function 
$P(\rho)$:
\beq
P = m\,+\,{c\over \rho^{{5\over 2}}}\,+\,\ldots\,\,,
\qquad\qquad
(\rho\to\infty)\,\,.
\eeq
Following similar steps as in Section~\ref{sec:BHembeddings}, it is straightforward to find the regulator:
\beq
 {\cal I}_{bdy}\,=\,-\int_0^{\rho_{max}}d\rho{\rho^3\,\tilde f_m^{{7\over 5}}\over
\big (\rho^2+m^2\big)^{{3\over 4}}}\Bigg(
\rho^2\,+\,{8\over 9}\,m^2\,-\,
{\rho^2\,-\,{8\over 9}\,m^2\over 
\big (\rho^2+m^2\big)^{{15\over 8}}}\,\Bigg)\,-\,{m\,c\over 9}\,\,.\label{eq:MNregulator}
\eeq
Alternatively, one can simply perform the change of variables on (\ref{eq:BHregulator}) to land on (\ref{eq:MNregulator}).
It follows that the free energy $F$ can be represented as:
\bea
{ F\over {\cal N}} & = & \int_0^{\infty}d\rho\,\rho^3\,\Bigg[\,{\tilde f^{{7\over 5}}\,f\,\over (\rho^2+P^2)^{{3\over 4}}}\, \Big(\rho^2+{8\over 9}\,P^2\Big)^{{1\over 2}}\,
 \Big[(\rho+P\,P')^2\,+\,{8\over 9}\,(\rho\,P'-P)^2\Big]^{{1\over 2}} \rc\rc
 & &  -\,{\tilde f_m^{{7\over 5}}\over \big (\rho^2+m^2\big)^{{3\over 4}}}\Bigg(\rho^2\,+\,{8\over 9}\,m^2\,-\,{\rho^2\,-\,{8\over 9}\,m^2\over \big (\rho^2+m^2\big)^{{15\over 8}}}\,\Bigg)\,\Bigg]-\,{m\,c\over 9}\ . \label{F_Min}
\eea
As a  highly non-trivial check of this expression,  we have verified numerically  that $ F$ vanishes when $m\to\infty$
(see Fig.~(\ref{fig:F_E_S_vsT})).  The other thermodynamic functions $S$, $E$, and $c_v$ can be obtained from the free energy $F$ by using (\ref{S_F}), (\ref{E_F}), and (\ref{cv_F}), respectively. The corresponding numerical values are presented in Fig.~\ref{fig:F_E_S_vsT}.

\section{Non-zero density}\label{non_zero_density}

Let us now explore the  D3-D5-D7 system at non-zero charge density. 
To simplify our analysis we restrict ourselves to the case in which the embedding function $\chi$ is zero which, according to the holographic dictionary, corresponds to having massless quarks. Recall that $\chi=0$ solves trivially the equation of motion (\ref{eom_chi_general}).  Moreover, when $\chi=0$ eq.~(\ref{At_d}) yields the following value for the radial derivative of the worldvolume gauge field potential $A_t$:
\beq
A_t' = {e^{{\phi\over 2}}\,d\over \sqrt{d^2+r^6\,e^{-\phi}}} \ .
\label{A_t_prime_massless}
\eeq
By the AdS/CFT dictionary, the chemical potential is given by the UV value of $A_t$:
\beq\label{eq:muintegral}
\mu \equiv A_t(r\to\infty) = \,d\,\int_{r_h}^{\infty}\,{e^{{\phi\over 2}}\over  \sqrt{d^2+r^6\,e^{-\phi}}}\ .
\eeq
The last equality follows for the black hole embeddings of the D7-brane, since the temporal part of the background gauge potential has to vanish at the horizon. Notice that in the current case, at finite density, the only possible embeddings are those that enter into the black hole. At zero density, but at finite chemical potential, there is the possibility of considering Minkowski embeddings in which case one needs to allow non-zero IR values of the gauge potential \cite{Bergman:2010gm,Jokela:2011eb}.

Using the value of the dilaton $\phi$ for our background (\ref{dilaton}), we can directly perform the integration in (\ref{eq:muintegral}), yielding
\be
\mu = \Big({3\over 4\,Q_f}\Big)^{{1\over 2}}\,\gamma\,d^{{1\over 2}}
-{3\over 4}\,\Big({3\over 4\,Q_f}\Big)^{{1\over 3}}\,r_h^{{4\over 3}}\,
F\Big({1\over 4}, {1\over 2}; {5\over 4}; -\Big({4\,Q_f\over 3}\Big)^{{2\over 3}} \ ,
{r_h^{{16\over 3}}\over d^2}\Big)\label{mu_rh}
\ee
where $\gamma$ is the following constant:
\beq\label{gamma_constant}
\gamma = {3\over \sqrt{\pi}}\,\Big[\Gamma\Big({5\over 4}\Big)\Big]^2\ .
\eeq
For later purposes we also define a function $H$ as follows
\beq
 H\equiv r^6\,e^{-\phi} \ .
\eeq

\subsection{Stiffness at zero temperature}
Before discussing the excitation spectrum of our system it is useful to start with computing the stiffness at zero temperature. This is given by the  following thermodynamic derivative at fixed entropy density
\be\label{eq:stiffness}
 u_i^2 = \frac{\partial p_i}{\partial \epsilon}\Big|_s\,\,,
\ee
where $p_i$ is the pressure along direction $x^i$. At zero temperature this typically agrees with $\frac{\partial p_i}{\partial \epsilon}\Big|_T$ \cite{Ecker:2017fyh} that we aim to compute. 
With an abuse of language we follow the existing literature and henceforth call (\ref{eq:stiffness}) the speed of first sound.

Let us now take $T=r_h=0$ in our equations and let us examine the thermodynamics of the probe at zero temperature.  When $r_h=0$ the second term in (\ref{mu_rh}) vanishes and $\mu$ is related to the density $d$ as follows:
\beq
\mu\,=\,\Big({3\over 4\,Q_f}\Big)^{{1\over 2}}\,\gamma\,d^{{1\over 2}}
\,\,.
\label{mu_zero_rh}
\eeq
Then, the on-shell action is:
\beq
S_{on-shell}\,=\,-{\cal N}\,V_4\,\int_0^{\infty}\,dr\,
{r^3\,\sqrt{H}\over \sqrt{d^2+H}}\,\,,
\eeq
where $V_4$ is the volume of our system in the four-dimensional   Minkowski space-time. Since the function $H$ is given by:
\beq
H\,=\,\alpha\,r^{{16\over 3}}\,\,,
\qquad\qquad\qquad
\alpha\equiv \Big({4Q_f\over 3}\Big)^{{2\over 3}}\,\,,
\label{H_alpha}
\eeq
the on-shell action is divergent and must be regulated by subtracting the action at zero density. We get:
\beq
S_{on-shell}^{reg}\,=\,-{\cal N}\,V_4\,\int_0^{\infty}\,dr\,
r^3\,\Big[{\sqrt{H}\over \sqrt{d^2+H}}-1\Big] = {1\over 3}\,{\cal N}\,V_4\,\gamma\,\Big({3\over 4\,Q_f}\Big)^{{1\over 2}}\,d^{{3\over 2}} \ .
\eeq
The grand potential $\Omega$ is just $\Omega=-S_{on-shell}^{reg}$. Therefore:
\beq
\Omega\,=\,-{1\over 3}\,{\cal N}\,V_4\,\gamma\,\Big({3\over 4\,Q_f}\Big)^{{1\over 2}}\,d^{{3\over 2}}\,=\,
-{1\over 3}\,{{\cal N}\,V_4\over \gamma^2}\,\Big({4Q_f\over 3}\Big)\,\mu^3\,\,.
\eeq
From $\Omega$ we get the density $\rho$ as:
\beq
\rho\,=\,-{1\over V_4}\,{\partial\Omega\over \partial\mu}\,=\,{\cal N}\,d\,\,,
\eeq
as well as the energy density $\epsilon$:
\beq
\epsilon\,=\,{\Omega\over V_4}\,+\,\mu\,\rho\,=\,{2\over 3}\,{\cal N}\,\gamma\,
\Big({3\over 4\,Q_f}\Big)^{{1\over 2}}\,d^{{3\over 2}}\,=\,-\,{2\,\Omega\over V_4}\,\,.
\eeq
In order to obtain the pressures along the different directions, let us put the system in a box of sides $L_1$, $L_2$, and $L_3$ in the directions of $x^1$, $x^2$, and $x^3$ respectively. Then, the pressure along the $i^{th}$ direction is given by:
\beq
p_i\,=\,-{L_i\over V_4}\,
{\partial \Omega\over \partial L_i}\,\,,
\qquad\qquad
(i=1,2,3)\,\,.
\label{p_i_s}
\eeq
The grand potential $\Omega$ is an extensive quantity which clearly depends linearly on $L_1$ and $L_2$. Therefore, if $p_{\parallel}$ refers to the pressure in the $x^1\,x^2$  plane, we have:
\beq
p_{\parallel}\,=\,-{\Omega\over V_4}\,=\,{\epsilon\over 2}\,\,.
\eeq
It follows that the corresponding in-plane speed of sound is:
\beq
u_{\parallel}^2\,=\,{\partial p_{\parallel}\over \partial \epsilon}\,=\,{1\over 2}\,\,.
\label{speed_first_sound}
\eeq
In section \ref{In-plane_zero_sound_section}, by studying the fluctuations of the probe D7-brane, we will show that this value of $u_{\parallel}$ coincides with the  value of the speed of the in-plane zero sound. 

The result in (\ref{speed_first_sound})  matches the value corresponding to a conformal theory in 2d and not that of 3d. Notice that at finite chemical potential there are plenty of examples in holography where one can exceed the conformal value \cite{Hoyos:2016cob,Ecker:2017fyh,Ishii:2019gta}, as long as the conformal symmetry is broken. Here the mechanism is quite different and can be associated with the presence of the defect instead of some intrinsic energy scale.  At vanishing chemical potential there are also other mechanisms that lead to stiff equations of state: brane intersections with non-AdS asymptotics \cite{Jokela:2015aha,Itsios:2016ffv,Kulaxizi:2008jx}, softly broken translational symmetry \cite{Burikham:2016roo}, having non-relativistic scaling symmetries \cite{HoyosBadajoz:2010kd,Jokela:2016nsv}, dynamical magnetic fields \cite{Grozdanov:2017kyl}, and those obtained from double trace deformations \cite{Anabalon:2017eri}.

The dependence of $\Omega$ on $L_3$ is  not clear since $Q_f$ is proportional to the density $N_f$ of D5-branes smeared along $x^3$. Therefore, the calculation of the thermodynamic derivative (\ref{p_i_s}) is far from obvious in this case. Actually, we will verify in section (\ref{off-plane_zero_sound_section}) that the dispersion relation of the off-plane modes in the zero sound channel is quite different from the ones corresponding to the sound modes propagating along the plane directions.

\subsection{Zero sound}

Let us now turn to discussing the collective phenomena that arise from fluctuating the fields on the worldvolume of the D7-brane.  
In the field theory dual, these excitations are density waves which correspond to quasinormal modes on the gravity side. 
We are only discuss vector fields, focusing solely on s-waves. Thus, we suppress the dependence on the internal part of the geometry. We start the analysis at zero temperature, where we can obtain analytic results. In the subsequent section, we instead focus on the high temperature result, which is also amenable for analytic treatment. 
In between there is the regime of finite temperature where we need to resort to numerical analysis, which accurately interpolates the limiting cases. 

Let us thus begin by considering our system at exactly zero temperature. Here the dominant mode is the so-called holographic zero sound \cite{Karch:2008fa} which is a pretty robust collective excitation mode persisting to finite temperature \cite{Bergman:2011rf}. Many other aspects of the holographic zero sound mode have been discussed in various contexts \cite{Kulaxizi:2008kv,Kim:2008bv,Kulaxizi:2008jx,Hung:2009qk,Edalati:2010pn,Lee:2010uy,HoyosBadajoz:2010kd,Lee:2010ez,Ammon:2011hz,Davison:2011ek,Jokela:2012vn,Goykhman:2012vy,Gorsky:2012gi,Brattan:2012nb,Jokela:2012se,Davison:2013bxa,Pang:2013ypa,Dey:2013vja,Edalati:2013tma,Davison:2013uha,DiNunno:2014bxa,Jokela:2015aha,Itsios:2015kja,Itsios:2016ffv,Jokela:2016nsv,Jokela:2017fwa,Itsios:2018hff}. To study these modes we perturb the worldvolume gauge potential as follows
\beq
A\,=\,A^{(0)}+\delta A \ ,
\eeq
where $A^{(0)}$ is the gauge field of (\ref{A_t_prime_massless}) and $\delta A$ is a small perturbation about the equilibrium value. We Fourier transform to momentum space:
\beq
\delta A_{\mu}\,=\,a_{\mu}(\,r,\omega, \vec k\,)\,e^{-i\omega t\,+\,i\vec k\cdot \vec x}\ .
\eeq
The equations of motion for the fluctuations can be obtained by perturbing the DBI action around the configuration with $\chi=0$ and $A=A^{(0)}$. 
The derivation of these equations is done in Appendix \ref{Fluctuation_appendix}. We work in the radial gauge by setting $a_r=0$. We also introduce a gauge-invariant combination, the electric field:
\beq
\vec E\,=\,\vec k\,a_t\,+\,\omega\,\vec a\ .
\eeq

In what follows we consider the case in which  $\vec a$ and $\vec k$ are parallel. There are then two possibilities depending on the direction of wave propagation. 
When $\vec k$ is oriented along the $(x^1,x^2)$ plane we call it ``in-plane'', whereas the ``off-plane'' waves propagate along $x^3$. 
We distinguish between these two cases in the subsequent sections.

\subsubsection{In-plane zero sound}
\label{In-plane_zero_sound_section}

Let us consider the in-plane fluctuations at zero temperature. For concreteness, we assume that $\vec k = k_\parallel\vec e_{x^1}$ is directed along $x^1\equiv x$.  The corresponding equation  for $E\equiv  k_\parallel\,a_t\,+\,\omega\, a_x$ is worked out in the appendix, c.f.~(\ref{eom_E_in_plane}), and takes the following form
\beq
E''\,+\,\partial_r\log\Bigg[{e^{-{\phi\over 2}}\,\big(H+d^2\big)^{{3\over 2}}\,B\over(\omega^2- k_\parallel^2\,B)\,H\,+\,\omega^2\,d^2}\Bigg]\,E'\,+\,{h\over B^2}\,{(\omega^2- k_\parallel^2\,B)\,H\,+\,\omega^2\,d^2\over H+d^2}\,E\,=\,0\,\,.
\label{E_eq_in_plane}
\eeq
Since in this section we are interested in the $T=0$ case, we take $B=1$ from now on. We first study (\ref{E_eq_in_plane}) near the horizon $r=0$, where it takes the form:
\beq
E''\,-\,{1\over 3\,r}\,E'\,+\,{R^4\omega^2\over r^4}\,E\,=\,0\,\,.
\eeq
The solution of this near-horizon equation with infalling boundary conditions at $r=0$ can be written in terms of a  Hankel function of the first type:
\beq
E(r)\sim r^{{2\over 3}}\,H_{{2\over 3}}^{(1)}\Big({R^2\,\omega\over r}\Big)\ .
\eeq
When the index $ \nu$ of $H_{\nu}^{(1)}$ is not an integer the Hankel function has the following expansion near the origin:
\beq
H_{\nu}^{(1)}(\alpha x)= -{2^{\nu}\,\Gamma(\nu)\over \pi\,\alpha^{\nu}}\,i\,
\Big[{1\over x^{\nu}}\,+\,{\pi\over \Gamma(\nu)\,\Gamma(\nu+1)}\,
\Big({\alpha\over 2}\Big)^{2\nu}\,
\Big(i\,-\,\cot (\pi\nu)\Big)\,x^{\nu}\,+\,\ldots\Big]\,\,.
\eeq
Let us then expand the near-horizon electric field at low frequency as:
\beq
E(r)\sim A\,\Big[1\,+\,{\cal C}\,\omega^{-{4\over 3}}\,r^{{4\over 3}}\,+\,\cdots\Big]\ ,
\label{E_nh_lowFreq}
\eeq
where ${\cal C}$ is a constant:
\beq
{\cal C}\,=\,{\sqrt{3}-3i\over 6\pi}\,\Gamma^2\Big({2\over 3}\Big)\,\Big({2\over R^2}\Big)^{{4\over 3}}\ .
\eeq

Let us now perform the two limits (near-horizon and low frequency) in opposite order. At low frequency and momentum we can neglect the last term in (\ref{E_eq_in_plane}) and the equation can be integrated once to give:
\beq
E'\,=\,c_E\,\omega^2\,{e^{{\phi\over 2}}\over (H+d^2)^{{1\over 2}}}\,-\,
c_E\,k_\parallel^2\,{e^{{\phi\over 2}}\,H\over (H+d^2)^{{3\over 2}}}\,\,,
\eeq
where $c_E$ is a constant. To perform a second integration, let us define the rescaled density $\tilde d$ as follows:
\beq
\tilde d^2\,=\,{d^2\over \alpha}\,\,,
\eeq
where $\alpha$ is the quantity defined in (\ref{H_alpha}). We also define the integrals $I(r)$ and $J(r)$ as:
\beq
I(r)\,\equiv\,\int_r^{\infty}\,{\rho^{{1\over 3}}\,d\rho\over
(\rho^{{16\over 3}}+\tilde d^{\,2})^{{1\over 2}}}\,\,,
\qquad\qquad
J(r)\,\equiv\,\int_r^{\infty}\,{\rho^{{17\over 3}}\,d\rho\over
(\rho^{{16\over 3}}+\tilde d^{\,2})^{{3\over 2}}}\,\,.
\label{I_J_integrals_def}
\eeq
Then, $E(r)$ at low frequency can be written as:
\beq
E(r)\,=\,E_0\,-\, {c_E\over \alpha}\,\omega^2\,I(r)\,+\,{c_E\over \alpha}\,k_\parallel^2\,J(r)\,\,,
\eeq
where $E_0$ is the value of $E$ at the UV boundary:  $E_0\,=\,E(r\to\infty)$. 

\begin{figure}[ht]
\center
 \includegraphics[width=0.45\textwidth]{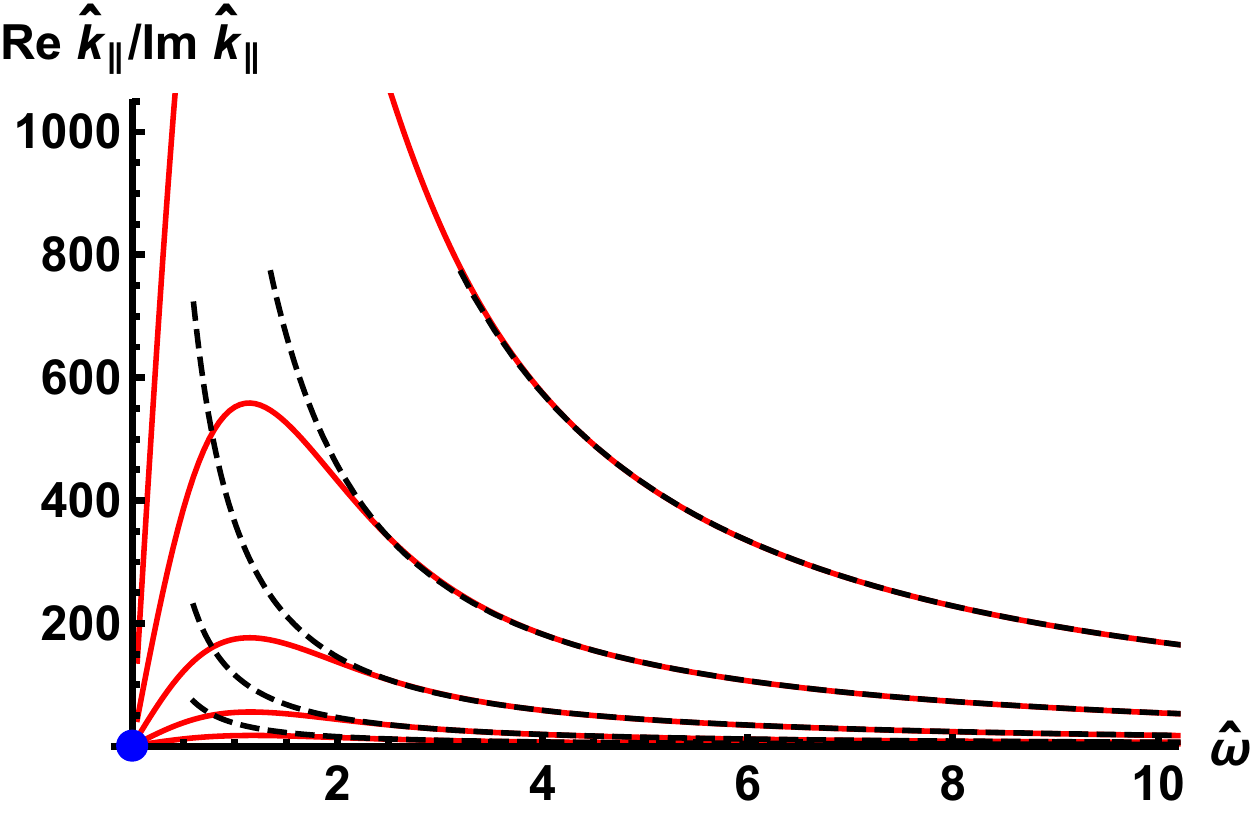}
 \qquad
 \includegraphics[width=0.45\textwidth]{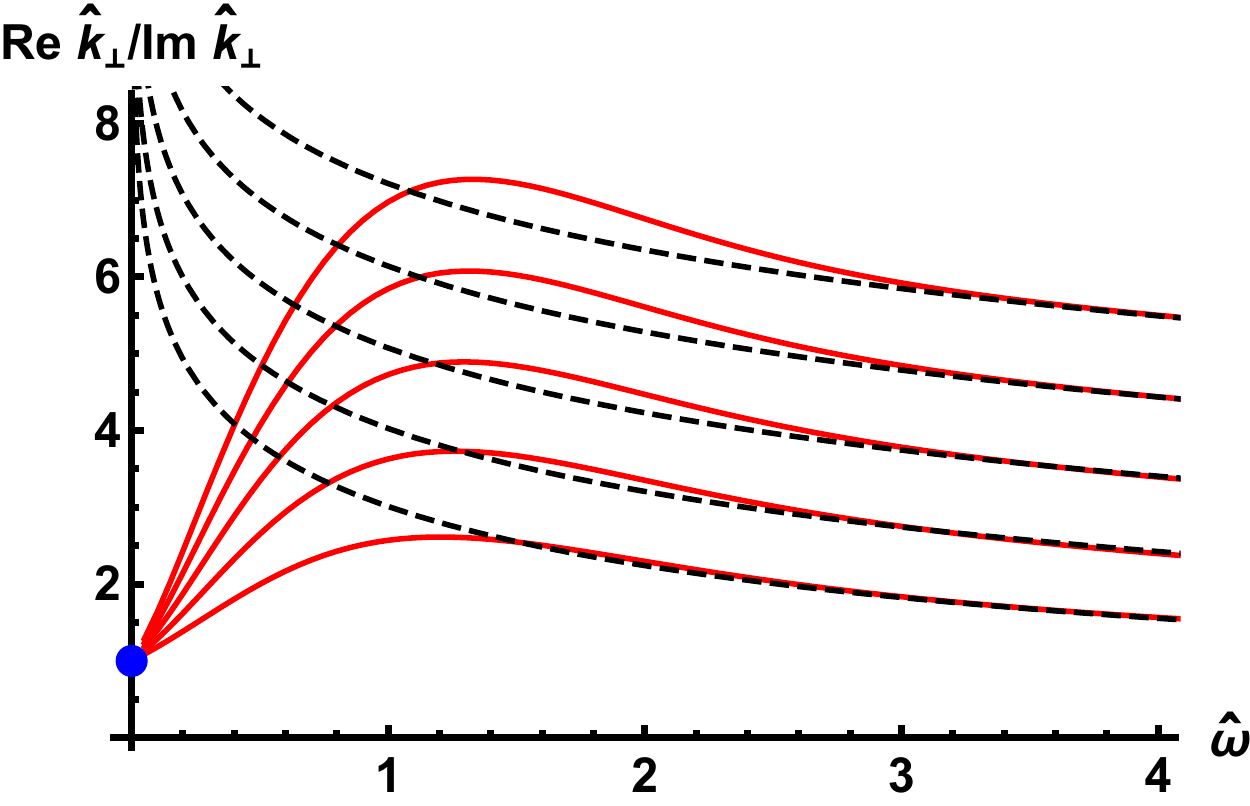}
   \caption{We display the numerical dispersions (red curves) and the approximated dispersions obtained analytically (black dashed curves). 
   On the left we show the results for the in-plane fundamental mode with analytics from (\ref{in_plane_dispersion}). 
   On the right we show the off-plane fundamental mode with analytics from (\ref{off_plane_dispersion}). 
   Different curves denote different temperatures: $\log_{10}\hat d=1,2,3,4,5,6$ (bottom-up), {\emph{i.e.}}, from highish to low temperature at fixed density; see (\ref{Diffusion_hattted_quantities}) for the definition of the quantities with hats.   All the numerical curves asymptote to unity on the vertical axis that correspond to diffusion poles, marked by blue points.    At high frequencies all the curves asymptote to $\sqrt 3$ and $0$ for the in-plane (left panel) and off-planes (right panel) cases, respectively, though this is not visible in the plotted range.} 
  \label{fig:kdisp}
\end{figure}

The integrals $I(r)$ and $J(r)$ are special cases of the integrals that we define and inspect in the Appendix \ref{integrals}, see (\ref{I_lambda12_def}) and (\ref{J_integral_definition}), respectively. More precisely, the relationships are:
\beq
I(r)\,=\,I_{{1\over 3}, {16\over 3}}(r)\,\,,
\qquad\qquad
J(r)\,=\,J_{{17\over 3}, {16\over 3}}(r) \ .
\eeq
According to the general formulas (\ref{I_lambda12_expansion}) and (\ref{J_lambda12_expansion}) these integrals behave near $r=0$ as:
\beq
I(r)\approx \gamma\,\tilde d^{-{1\over 2}}\,-\,{3\over 4\tilde d}\,r^{{4\over 3}}\,\,,
\qquad\qquad\qquad\qquad
J(r)\approx {\gamma\over 2}\,\tilde d^{-{1\over 2}}\,-\,{3\over 20}\,
{r^{{20\over 3}}\over \tilde d^{\,3}} \ , \label{I_J_integrals_low_r}
\eeq
where $\gamma$ is the constant defined in (\ref{gamma_constant}). Therefore, $E(r)$ behaves near the horizon as:
\beq
E(r)\,\approx \, E_0\,+\,{3\,c_E\over 4\alpha}\,{\omega^2\over \tilde d}\,r^{{4\over 3}}\,-\,
{c_E\,\gamma\over \alpha\,\sqrt{\tilde d}}\,\Big(\omega^2-{1\over 2}\,k_\parallel^2\Big)\,\,. \label{E_lowFreq_nh}
\eeq
Let us now compare (\ref{E_nh_lowFreq}) and (\ref{E_lowFreq_nh}).  By looking at the terms depending on $r$ we get:
\beq
A\,=\,{3\,c_E\over 4\,\alpha\,{\cal C}}\,{\omega^{{10\over 3}}\over \tilde d}\,\,.
\label{A_cE_inPlane}
\eeq
By comparing the constant terms and using (\ref{A_cE_inPlane}) we find:
\beq
E_0\,=\,{c_E\,\gamma\over \alpha\,\sqrt{\tilde d}}\,\Big(\omega^2-{1\over 2}\,k_\parallel^2\Big)\,+\,
{3\,c_E\over 4\,\alpha\,{\cal C}}\,{\omega^{{10\over 3}}\over \tilde d}~.
\eeq
By imposing Dirichlet conditions at the UV boundary, i.e.~$E_0=0$, we get the following dispersion relation:
\beq
{1\over 2}\,k_\parallel^2\,-\,\omega^2\,=\,{3\over 4\,\gamma\,{\cal C}}\,{\omega^{{10\over 3}}\over \sqrt{\tilde d}}\,\,.
\label{in_plane_dispersion}
\eeq
In Fig.~\ref{fig:kdisp}, where we have used the reduced quantities defined in \eqref{Diffusion_hattted_quantities}, we compare this to the numerical results and find that they are accurately captured at small temperature. We choose to represent the dispersion in the case where we keep frequency real and take complex momentum. Then a convenient way to display the result is to plot the ratio
\be
 \frac{{\rm{Re}}k_\parallel}{{\rm{Im}}k_\parallel} = \frac{\sqrt 3 g}{g+\omega^{-4/3}}\left[\sqrt{1+\frac{3g^2\omega^{20/3}}{(\om^2+g \omega^{10/3})^2}}-1\right]^{-1} \ ,
 \ g\equiv \frac{3\sqrt 3}{8\gamma}\frac{\pi}{\sqrt{\tilde d}\; \Gamma(2/3)^2}\left(R^2/2\right)^{4/3} \ ,
\ee
which readily follows from (\ref{in_plane_dispersion}). In the limit of high frequency this ratio asymptotes to $\sqrt 3$. We note that at any non-zero temperature, the zero sound is not the dominant mode at small frequency, but the relevant physics is dominated by the diffusion pole discussed in the subsequent section.

Let us then study the dispersion relation at small frequency. At leading order in $\omega$ (\ref{in_plane_dispersion}) becomes:
\beq
\omega\,=\,\pm{1\over \sqrt{2}}\,k_\parallel \ .
\eeq
We note that this result equals the first sound result (\ref{speed_first_sound}) obtained in the previous section. Note also that we could write the dispersion relation (\ref{in_plane_dispersion}) in terms of the chemical potential $\mu$ by using the relation:
\beq
\gamma\,\sqrt{\tilde d}\,=\,\Big({4Q_f\over 3}\Big)^{{1\over 3}}\,\mu\,=\,\sqrt{\alpha}\,\mu\,\,.
\eeq
Let us now find the next order dispersion relation by writing
\beq
\omega\,=\,{1\over \sqrt{2}}\,k_\parallel\,+\,\delta \omega\,\,,
\eeq
and plugging this ansatz in (\ref{in_plane_dispersion}), we get at first order in $\delta \omega$:
\beq
\delta\omega\,\approx \,-{3\over 4\sqrt{\alpha}\,{\cal C}\,\mu}\,
\Bigg[{k_\parallel^{{7\over 3}}\over 4\,\cdot 2^{{1\over 6}}}\,+\,{5\cdot 2^{{1\over 3}}\over 6}\,k_\parallel^{{4\over 3}}\,\delta\omega
\Bigg]\,\,.
\eeq
We can now solve for $\delta\omega$ in powers of $k_\parallel$. The second term on the RHS is clearly of higher order. Therefore:
\beq
\delta\omega\,\approx -{3\over 16\cdot 2^{{1\over 6}}}\,{1\over \sqrt{\alpha}\,{\cal C}\,\mu}\,k_\parallel^{{7\over 3}}\,=\,
-{3\pi\over 64\,\sqrt{2}\,\,\Gamma^2\Big({2\over 3}\Big)}\,
(\sqrt{3}+3\,i)\,
{R^{{8\over 3}}\over \sqrt{\alpha}\,\mu}\,k_\parallel^{{7\over 3}}\,\,.
\eeq
The imaginary part of this equation gives the attenuation of the in-plane zero sound, namely:
\beq
{\rm Im}\,\omega\,=\,-{9\,\pi\over 64\,\sqrt{2}\,\,\Gamma^2\Big({2\over 3}\Big)}\,
{R^{{8\over 3}}\over \sqrt{\alpha}\,\mu}\,k_\parallel^{{7\over 3}}\,\,.
\eeq
Moreover, the higher order correction to ${\rm Re}\, \omega$ is:
\beq
{\rm Re}\,\delta\omega\,=\,-{3\sqrt{3}\,\pi\over 
 64\,\sqrt{2}\,\,\Gamma^2\Big({2\over 3}\Big)}\,
{R^{{8\over 3}}\over \sqrt{\alpha}\,\mu}\,k_\parallel^{{7\over 3}}\,\,.
\eeq

\subsubsection{Off-plane zero sound}
\label{off-plane_zero_sound_section}

We now analyze the case in which $\vec k=k_\perp\vec e_{x^3}$ lies in the direction of $x^3\equiv z$. The equation for the gauge-invariant combination $E=k_\perp\,a_t+\omega a_z$ has been obtained in (\ref{eom_E_off_plane})  and can be written as:
\beq
E''\,+\,\partial_r\log\Bigg[{e^{{3\phi\over 2}}\,\big(H+d^2\big)^{{3\over 2}}\,B\over
(\omega^2- k_\perp^2\,B\,e^{2\phi})\,H\,+\,\omega^2\,d^2}\Bigg]\,E'\,+\,
{h\over B^2}\,{(\omega^2- k_\perp^2\,B\,e^{2\phi})\,H\,+\,\omega^2\,d^2\over H+d^2}\,E\,=\,0\,\,.
\label{E_eq_off_plane}
\eeq
We first analyze (\ref{E_eq_off_plane})  with $B=1$ near the horizon $r=0$, where it takes the form:
\beq
E''\,+\,{1\over r}\,E'\,+\,{R^4\omega^2\over r^4}\,E\,=\,0\,\,.
\eeq
The solution of this near-horizon equation with infalling boundary conditions is:
\beq
E(r)\sim \,H_{0}^{(1)}\Big({R^2\,\omega\over r}\Big)\,\,.
\eeq
In order to expand $E$ at low frequencies, we use the expansion of the Hankel function of index zero near the origin:
\beq
H_{0}^{(1)}(x)\sim {2i\,\gamma_E\,+\,\pi\,-\,2i\log 2\over \pi}\,+\,{2i\over \pi}\log x\,\,,
\eeq
where $\gamma_E$ is the Euler-Mascheroni constant, yielding
\beq
E(r)\approx A\,\Big[i{\pi\over 2}-\gamma_E\,-\,\log{R^2\,\omega\over 2}\,+\,\log r\Big]\,\,.
\label{E_nh_low_off}
\eeq

Let us now redo the computation taking the limits in the opposite order as in the previous subsection. At low frequency we neglect the last term in (\ref{E_eq_off_plane}), which leads to an equation that can be integrated once as:
\beq
E'\,=\,c_E\,\omega^2\,{e^{-{3\,\phi\over 2}}\over (H+d^2)^{{1\over 2}}}\,-\,
c_E\,k_\perp^2\,{e^{{\phi\over 2}}\,H\over (H+d^2)^{{3\over 2}}}\,\,,
\eeq
where $c_E$ is a constant. A second integration yields:
\beq
E(r)\,=\,E_0\,-\,c_E\,\alpha\,\omega^2\,\tilde I(r)\,+\,{c_E\over \alpha}\,k_\perp^2\,J(r)\,\,,
\eeq
where $\alpha$ has been defined in  (\ref{H_alpha}),  $J(r)$ is the integral defined in (\ref{I_J_integrals_def}) and $\tilde I(r)$ is a new integral defined as:
\beq
\tilde I(r)\,\equiv\,\int_r^{\infty}\,
{d\rho\over \rho\,\big(\rho^{{16\over 3}}\,+\,\tilde d^2\big)^{{1\over 2}}} = {3\over 8\tilde d}\,\,{\rm arcsinh}\,\Big({\tilde d\over r^{{8\over 3}}}\Big) \ .
\eeq
For small $r$ this function behaves as:
\beq
\tilde I(r)\approx {1\over \tilde d}\,\Big[\log (2\,\tilde d)^{{3\over 8}}\,-\,\log r\Big]\ .
\eeq
Using this result, together with the expansion  of $J(r)$ written in (\ref{I_J_integrals_low_r}), we get:
\beq
E(r)\approx E_0\,+\,{c_E\,\alpha\,\omega^2\over \tilde d}\,\log r\,-\,
c_E\,\Big[{\alpha\,\log (2\,\tilde d)^{{3\over 8}}\over \tilde d}\,\omega^2\,-\,
{\gamma\over 2\,\alpha\,\sqrt{\tilde d}}\,k_\perp^2\Big]\,\,.
\label{E_low_nh_off}
\eeq
Let us now match (\ref{E_nh_low_off}) and (\ref{E_low_nh_off}). First we identify the terms containing 
$\log r$, which gives:
\beq
A\,=\,{c_E\,\alpha\,\omega^2\over \tilde d}\,\,.
\eeq
Using this result, we identify the constant terms and write the UV value of $E$ as:
\beq
E_0\,=\,{c_E\,\alpha\,\omega^2\over \tilde d}\,
\Big[i{\pi\over 2}-\gamma_E\,-\,\log{R^2\,\omega\over 2^{{11\over 8}}\,\tilde d^{{3\over 8}}}\Big]\,-\,
{c_E\,\gamma\over 2\alpha\,\sqrt{\tilde d}}\,k_\perp^2\,\,.
\eeq
By requiring Dirichlet boundary condition $E_0=0$ we arrive at the desired dispersion relation. To simplify the final expression we define the constant 
${\cal D}$ as:
\beq
{\cal D}\equiv i{\pi\over 2}-\gamma_E\,-\,\log{R^2\over 2^{{11\over 8}}\,\tilde d^{{3\over 8}}}\ ,
\eeq
so that the dispersion relation can be neatly written as
\beq\label{off_plane_dispersion}
k_\perp^2\,=\,{2\alpha^2\over \gamma\,\sqrt{\tilde d}}\,\,\Big[{\cal D}\,-\,\log\omega\Big]\,\,\omega^2\ .
\eeq
In Fig.~\ref{fig:kdisp} we compare this to the numerical results and find that they are accurately captured at small temperature. Again, we have chosen to represent the dispersion in the case where momentum is complex while frequency is real. The ratio of the real and imaginary parts of the momentum in this case reads
\be
 \frac{{\rm{Re}}k_\perp}{{\rm{Im}}k_\perp} = \frac{\pi}{2}\left[\log(f\omega)+\sqrt{\log^2(f\omega)+\frac{\pi^2}{4}}\right]^{-1} \ , \ f\equiv \frac{R^2}{2^{11/8}\tilde d^{3/8}}e^{\gamma_E} \ .
\ee
In the limit of high frequency this ratio asymptotes to zero.

Let us also study the small frequency limit of the dispersion relation just obtained. At leading order in small $\omega$ (\ref{off_plane_dispersion}) becomes:
\beq
k_\perp\,=\,\pm\,\Big({2\alpha^2\over \gamma\,\sqrt{\tilde d}}\Big)^{{1\over 2}}\,\,
\omega\,\Big[\log\big(1/ \omega\big)\Big]^{1\over 2}\,\,,
\eeq
which differs from the usual one by logarithmic terms.  Similar logarithmic terms appear in the Lifshitz spacetime when the dynamical exponent $z$ is equal to 2 \cite{HoyosBadajoz:2010kd}.

\subsection{Diffusion modes}

We now consider the fluctuation modes at non-zero temperature in the diffusive channel, {\emph{i.e.}}, solve exactly 
the same fluctuation equations as before but consider the other limiting case, that of very high temperature. Thus, the corresponding equations for these modes are (\ref{E_eq_in_plane}) and (\ref{E_eq_off_plane}) for the in-plane  and off-plane diffusion, respectively. These two cases are considered separately in the two subsections that follow. The goal is to find diffusive modes in which $\omega$ is purely imaginary and related to the momentum as $\omega=-i\, D\,k^2+\ldots$, with $D$ being the diffusion constant. In what follows we get an analytic expression of $D$ for the two types of waves.

\subsubsection{In-plane diffusion}

Let us begin our analysis by expanding the in-plane fluctuation equation (\ref{E_eq_in_plane}), with $B\not=1$,  around the horizon $r=r_h$. The expansion of the emblackening factor $B$ is:
\beq
B(r) = {10\over  3\,r_h}\,(r-r_h)+\ldots \ .
\eeq
The functions multiplying $E'$ and $E$ in (\ref{E_eq_in_plane}) can be expanded as:
\bear
\partial_r\log\Bigg[{e^{-{\phi\over 2}}\big(H+d^2\big)^{{3\over 2}}B\over(\omega^2- k_\parallel^2\,B)\,H\,+\,\omega^2\,d^2}\Bigg] & = & {1\over r-r_h}\,+\,c_1+\ldots \nonumber\\
{h\over B^2}{(\omega^2- k_\parallel^2\,B)\,H\,+\,\omega^2\,d^{\,2}\over H+d^{\,2}} & = & {A\over (r-r_h)^2}\,+\,{c_2\over  r-r_h}+\ldots \ ,
\eear
where the constants $A$, $c_1$, and $c_2$ are given by:
\bear
&&A\,=\,{9\,R^4\over 100\,r_h^2}\,\omega^2\ \rc\rc
&&c_1\,=\,{10\over 3}\,{\alpha\,r_h^{{13\over 3}}\over d^{\,2}+\alpha\,r_h^{{16\over 3}}}\,\,{k_\parallel^2\over \omega^2}\,+\,
{1\over  6 r_h}\,-\,{8\over 3}\,{d^{\,2}\over r_h\,\Big( d^{\,2}+\alpha\,r_h^{{16\over 3}}\Big)}\ \rc\rc
&&c_2\,=\,-{3\,\alpha\,R^4\over 10}\,{r_h^{{7\over 3}}\over d^{\,2}+\alpha\,r_h^{{16\over 3}}}\,k_\parallel^2\,+\,
{3\,R^4\over 100\, r_h^3}\,\omega^2\,\,.
\label{Diffusion_A_c_values}
\eear
Plugging these expansions in  (\ref{E_eq_in_plane}), we get the following near-horizon fluctuation equation:
\beq
E''\,+\,\Big({1\over r-r_h}\,+\,c_1\Big)\,E'\,+\,\Big({A\over (r-r_h)^2}\,+\,{c_2\over  r-r_h}\Big)\,E\,=\,0\,\,.
\label{Diffusion_E_nh_eq}
\eeq
We solve this equation in  Frobenius series near the horizon:
\beq
E(r)\,=\,E_{nh}\,(r-r_h)^a\,\big[1+\beta (r-r_h)+\ldots\big]\ ,
\label{Diffusion_E_nh}
\eeq
where $a$ is determined by solving the indicial equation: $a^2=-A$. Choosing the infalling solution, and determining $\beta$ by plugging (\ref{Diffusion_E_nh}) into (\ref{Diffusion_E_nh_eq}) yields
\beq\label{Diffusion_a_value}
a\,=\,-i\,\sqrt{A} = -{3\,R^2\over 10\,r_h}\,i\,\omega \ ,
\qquad\qquad
\beta\,=\,-{a\,c_1+c_2\over 1+2\,a}\,\,.
\eeq
We now perform a low frequency expansion in which $k\sim {\mathcal O}(\epsilon)$ and 
$\omega\sim {\mathcal O}(\epsilon^2)$, as it corresponds to a diffusion mode. Then, from (\ref{Diffusion_a_value})  and (\ref{Diffusion_A_c_values}) we have:
\beq
a\sim {\mathcal O}(\epsilon^2)\,\,,
\qquad\qquad
c_1\sim {\mathcal O}(\epsilon^{-2})\,\,,
\qquad\qquad
c_2\sim {\mathcal O}(\epsilon^{2})\,\,.
\eeq
Then,  $a\,c_1\sim {\mathcal O}(1)$ and, at leading order:
\beq
\beta\sim -a\,c_1\,\,.
\eeq
Therefore, also at leading order, we have:
\beq
\beta\,=\,i\,{k_\parallel^2\over \omega}\,
{\alpha\,R^2\,r_h^{{10\over 3}}\over d^{\,2}+\alpha\,r_h^{{16\over 3}}} + \ldots \ .
\eeq
Notice that $\beta \sim {\mathcal O}(1)$. 
Moreover, as $a\sim\omega\sim {\mathcal O}(\epsilon^2)$ we can take $a=0$ in (\ref{Diffusion_E_nh}) at leading order and write:
\beq
E(r)\approx E_{nh}\,\big[1\,+\,\beta (r-r_h)\big]\,\,,
\label{Diffusion_E_nh_leading}
\eeq
with $E_{nh}=E(r=r_h)$. 

We now perform the limits in the opposite order. First, we take the limit of low frequency. The fluctuation equation 
(\ref{E_eq_in_plane}) takes the form:
\beq
E''-\partial_r\,\log\,{r^{{17\over 3}}\over \big(
d^{\,2}+\alpha\,r^{{16\over 3}}\big)^{{3\over 2}}}\,E'\,=\,0\,\,,
\label{E_eq_low_freq}
\eeq
which can be integrated as:
\beq
E(r)= E^{(0)}\,+\,c_E\,\int_r^{\infty}d\rho\,{\rho^{{17\over 3}}\over \big(
\rho^{{16\over 3}}+{d^{\,2}\over \alpha} \big)^{{3\over 2}}}\,\,,
\qquad\qquad
E^{(0)}\,=\,E(r\to\infty)\,\,,
\label{Diffusion_E_low_freq}
\eeq
where $c_E$ is an integration constant. Let us now expand (\ref{Diffusion_E_low_freq})  near $r=r_h$. We have:
\beq
E(r) =  E^{(0)}\,+\,c_E\,{\mathcal I}\,-\,{r_h^{{17\over 3}}\,c_E\over \big(r_h^{{16\over 3}}+{d^{\,2}\over \alpha} \big)^{{3\over 2}}}\,(r-r_h)\,+\,\ldots \ ,\label{Diffusion_E_low_freq_nh}
\eeq
where ${\mathcal I}$ is the integral:
\beq
{\mathcal I}\,=\,\int_{r_h}^{\infty}\,
{\rho^{{17\over 3}}\over \big(
\rho^{{16\over 3}}+{d^{\,2}\over \alpha} \big)^{{3\over 2}}}\,=\,
{3\over 4\,r_h^{{4\over 3}}}\,F\Big({1\over 4}\,, \,{3\over 2}\,;{5\over 4}\,;\,-{d^{\,2}\over \alpha\,r_h^{{16\over 3}}}\Big)\,\,.
\label{Diffusion_cal_I_integral}
\eeq
Clearly,
\beq
E_{nh}\,=\,E^{(0)}\,+\,c_E\,{\mathcal I}\,\,.
\eeq
By imposing the Dirichlet boundary condition $E^{(0)}=0$ at the UV boundary $r\to\infty$, we get:
\beq
E_{nh}\,=\,c_E\,{\mathcal I}\,\,.
\label{Enh_cE_I}
\eeq
Moreover, by comparing the linear terms in $r-r_h$ in (\ref{Diffusion_E_nh_leading}) and (\ref{Diffusion_E_low_freq_nh}) we obtain the following relation between 
$c_E$ and $E_{nh}$:
\beq
c_E\,=\,-\beta\,{\big(
r_h^{{16\over 3}}+{d^{\,2}\over \alpha} \big)^{{3\over 2}}\over r_h^{{17\over 3}}}\,E_{nh}\,=\,
-i\,{k_\parallel^2\over \omega}\,{R^2\over  r_h^{{7\over 3}}}\,
\Big(
r_h^{{16\over 3}}\,+\,
{d^{\,2}\over \alpha} \Big)^{{1\over 2}}\, E_{nh}\,\,.
\label{cE_Enh}
\eeq
Plugging (\ref{cE_Enh}) into (\ref{Enh_cE_I}) we can eliminate $E_{nh}$ and finally obtain the dispersion relation
\beq
\omega\,=\,-i\,D_{\parallel}\,k_\parallel^2 + \ldots \ ,
\eeq
with the diffusion constant $D_{\parallel}$ equal to:
\beq
D_{\parallel}\,=\,{R^2\over  r_h^{{7\over 3}}}\,
\Big(
r_h^{{16\over 3}}\,+\,
{d^{\,2}\over \alpha} \Big)^{{1\over 2}}\,{\mathcal I}\,=\,
{3\,R^2\over 4\,r_h}\,\Big(
1\,+\,
{d^{\,2}\over \alpha\,r_h^{{16\over 3}}} \Big)^{{1\over 2}}\,
F\Big({1\over 4}\,, \,{3\over 2}\,;{5\over 4}\,;\,-{d^{\,2}\over \alpha\,r_h^{{16\over 3}}}\Big)\,\,.
\eeq
Let us now introduce the reduced quantities $\hat d$, $\hat \omega$, and $\hat k_{\parallel}$ as:
\beq
\hat d\,=\,{d\over \sqrt{\alpha}\,r_h^{{8\over 3}}}\propto 
{d\over Q_f^{{1\over 3}}\,Q_c^{{4\over 3}}\,T^{{8\over 3}}}\,\,,
\qquad
\hat \omega\,=\,{R^2\over r_h}\,\omega\,=\,{5\over 6\pi}\,{\omega\over T}\,\,,
\qquad
\hat k_{\parallel}\,=\,{R^2\over r_h}\,k_\parallel\,=\,{5\over 6\pi}\,{k_\parallel\over T}\,\,. 
\label{Diffusion_hattted_quantities}
\eeq
If we change in (\ref{E_eq_in_plane}) to a new radial  variable $\hat r=r/r_h$, the resulting equation expressed in terms of 
$\hat d$, $\hat \omega$, and $\hat k_{\parallel}$ does not contain the constants $R$, $\alpha$, and $r_h$. The dispersion relation in this reduced frequency and momentum takes the form:
\beq
\hat \omega\,=\,-i\,\hat D_{\parallel}\,\hat k_{\parallel}^{\,2} + \ldots \ ,
\eeq
where $\hat D_{\parallel}$ reads 
\bea
\hat D_{\parallel} & = & {r_h\over R^2}\,D_{\parallel}\,=\,{6\pi\over 5}\,T\,D_{\parallel} \\ \label{hat_D_parallel_def}
 & = & {3\over 4}\big(1+\hat d^{\,2}\big)^{{1\over 2}} F\Big({1\over 4}\,, \,{3\over 2}\,;{5\over 4}\,;\,-\hat d^{\,2}\Big) \ .\label{hat_D_parallel}
\eea

At high $T$ the reduced density $\hat d$ becomes very small and the hypergeometric function in (\ref{hat_D_parallel}) is approximately equal to one. At leading order, we have:
\beq
\hat D_{\parallel}\,\approx\,{3\over 4}\,\,,
\qquad\qquad
(T\to\infty\,\,,\,\,\hat d\to\infty)\,\,.
\eeq
Taking into account the relation between $\hat D_{\parallel}$ and $D_{\parallel}$ written in (\ref{hat_D_parallel_def}), we get that at large $T$ the parallel diffusion constant behaves as:
\beq
 D_{\parallel}\approx {5\over 8\,\pi\,T}\,\,, \qquad\qquad (T\to\infty) \label{high_T_inplane_dif} \ .
 \eeq

Let us now study the opposite regime of small $T$ (large $\hat d$) and we can approximate hypergeometric function in (\ref{hat_D_parallel}) as
\beq
F\Big({1\over 4}\,, \,{3\over 2}\,;{5\over 4}\,;\,-
\hat d^{\,2}\Big)\,\approx\,{2\over \sqrt{\pi}}
\Big[\Gamma\Big({5\over 4}\Big)\Big]^2\,
\hat d^{-{1\over 2}} \ .
\eeq
Therefore $\hat D_{\parallel}$ can be approximated as:
\beq
\hat D_{\parallel}\,\approx\,{3\over 2\sqrt{\pi}}\,\Big[\Gamma\Big({5\over 4}\Big)\Big]^2\,
\hat d^{{1\over 2}}\,\,.
\eeq
Using now that $D_{\parallel}\sim T^{-1}\,\hat D_{\parallel}$ and that 
$\hat d^{{1\over 2}}\sim T^{-{4\over 3}}$,  we find 
\beq
D_{\parallel}\sim T^{-{7\over 3}}\ , \qquad\qquad (T\to 0) \ .
\label{low_T_inplane_dif}
\eeq

\begin{figure}[ht]
\center
 \includegraphics[width=0.45\textwidth]{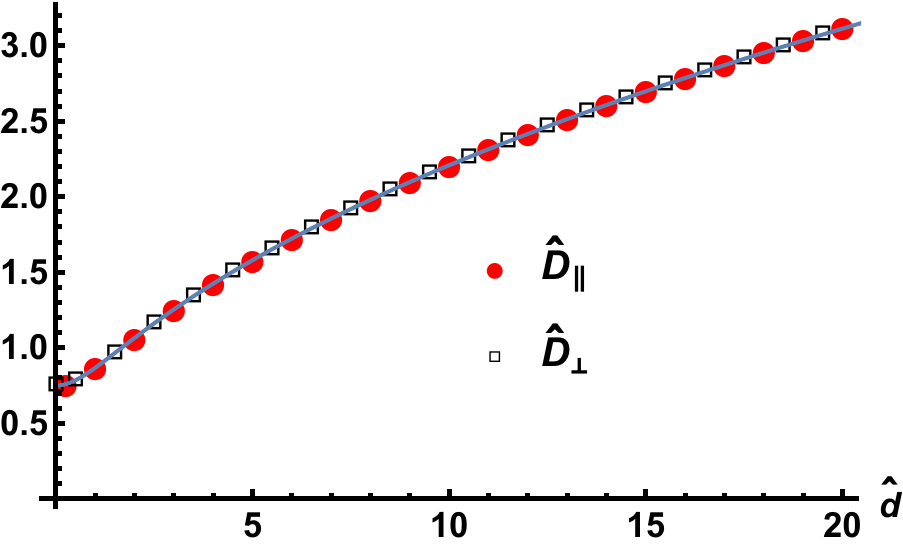}
 \qquad
 \includegraphics[width=0.45\textwidth]{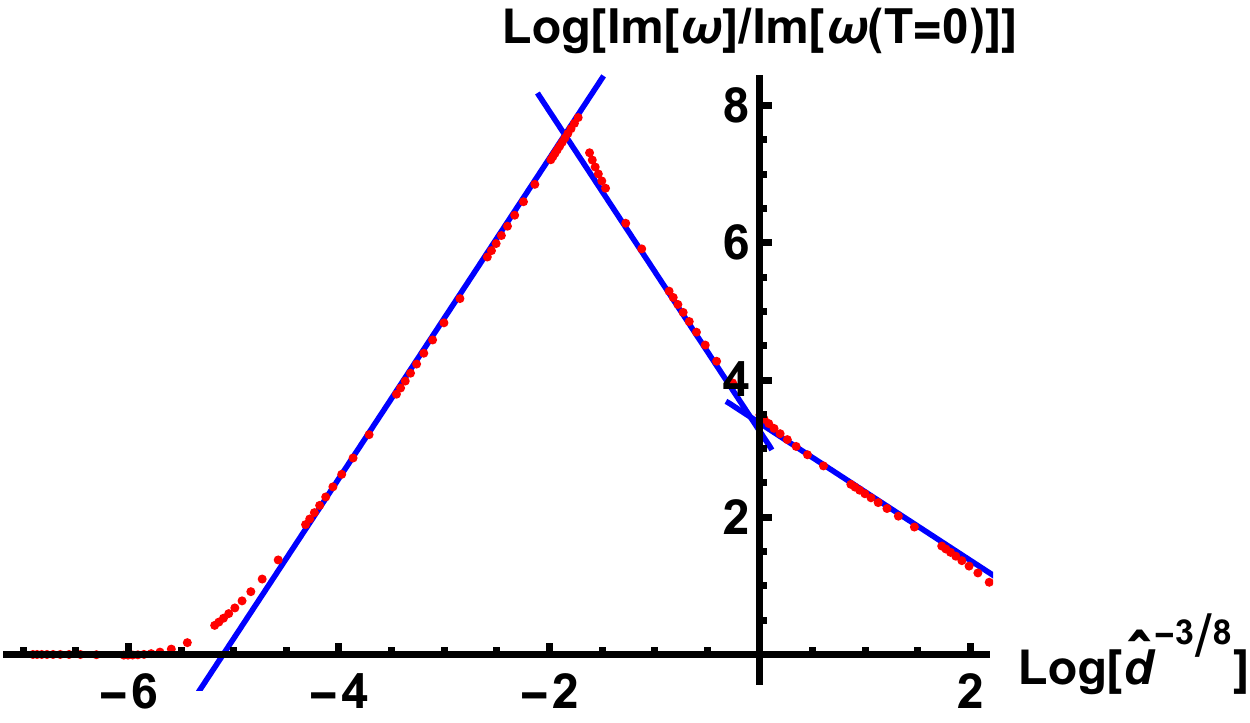}
   \caption{On the left panel we compare the numerical results for the reduced in-plane and off-plane diffusion constants $\hat D_{\parallel}$ and   $\hat D_{\perp}$ (discrete points) and the analytic results (\ref{hat_D_parallel}) and (\ref{hat_D_perp}) (continuous curve). On the right panel, we plot the imaginary part of the lowest in-plane excitation mode, keeping $\hat k_\parallel/\hat d^{3/8} = 0.01$, so that both the density and the wavevector are kept fixed and only the temperature varies. The red points are numerical results and the line segments have slopes $0$, $7/3$, -$7/3$, -$1$ from left to right. The latter two slopes correspond to the low and high temperature behavior of the diffusion constant (see (\ref{low_T_inplane_dif}) and (\ref{high_T_inplane_dif})).  } 
  \label{fig:parallel_dif}
\end{figure}

We have solved (\ref{E_eq_in_plane}) numerically and checked that, indeed, for high enough temperature the diffusive mode is the dominant one. This numerical analysis allows us to extract the diffusion constant and to compare the result with the analytic formula (\ref{hat_D_parallel}). This comparison is performed in Figure \ref{fig:parallel_dif}, where we see that the agreement between (\ref{hat_D_parallel}) and the numerical results is very good. 

When the temperature is low enough, the zero sound mode is the dominant one and the system enters a collisionless regime. 
This collisionless/hydrodynamic crossover is illustrated in Figure \ref{fig:parallel_dif}, where we notice that the zero sound persists at $T\not=0$ if $T$ is small enough. 
When $T$ is increased, the imaginary part of the zero sound grows as $T^{{7\over 3}}$ until the crossover to the hydrodynamic diffusive regime takes place. 
The frequency $\omega_{cr}$ and momentum $k^\parallel_{cr}$ at which this transition occurs depend on the temperature and chemical potential. Our numerical analysis has allowed us to determine that $\omega_{cr}$ and  $k^\parallel_{cr}$ scale with  $T$ and $\mu$ as:
\beq
\omega_{cr}\sim {T^{{7\over 3}}\over \mu}\,\,,
\qquad\qquad
k^\parallel_{cr}\sim {T^{{7\over 3}}\over \mu}\,\,.
\eeq

\begin{figure}[ht]
\center
 \includegraphics[width=0.45\textwidth]{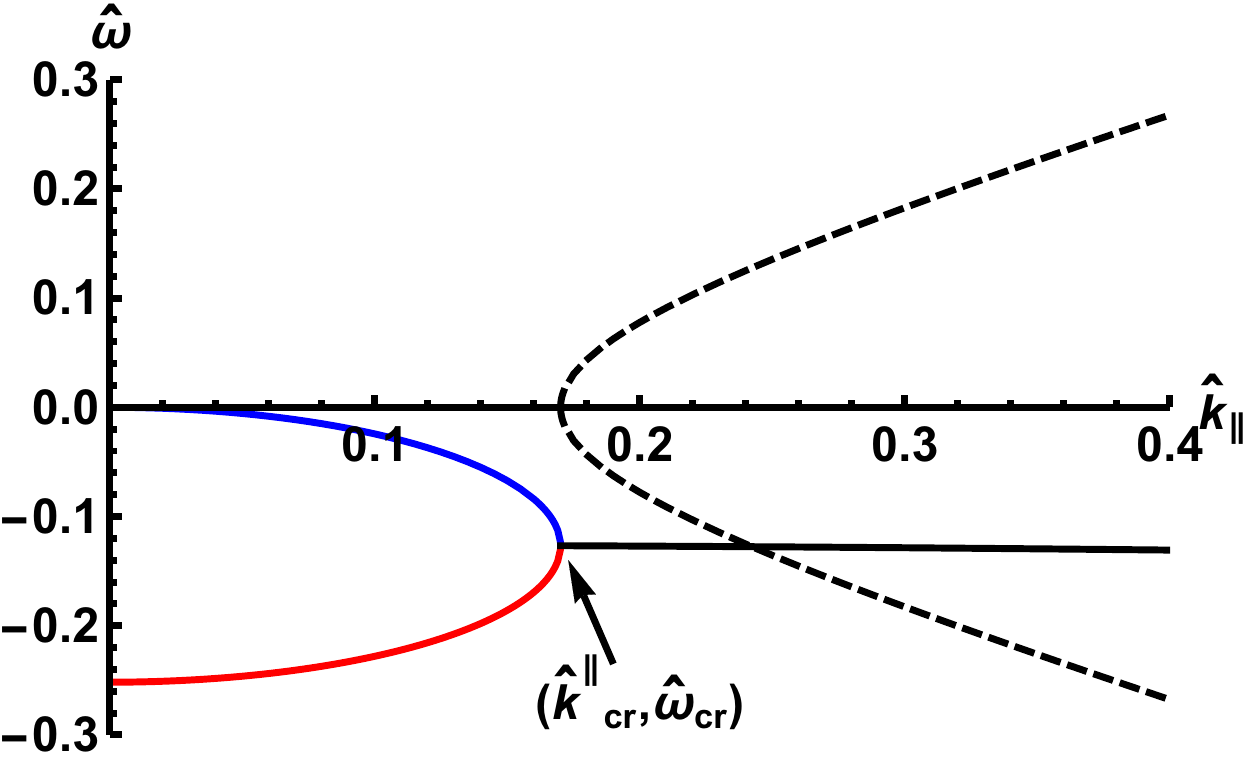}
 \qquad
 \includegraphics[width=0.45\textwidth]{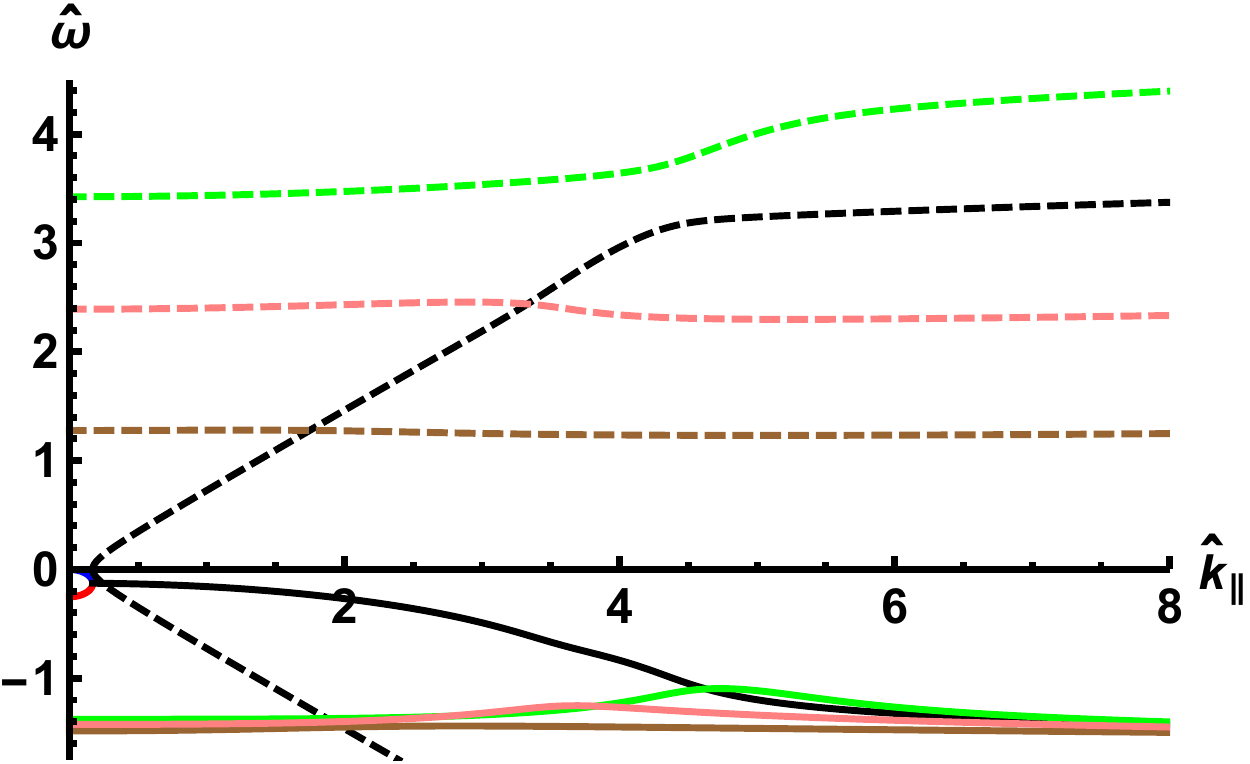}
   \caption{We display the typical dispersion relations for the in-plane case (off-plane is qualitatively the same) at $\hat d=10$. The solid curves denote imaginary parts while dashed curves are real parts of the modes. On the right panel we  extend the range of $\hat k_{\parallel}$ and notice that the zero sound collides with complex modes, eventually leading to saturation.} 
  \label{fig:omega_parallel}
\end{figure}

In Figure \ref{fig:omega_parallel} we display typical dispersion relations of several excitation modes for the in-plane case. 
The off-plane dispersion relations are qualitatively the same. We notice that the zero sound dispersion saturates at high frequencies. 
This is due to interactions between complex modes, as is clear from the figure. 
It would be tempting to identify this in-plane zero sound mode with a surface plasmon polariton, whose dispersion relation has close resemblance. 
However, we cannot clearly separate the underlying physics and associate the saturation directly with surface phenomena due to the following observation: 
We found out that for {\emph{any}} flavor D$q$-brane configuration that we have tested, the zero sound saturates if one reaches high enough frequency. 
This in itself is remarkable and had gone unnoticed in all the previous works in this field. It would be very interesting to understand why this happens and in particular discern if this phenomenon is due to non-linearities of the DBI action.

\subsubsection{Off-plane diffusion}

Let us now study the off-plane fluctuation equation (\ref{E_eq_off_plane}) in the diffusive regime. We first expand the coefficients of $E'$  and $E$ near the horizon as:
\bea
\partial_r\log\Bigg[{e^{{3\phi\over 2}}\,\big(H+d^2\big)^{{3\over 2}}\,B\over(\omega^2- k_\perp^2\,B\,e^{2\phi})\,H\,+\,\omega^2\,d^2}\Bigg] & = & {1\over r-r_h}\,+\,\tilde c_1+\ldots \nonumber\\
{h\over B^2}\,{(\omega^2- k_\perp^2\,B\,e^{2\phi})\,H\,+\,\omega^2\,d^2\over H+d^2} & = & {\tilde A\over (r-r_h)^2}\,+\,{\tilde c_2\over  r-r_h}+\ldots \ ,
\eea
where $\tilde A$, $\tilde c_1$, and $\tilde c_2$ are given by:
\bear
\tilde A & = & {9\,R^4\over 100\,r_h^2}\,\omega^2    \rc\rc
\tilde c_1 & = & {10\over 3\alpha}\,\,{r_h^{{17\over 3}}\over d^{\,2}+\alpha\,r_h^{{16\over 3}}}\,\,{k^2\over \omega^2}\,+{3\over  2 r_h}\,-\,{8\over 3}\,{d^{\,2}\over r_h\,\Big( d^{\,2}+\alpha\,r_h^{{16\over 3}}\Big)}    \rc\rc
\tilde c_2 & = & -{3\,\,R^4\over 10\alpha}\,{r_h^{{11\over 3}}\over d^{\,2}+\alpha\,r_h^{{16\over 3}}}\,k^2+{3\,R^4\over 100\, r_h^3}\,\omega^2 \ .
\label{Diffusion_A_c_values_off}
\eear
Following the same procedure as in the in-plane case, we next expand $E(r)$ at low frequency and write $E(r)$ near the horizon as in (\ref{Diffusion_E_nh_leading}), where now $\beta$ is given at leading order by:
\beq
\beta\,=\,i\,{k_\perp^2\over \omega}\,{R^2\,r_h^{{14\over 3}}\over \alpha\,\Big(d^{\,2}+\alpha\,r_h^{{16\over 3}}\Big)}\ .
\eeq
Taking the limits in the opposite order, we arrive exactly at (\ref{E_eq_low_freq}). Thus, we can expand $E(r)$ as in (\ref{Diffusion_E_low_freq_nh}) with ${\cal I}$ being the integral written in (\ref{Diffusion_cal_I_integral}). The constant $c_E$ can be obtained from the consistency of both expansions. This procedure yields
\beq
c_E=-i\,{k_\perp^2\over \omega}\,{R^2\over \alpha^2\, r_h}\,\Big(r_h^{{16\over 3}}\,+\,{d^{\,2}\over \alpha} \Big)^{{1\over 2}}\, E_{nh}\ ,
\label{cE_Enh_off}
\eeq
and, by again imposing Dirichlet boundary conditions at the UV we get the diffusion dispersion relation $\omega\,=\,-i\,D_{\perp}\,k_\perp^2+\ldots$, where $D_{\perp}$ is given by:
\beq
D_{\perp} = {3\,r_h^{{1\over 3}}\,R^2\over 4\,\alpha^2}\,\Big(1\,+\,{d^{\,2}\over \alpha\,r_h^{{16\over 3}}}\Big)^{{1\over 2}}F\Big({1\over 4}\,, \,{3\over 2}\,;{5\over 4}\,;\,-{d^{\,2}\over \alpha\,r_h^{{16\over 3}}}\Big)\ .
\label{Diffusion_D_perp}
\eeq
Let us now reintroduce reduced quantities $\hat d$, $\hat \omega$, and $\hat k_\perp$ which allow to absorb $R$, $r_h$, and $\alpha$ in (\ref{E_eq_off_plane}). Due to different scaling in the off-plane direction one needs to be a bit more careful with the momentum. The $\hat d$ and $\hat \omega$ are defined as in (\ref{Diffusion_hattted_quantities}), whereas $\hat k_\perp$ must be defined as:
\beq
\hat k_{\perp}\,=\,{R^2\over \alpha\,r_h^{{1\over 3}}}\,k_\perp\ .
\eeq
Using these definitions the dispersion relation for the off-plane diffusion can be written as $\hat\omega\,=\,-i\,\hat D_{\perp}\,\hat k_{\perp}^2+\ldots$, where $\hat D_{\perp}$ is related to $D_{\perp}$  as:
\beq
\hat D_{\perp}\,=\,{\alpha^2\over r_h^{{1\over 3}}\,R^2}\,D_{\perp}\,=\,
\Big({5\over 6\pi}\Big)^{{1\over 3}}\,{\alpha^2\over R^{{8\over 3}}}\,
{D_{\perp}\over T^{{1\over 3}}}\,\,.
\eeq
Moreover, from the expression of $D_{\perp}$ in (\ref{Diffusion_D_perp}) we get that $\hat D_{\perp}$ is given by:
\beq
\hat D_{\perp}\,=\,{3\over 4}\big(1+\hat d^{\,2}\big)^{{1\over 2}}\,
F\Big({1\over 4}\,, \,{3\over 2}\,;{5\over 4}\,;\,-
\hat d^{\,2}\Big)\,\,,
\label{hat_D_perp}
\eeq
which means that $\hat D_{\perp}=\hat D_{\parallel}$.  The fact that the reduced, non-physical,  in-plane and off-plane diffusion constants can be made equal by a temperature-dependent rescaling is a reflection of the scaling symmetry of the background metric. However, we should emphasize that the physical diffusion constants $D_{\perp}$ and $D_{\parallel}$ behave rather differently due to the different scalings used to define the corresponding hatted quantities. For example, at high temperature $\hat D_{\perp}=\hat D_{\parallel}\approx{3\over 4}$ and thus:
\beq
D_{\perp}={3\over 4}\Big({6\pi\over 5}\Big)^{{1\over 3}}\,{R^{{8\over 3}}\over \alpha^2}\,T^{{1\over 3}}\,\,,
\qquad\qquad
(T\to \infty)\,\,,
\eeq
which means that, contrary to $D_{\parallel}$, the off-plane diffusion constant grows as $T^{{1\over 3}}$ for large $T$. Moreover, at low temperature, $\hat D_{\perp}\,\sim\,\hat d^{{1\over 2}}\sim
 T^{-{4\over 3}}$ and, since 
$D_{\perp}\sim T^{{1\over 3}}\,\hat D_{\perp}$, we have:
\beq
D_{\perp}\sim T^{-1}\,\,.
\qquad\qquad
(T\to 0)\,\,.
\eeq

\section{Discussion}\label{sec:discussion}

Holography offers a possibility to investigate various phases of dense matter in the strong coupling regime. Intense effort has been put to study inhomogeneous phases,\footnote{A field theory model to study spontaneous inhomogeneous phases has been recently described in \cite{Musso:2019kii}.} starting with the early works of \cite{Domokos:2007kt,Nakamura:2009tf} with the first explicit example \cite{Bergman:2011rf} showing striped phases for fundamental matter realized with defect flavor D7'-branes in $AdS_5\times S^5$ spacetime \cite{Bergman:2010gm}. The construction and analysis of such modulated ground states has led to increasingly demanding numerics \cite{Jokela:2014dba,Jokela:2016xuy,Jokela:2017ltu} due to the highly non-linear DBI action. 

Another important merit of holographic techniques is to yield useful toy models that keep computations simple.\footnote{In this regard, the smearing technique adopted to describe a layered system has some conceptual analogy with the homogeneous breaking of translations realized by Q-lattices and helical models, see for example \cite{Donos:2013eha,Donos:2011ff,Amoretti:2016bxs} and \cite{Musso:2018wbv} for a model in field theory. In both cases, under a simplifying hypothesis, a spatial structure does not backreact on the densities, which remain spatially independent.} The D3-D5 geometry \cite{Conde:2016hbg} has provided us with a powerful framework to study anisotropic media; in the present paper, we probe this medium with fundamental matter and explore both the thermodynamics and the excitation spectra along and across the anisotropic direction. We believe that our work sets the standard and sparks many other investigations in the future. One particularly interesting case is in the context of the astrophysics of compact objects where anisotropy seems to offer clues regarding the ``universal relations'' for neutron stars \cite{Alexander:2018wxr} and the already established holographic modeling of isotropic matter \cite{Hoyos:2016zke,Annala:2017tqz,Jokela:2018ers,Ishii:2019gta,Hirayama:2019vod,Chesler:2019osn,Ecker:2019xrw}.

While the D3-D5 geometry enables a multitude of investigations of anisotropic matter, it is not at all perfect. 
The real life anisotropic systems in laboratories consist of a finite number of layers of perhaps different materials with finite separation. 
The D3-D5 system is an idealization: The anisotropic direction spans an infinite range and the separation between the layers is strictly vanishing. 
Interestingly, contrary to common beliefs, relaxing these approximations in a controlled fashion may not be a major hurdle and can be obtained by a 
judicious selection of the smearing form for the D5-branes. Reaching this milestone would be very rewarding and will teach us further lessons on surface phenomena. 
In particular, it would allow us to dissect the issues of the surface plasmon left open in this paper. Besides, the shear-viscosity over entropy density bound \cite{Policastro:2001yc,Kovtun:2004de} is known to be affected by anisotropy \cite{Erdmenger:2010xm,Link:2017ora,Rebhan:2011vd,Jain:2015txa,Ge:2014aza,Inkof:2019gmh};  it would be interesting to consider the status of the bound in the presence of a further scale introduced by a finite inter-layer spacing. Furthermore, a finite-spacing would break the translations along the orthogonal direction to the layers.


\addcontentsline{toc}{section}{Acknowledgements}
\paragraph{Acknowledgements}

\noindent

We would like to thank Daniel Are\'an, Carlos Hoyos, and Javier Tarr\'io for relevant and interesting discussions. The research of U.~G. and M.~T. has been funded by the Swedish Research Council. The research of N.~J. has been supported in part by the Academy of Finland grant no. 1322307. The research of D.~M. and A.~V.~R. has been funded by the Spanish grants FPA2014-52218-P and FPA2017-84436-P by Xunta de Galicia (ED431C-2017/07), by FEDER, and by the Mar\'ia de Maeztu Unit of Excellence MDM-2016-0692.

\vskip 1cm
\renewcommand{\theequation}{\rm{A}.\arabic{equation}}
\setcounter{equation}{0}

\appendix

\section{Kappa symmetry} \label{kappa}

Let us verify that the embeddings we found  in Section~\ref{SUSY} are supersymmetric.  We will  work in the following vielbein basis for the metric presented in Section~\ref{sec:setup} for $B=1$:
\bear
&&E^{x^{\mu}}\,=\,h^{-{1\over 4}}\,dx^{\mu}\,\,,\qquad\qquad (\mu=0, 1,2),
\qquad\qquad
E^{x^{3}}\,=\,h^{-{1\over 4}}\,e^{-\phi}\,dx^{3}\,\,,\rc\rc
&&E^{r}\,=\,h^{{1\over 4}}\,dr\,\,,
\qquad\qquad E^{i}\,=\,{1\over 2}\,h^{{1\over 4}}\,e^{g}\,\cos\chi\,\,\omega^{i}\,\,,
\qquad \qquad (i=1,2)\,\,,\rc\rc
&&E^3\,=\,{1\over 2}\,h^{{1\over 4}}\,e^{g}\,\cos\chi\,\sin\chi\,\omega^{3}\,\,,
\qquad\qquad
E^4\,=\,h^{{1\over 4}}\,e^{g}\,d\chi\,\,,\rc\rc
&&E^5\,=\,h^{{1\over 4}}\,e^{f}\,\big(d\tau+{1\over 2}\cos^2\chi\,\, \omega^3\big)\,\,,
\label{10d_vielbein_basis}
\eear
where  the warp factor $h$ is the function written in (\ref{warp_blackening}) and $g$ and $f$ are given by:
\beq
e^{g}\,=\,{r\over \sqrt{b}}\,\,,
\qquad\qquad
e^{f}\,=\,{r\over b}\,\,.
\label{background_functions}
\eeq
We will write the SU(2) left-invariant one-forms $\omega^1$, $\omega^2$, and $\omega^3$  in terms of three angles $(\theta, \varphi, \psi)$ as follows:
\bear
&&\omega^1\,=\,\cos\psi\,d\theta\,+\,\sin\psi\,\sin\theta\,d\varphi \rc\rc
&&\omega^2\,=\,\sin\psi\,d\theta\,-\,\cos\psi\,\sin\theta\,d\varphi \rc\rc
&&\omega^3\,=\,d\psi\,+\,\cos\theta\,d\varphi\,\,,
\label{SU2_oneforms}
\eear
where $0\le \theta\le \pi$,  $0\le \varphi<2 \pi$, and $0\le \psi<4 \pi$.

The supersymmetric embeddings of the D7-brane  are those that  satisfy the kappa symmetry condition:
\beq
\Gamma_{\kappa}\,\epsilon\,=\,\pm\epsilon\,\,,
\label{kappa_epsilon}
\eeq
where $\epsilon$ is a Killing spinor of the background. For a D7-brane without any worldvolume gauge field, the matrix $\Gamma_{\kappa}$ is given by:
\beq
\Gamma_{\kappa}\,=\,{1\over 8!\sqrt{-g_8}}\,\,
\epsilon^{\alpha_1\,\cdots\,\alpha_8}\,i\sigma_2\,\gamma_{\alpha_1\cdots\alpha_8}\,\,,
\eeq
where $\gamma_{\alpha_1\cdots\alpha_8}$ is the antisymmetrized product of induced Dirac matrices and $\sigma_2$ is the Pauli matrix. We will take the following set of worldvolume coordinates:
\beq
\xi^{\alpha}\,=\,(x^0, x^1, x^2, x^3, r, \theta, \psi, \varphi)\,\,,
\eeq
and the embedding will be determined by  two embedding equations of the type:
\beq
\tau\,=\,{\rm constant}\,\,,
\qquad\qquad
\chi\,=\,\chi(r)\,\,.
\eeq
The kappa symmetry matrix for this ansatz  is:
\beq
\Gamma_{\kappa}\,=\,{i\sigma_2\over \sqrt{-g_8}}\,\,
\gamma_{x^0 x^1 x^2 x^3 r \theta\psi \varphi}\,\,.
\label{G_k_ansatz}
\eeq
To obtain the induced $\gamma$-matrices, let us write the vielbein one-forms (\ref{10d_vielbein_basis}) in a coordinate basis as:
\beq
E^{\underline{M}}\,=\, {\cal E}^{\underline{M}}_{N}\,dX^{N}\,\,.
\eeq
Then, the $\gamma$'s are:
\beq
\gamma_{\alpha}\,=\,\partial_{\alpha}X^{N}\,{\cal E}^{\underline{M}}_{N}\,\Gamma_{\underline{M}}\,\,,
\eeq
where the $\Gamma$'s are flat constant matrices for the vielbein (\ref{10d_vielbein_basis}). Specifically, 
the induced $\gamma$-matrices on the worldvolume  for our configuration are:
\bear
&&\gamma_{x^\mu}\,=\,h^{-{1\over 4}}\,\Gamma_{x^{\mu}}\,\,,
\qquad\qquad (\mu=0,1,2)\,\,,\rc\rc
&&\gamma_{x^3}\,=\,h^{-{1\over 4}}\,e^{-\phi}\,\Gamma_{x^{3}}\,,\,
\qquad\qquad
\gamma_{r}\,=\,h^{{1\over 4}}\,\big[
\Gamma_{r}\,+\,
e^{g}\,\chi'\,\Gamma_{4}\big]\,\,,\rc\rc
&&\gamma_{\theta}\,=\,{h^{{1\over 4}}\over 2}\,\cos\chi\,e^g
\big[\cos\psi\Gamma_1\,+\,\sin\psi\Gamma_2\big]\,\,,
\quad
\gamma_{\psi}\,=\,{h^{{1\over 4}}\over 2}\,\cos\chi\,\big[
\sin\chi\,e^{g}\,\Gamma_3\,+\,\cos\chi\,e^{f}\,\Gamma_5\big]\,\,,
\rc\rc
&&\gamma_{\varphi}=
{h^{{1\over 4}}\over 2}\cos\chi\big[
\sin\psi\sin\theta e^g\Gamma_1-\cos\psi\sin\theta e^{g}\Gamma_2+
\sin\chi\cos\theta e^{g} \Gamma_3+\cos\chi\,\cos\theta e^f\Gamma_5\big]\,\,.\qquad
\eear
Clearly, we can factorize the antisymmetrized product in (\ref{G_k_ansatz}) as:
\beq
\gamma_{x^0 x^1 x^2 x^3 r \theta\psi \varphi}\,=\,
\gamma_{x^0 x^1 x^2 x^3}\,
\gamma_{r \theta\psi \varphi}\,\,,
\eeq
where the first factor is:
\beq
\gamma_{x^0 x^1 x^2 x^3}\,=\,
h^{-1}\,e^{-\phi}\,
\Gamma_{x^0 x^1 x^2 x^3}\,\,.
\eeq
The second factor can be written as:
\beq
\gamma_{r \theta\psi \varphi}\,=\,{h\over 8}\,e^{2g}\,\cos^3\chi\,\sin\theta\,\Gamma_{12}\,
\big[\,c_1\,\Gamma_{r3}\,+\,c_2\,\Gamma_{r5}\,+\,c_3\,\Gamma_{34}\,+\,c_4\,\Gamma_{45}\big]\,,
\eeq
where the $c_i$ coefficients are given by:
\bear
&&c_1\,=\,e^g\,\sin\chi\,\,,
\qquad\qquad\qquad
c_2\,=\,e^{f}\,\cos\chi\,\,,\rc\rc
&&c_3\,=\,-e^{2g}\,\sin\chi\,\chi'\,\,,
\qquad\qquad
c_4\,=\,e^{g+f}\cos\chi\,\chi'\,\,.
\label{c_coefficients}
\eear
Putting everything together, we can write $\Gamma_{\kappa}$ as:
\beq
\Gamma_{\kappa}\,=\,{e^{-\phi+2g}\over 8\,\sqrt{-g_8}}\,
\cos^3\chi\,\sin\theta\,(i\sigma_2)\,
\Gamma_{x^0 x^1 x^2 x^3}\,\Gamma_{12}\,
\big[c_1\,\Gamma_{r 3}\,+\,c_2\,\Gamma_{r5}\,+\,c_3\,\Gamma_{34}\,+\,c_4\,\Gamma_{45}\big]\,\,.
\label{Gamma_kappa_simplified}
\eeq
It was proven in  \cite{Conde:2016hbg} that  the Killing spinor of the background can be written as:
\beq
\epsilon\,=\,h^{-{1\over 8}}\,e^{{3\over 2}\,\Gamma_{12}\,\tau}\,\,\eta\,\,,
\label{spinor_sol}
\eeq
where $\eta$ is a doublet of constant Majorana-Weyl spinor satisfying the following projection conditions:
\bear
&&\Gamma_{x^0\,x^1\,x^2\,x^3}\,(i\sigma_2)\,\eta\,=\,\eta\rc\rc
&&\Gamma_{rx^314}\,\sigma_1\,\eta\,=\,\eta\rc\rc
&&\Gamma_{12}\,\eta\,=\,\Gamma_{34}\,\eta\,=\,\Gamma_{r5}\,\eta\,=\,i\sigma_2\,\eta\,\,.
\label{eta_projections}
\eear
As $\Gamma_{12}$ commutes with the products of matrices of the rhs of (\ref{Gamma_kappa_simplified}), we can rewrite the kappa symmetry condition (\ref{kappa_epsilon}) as:
\beq
\Gamma_{\kappa}\,\eta\,=\,\pm\eta\,\,.
\label{kappa_eta}
\eeq
Moreover, from the last equation in (\ref{eta_projections}) one can easily demonstrate that:
\beq
\Gamma_{45}\,\eta\,=\,\Gamma_{r3}\,\eta\,\,.
\eeq
Thus, $\Gamma_{\kappa}$ acting on $\eta$ can be written as:
\beq
\Gamma_{\kappa}\,\eta\,=\,
{e^{-\phi+2g}\over 8\,\sqrt{-g_8}}\,
\cos^3\chi\,\sin\theta\,
\big[ (c_1+c_4)\Gamma_{45}\,+\,(c_2+c_3) i\sigma_2
\big]
(i\sigma_2)\,
\Gamma_{x^0 x^1 x^2 x^3}\,\Gamma_{12}\,\eta\,\,.
\label{G_k_eta_simplified}
\eeq
Moreover, since from (\ref{eta_projections}) we have:
\beq
(i\sigma_2)\,
\Gamma_{x^0 x^1 x^2 x^3}\,\Gamma_{12}\,\eta\,=\, (i\sigma_2)\,\eta\,\,,
\eeq
we  can write $\Gamma_{\kappa}\eta$ simply as:
\beq
\Gamma_{\kappa}\,\eta\,=\,
{e^{-\phi+2g}\over 8\,\sqrt{-g_8}}\,
\cos^3\chi\,\sin\theta\,
\big[-(c_2+c_3)\,+\,(c_1+c_4)\,\Gamma_{45}\,(i\sigma_2)\big]\,\eta\,\,.
\label{G_k_eta_fullsimplified}
\eeq
According to (\ref{kappa_eta}), we are looking for embeddings such  that $\Gamma_k$ acts on $\eta$ as plus/minus the identity. Inspecting the rhs of (\ref{G_k_eta_fullsimplified})  we notice that the last two terms 
act on $\eta$ as a non-trivial matrix. Therefore, we should require that the coefficient of these terms is zero, namely:
\beq
c_1+c_4=0\,\,,
\label{BPS_c}
\eeq
which is the BPS equation for the embedding. Taking into account the values of $c_1$ and $c_4$ written in (\ref{c_coefficients}), we can recast (\ref{BPS_c}) as the following first-order differential equation:
\beq
e^{f}\,\chi'\,=\,-\tan\chi\,\,.
\label{BPS_ef}
\eeq
Using the value of $e^f$ displayed in (\ref{background_functions}), we can rewrite this BPS equation as:
\beq
{r\over b}\,{d\chi\over dr}\,=\,-\tan \chi\,\,,
\label{BPS_r}
\eeq
which is the same as (\ref{BPS_ODE}). 
Notice that the general solution of (\ref{BPS_r}) is indeed (\ref{sin_chi_BPS}).  To finish with the proof of kappa symmetry, let us compute the terms of $\Gamma_{\kappa}\,\eta$ which are proportional to the unit matrix. One can readily prove from (\ref{DBI_det_general}) and (\ref{c_coefficients}) that, when the BPS condition (\ref{BPS_r}) holds, we have:
\bea
\sqrt{-g_8}\big|_{BPS} & = & {r^3\,e^{-\phi}\,\cos^2\chi\,\sin\theta\over 8\, b^2}\,\big[\cos^2\chi+b\,\sin^2\chi\big]\rc\rc
(c_2+c_3)\big|_{BPS} & = & {r\over b\,\cos\chi}\,\big[\cos^2\chi+b\,\sin^2\chi\big]\ .
\eea
From these two equations we get:
\beq
\sqrt{-g_8}\big|_{BPS}\,=\,{e^{2g-\phi}\,\cos^3\chi\,\sin\theta\over 8}\,
(c_2+c_3)\big|_{BPS}\,\,,
\eeq
and, as a consequence, 
\beq
\Gamma_{\kappa}\,\eta\,=\,-\eta\,\,.
\eeq

\vskip 1cm
\renewcommand{\theequation}{\rm{B}.\arabic{equation}}
\setcounter{equation}{0}

\section{The dictionary}\label{sec:dictionary}

The bare quark mass $m_q$ is obtained from the Nambu-Goto action of a fundamental string hanging from the boundary to the horizon.  We will obtain $m_q$  as the limit of the constituent quark mass $M_c$, which contains the effects of the thermal screening of the quarks and is also obtained from the Nambu-Goto action. 
Let us consider a  fundamental string extended in $t$, $P$ at constant $\rho=0$. The induced metric is:
\beq
 ds_2^2\,=\,-{r_h^2\over 2^{{6\over 5}}\,R^2}\,P^{{9\over 4}}\,(1+P^{-{15\over 4}})^{-{4\over 5}}\,
 (1-P^{-{15\over 4}})^{2}\,dt^2\,+\,\Big({9\over 8}\Big)^2\,R^2\,{dP^2\over P^2}\,\,,
\eeq
whose determinant is:
\beq
 \sqrt{-\det g_2}\,=\,{9\over 8}\,{r_h\over 2^{{3\over 5}}}\,P^{{1\over 8}}\,
 {1-P^{-{15\over 4}}\over (1+P^{-{15\over 4}})^{{2\over 5}}}\,\,.
\eeq
The constituent quark mass $M_c$ is minus the action per unit of time of the Nambu-Goto action:
\beq
 M_c\,=\,{1\over 2\pi}\,\int_1^{P_0}\,e^{{\phi\over 2}}\,\sqrt{-\det g_2}\,\,dP\,\,,\label{const_mass_def}
\eeq
where the $e^{{\phi\over 2}}$ factor is due the fact that our metric is written in Einstein frame ($g_{string}=e^{{\phi\over 2}}\,g_{Einstein}$). When $\rho=0$, we have:
\beq
 e^{{\phi\over 2}}\,=\,\Big({3\over 4\,Q_f}\Big)^{{1\over 3}}\,
 {r_h^{{1\over 3}}\over 2^{{1\over 5}}}\,\,P^{{3\over 8}}\,(1+P^{-{15\over 4}})^{{1\over 5}}\,\,.
 \eeq
Plugging this into (\ref{const_mass_def}), we get:
\beq
 M_c\,=\,{9\over 16\,\pi}\,\,\Big({3\over 4\,Q_f}\Big)^{{1\over 3}}\,{r_h^{{4\over 3}}\over 2^{{4\over 5}}}\,J(P_0)\ , \label{eq:constituentmass}
\eeq
where  $J(P_0)$ is defined as the following integral:
\beq
  J(P_0)\equiv \int_1^{P_0}\,dP\,{P^{{1\over 2}}(1-P^{-{15\over 4}})\over  (1+P^{-{15\over 4}})^{{1\over 5}}}\,\,.
\eeq
This integral can be obtained in analytic form and is given by:
\beq
 J(P_0)\,=\,{2\over 3}\,\Big[P_0^{{3\over 2}}\Big(1+{1\over P_0^{{15\over 4}}}\Big)^{{4\over 5}}\,-\,2^{{4\over 5}}\Big]\,\,.
\eeq
In order to get the bare quark mass $m_q$, let us write $P_0=m$ and consider the limit $m\to\infty$, where $M_c$ equals $m_q$.  We get:
\beq
 m_q\,=\,{3\over 8\pi\,2^{{4\over 5}}}\,\Big({3\over 4\,Q_f}\Big)^{{1\over 3}}\,r_h^{{4\over 3}}\,m^{{3\over 2}}\,\,.\label{mass_dictionary}
\eeq
Notice that,  for fixed $m_q$ the mass parameter $m$ depends on the temperature as $m\approx T^{-{8\over 9}}=T^{-b}$. 
 
The condensate  $\langle {\cal O}_m\rangle$  is obtained by computing the derivative of the free energy $F$ with respect  to the bare quark mass $m_q$:
\beq
 \langle {\cal O}_m\rangle\,=\,{\partial F\over \partial m_q} = {2\over 3}{m\over m_q} {\partial F\over \partial m} \ .
\eeq
The free energy $F$ can be written as:
\beq
 F\,=\,{\cal N}\,\big({\cal I}_{bulk}+{\cal I}_{bdy}\big)\,\,,
\eeq
where ${\cal N}$ is written in (\ref{calN_eta}), ${\cal I}_{bulk}$ is the integral (\ref{I_bulk}) evaluated on-shell and  ${\cal I}_{bay}$  is given in (\ref{I_bdy_explicit}). Since  the normalization factor ${\cal N}$ does not depend on the mass parameter $m$, we have:
 \beq
{\partial F\over \partial m}\,=\,{\cal N}\,\Big(
{\partial {\cal \cal I}_{bulk}\over \partial m}\,+\,
{\partial {\cal \cal I}_{bdy}\over \partial m}\Big)
\eeq

 In order to compute its derivative with respect to the mass parameter $m$, let us write the bulk action as:
 \beq
 {\cal \cal I}_{bulk}\,=\,\int_{u_{min}}^{u_{max}}\,du\,{\cal J}(u,\eta, \dot\eta)\,\,,
 \eeq
 where $ {\cal J}$ is given by:
 \beq
 {\cal J}\,=\,u^{{7\over 2}}\,f\,\tilde f^{{7\over 5}}\,(1-\eta^2)\,\sqrt{1-{\eta^2\over 9}}\,
  \sqrt{1-\eta^2+{8u^2\over 9}\,\dot\eta^2}\,\,.
 \eeq
 Then:
  \beq
 {\partial {\cal \cal I}_{bulk}\over \partial m}\,=\,
 \int_{u_{min}}^{u_{max}}\,du\Big[
 {\partial {\cal J}\over \partial\dot\eta}\,{d\over du}\Big({\partial\eta\over\partial m}\Big)\,+\,
  {\partial {\cal J}\over \partial\eta}\,{\partial\eta\over\partial m}\,\Big]\,\,.
 \eeq
 Taking into account that the on-shell action satisfies the Euler-Lagrange equation 
 $ {\partial {\cal J}\over \partial\eta}\,=\,{d\over du}\, \Big({\partial {\cal J}\over \partial\dot\eta}\Big)$, we can integrate this equation and write:
 \beq
  {\partial {\cal \cal I}_{bulk}\over \partial m}\,=\,
{\partial {\cal J}\over \partial\dot\eta}\,{\partial\eta\over\partial m}
\Big|_{u_{min}}^{u_{max}}\,\,.
\label{variation_bulkI}
\eeq
The momentum appearing on the right-hand side of this equation is:
 \beq
 {\partial {\cal J}\over \partial\dot\eta}\,=\,{8\over 9}\,u^{{11\over 2}}\,f\,\tilde f^{{7\over 5}}\,
 {(1-\eta^2)\,\sqrt{1-{\eta^2\over 9}}\over
  \sqrt{1-\eta^2+{8u^2\over 9}\,\dot\eta^2}}\,\,\dot\eta\,\,.
  \label{momentum}
 \eeq
 Let us consider a black hole embedding, in which $u_{min}=1$. In this case the momentum (\ref{momentum})  vanishes at the lower value in (\ref{variation_bulkI}) and the  only non-vanishing contribution comes from the upper limit $u_{\max}\to\infty$.  To evaluate this contribution we use the asymptotic UV expansion (\ref{eta_UV}), as well as:
  \beq
 \dot\eta\,=\,-{m\over u^2}\,-\,{7\over 2}\,{c\over u^{{9\over 2}}}\,+\,\ldots\,\,,
 \qquad\qquad
 {\partial\eta\over \partial m}\,=\,{1\over u}\,+\,{1\over u^{{7\over 2}}}\,{\partial c\over \partial m}\,+\,\ldots\,\,.
 \eeq
 We get:
  \beq
  {\partial {\cal \cal I}_{bulk}\over \partial m}\,=\,
 {\partial {\cal J}\over \partial\dot\eta}\,{\partial\eta\over\partial m}
\Big|_{u_{max}}\,=\,{8\over 9}\,m\,u_{max}^{{1\over 2}}(m^2-u_{max}^2)\,-\,{28\over 9}\,c\,-\,{8\over 9}\,m\,
{\partial c\over \partial m}\,\,,
\label{d_Ibulk}
\eeq
where we have only included terms that are non-vanishing when $u_{max}\to\infty$. 
Moreover, by computing the derivative with  respect to $m$ of the explicit expression of ${\cal \cal I}_{bdy}$ written in 
(\ref{I_bdy_explicit}), we obtain: 
 \beq
  {\partial {\cal \cal I}_{bdy}\over \partial m}\,=\,-{8\over 9}\,m\,u_{max}^{{1\over 2}}(m^2-u_{max}^2)\,+\,{8\over 9}\,c\,+\,
  {8\over 9}\,m\,{\partial c\over \partial m}\,\,.
  \label{d_Ibdy}
 \eeq
 Adding (\ref{d_Ibulk}) and (\ref{d_Ibdy}) we notice that the result is finite when $u_{max}\to\infty$ and given by
  \beq
 {\partial F\over \partial m}\,=\,-{20\over 9}\,{\cal N}\,c\,\,.
 \label{mass_derivative_F}
 \eeq
 Notice also that the terms containing $\partial c/\partial m$  vanish and the VEV $ \langle {\cal O}_m\rangle$ is proportional to $c$, as expected:
 \beq
\langle {\cal O}_m\rangle\,=\,-{40\over 27}\,{m\,{\cal N}\over m_q}\,c\,\,.
\label{Om_c}
 \eeq
 
 \vskip 1cm
\renewcommand{\theequation}{\rm{C}.\arabic{equation}}
\setcounter{equation}{0}

\section{Critical embeddings}
\label{CE}

In this appendix we  perform an analysis of critical embeddings, at zero density  and non-zero temperature,  of the probe D7-brane along the lines of \cite{Mateos:2007vn,Jokela:2012dw}. In these configurations the D7-brane touches the horizon near $\chi=\pi/2$. It is then useful to define the following  new coordinates $(y,z)$ as
\begin{equation}\label{cry_cor}
 \chi = \frac{\pi}{2} - \frac{\sqrt{b}}{R}\, y\ , 
 \qquad r = r_h + C z^\alpha\ ,
\end{equation}
to zoom on the vicinity of the contact point. In (\ref{cry_cor}) $C$ and $\alpha$ are constants to be determined in order to cast the induced metric into a Rindler form. In the new coordinates \eqref{cry_cor}, the emblackening factor $B(r)$ near the horizon reads
\begin{equation}
 B(r) = 1 - \left(\frac{r_h}{r}\right)^{\frac{10}{3}}
      = \frac{10}{3} \frac{C}{r_h} z^\alpha + \ldots \ .
\end{equation}
Choosing
\begin{equation}
 \alpha = 2\ , \qquad 
 C = \frac{5}{6} \frac{r_h}{R^2} = \pi T\ ,
 \label{alpha_C}
\end{equation}
where $T$ is the temperature defined in (6.111), the radial and temporal pieces of the induced metric take the desired Rindler form, namely
\begin{equation}
 \left.ds^2\right|_{(t,z)} = -(2\pi T)^2 z^2 dt^2 + dz^2\ .
\end{equation}
In terms of the $z$ and $y$ coordinates, 
the  metric in the vicinity of the horizon, at leading order, becomes
\begin{equation}\label{ind_hor}
 \begin{split}
 ds^2_8 =& - (2\pi T)^2 z^2 dt^2 
          +\frac{r_h^2}{R^2}\left[(dx^1)^2+(dx^2)^2+e^{-2\phi}(dx^3)^2\right]\\
         &\qquad + dz^2 + dy^2 + \frac{y^2}{4}  \left[(\omega_1)^2+(\omega_2)^2+(\omega_3)^2\right]\,+\,\ldots\,\,,
 \end{split}
\end{equation}
where the ellipsis represents terms that do not contribute to the embedding of the probe. Describing the D7 embedding with the parametrization $y=y(z)$, one has $dz^2+dy^2 = (1+\dot y^2)dz^2$, where the dot denotes derivation with respect to $z$. Substituting into the metric \eqref{ind_hor}, one obtains the associated DBI Lagrangian density
\begin{equation}\label{DBI_hor_z}
 {\cal L}_{\text{DBI}} \propto \, z\, y^3\, \sqrt{1+\dot y^2}\ .
\end{equation}
The equation of motion descending from \eqref{DBI_hor_z} is
\begin{equation}\label{bh_eom}
 z y \ddot y + (y\dot y - 3 z)(1+\dot y^2) = 0\ .
\end{equation}
The Lagrangian density \eqref{DBI_hor_z} and the associated equation of motion \eqref{bh_eom}
correspond to the case $n=3$ in the generic analysis performed in \cite{Mateos:2007vn}, according to the fact 
that three internal directions wrapped by the D7-brane collapse in the critical embedding at $y=0$. More generically, the parametrization $y=y(z)$ is suitable to describe near-to-critical black hole embeddings where the D7-brane intersects the horizon at an angle 
\begin{equation}
 \chi_h = \frac{\pi}{2} - \frac{\sqrt{b}}{R}\, y_h\ , 
\end{equation}
according to \eqref{cry_cor} where $y_h=y(z=0)$. More precisely, the equation of motion \eqref{bh_eom} must be solved imposing the boundary conditions
\begin{equation}\label{nc_bh_bc}
 y(z=0) = y_h \ .
\end{equation}
By consistency
\begin{equation}
 \dot y(z=0) = 0\ ,
\end{equation}
which can be seen solving the equation order by order.

To study the near-to-critical Minkowski embeddings instead, one has to parameterize the embeddings as $z=z(y)$, where $z_h=z(y=0)$ represents the IR radial distance of the D7-brane from the horizon. From \eqref{ind_hor} one gets the DBI Lagrangian
\begin{equation}\label{DBI_hor_y}
 {\cal L}_{\text{DBI}} \propto \, z\, y^3\, \sqrt{1+z'^2}\ ,
\end{equation}
where the prime denotes differentiation with respect to $y$. The associated equation of motion is
\begin{equation}\label{Min_eom}
 yzz''+(3zz'-y)(1+z'^2) = 0\ ,
\end{equation}
to be solved imposing the boundary conditions
\begin{equation}
 z(y=0) = z_h\ , 
\end{equation}
that implies
\begin{equation}
 z'(y=0) = 0 \ .
\end{equation}
The critical solution 
\begin{equation}
 y = \sqrt{3} z\ ,
 \label{y_critical}
\end{equation}
solves both \eqref{bh_eom} and \eqref{Min_eom} with boundary conditions
\begin{equation}
 y_h = z_h = 0\ .
\end{equation}
The critical solution \eqref{y_critical} can be seen as a limiting case  of both near-critical black hole and Minkowski embeddings. The limit involves appropriate scaling of $y_h$ and $z_h$ in the pinching direction, $z$ and $y$, respectively.

Let us now write the critical solution in terms of the isotropic variable $u$ defined in  (\ref{u_r}), rewritten as
\begin{equation}
 u^{\frac{5}{3b}} = \left(\frac{r}{r_h}\right)^{\frac{5}{3}} \left(1+\sqrt{B(r)}\right)
                          = \left(\frac{r}{r_h}\right)^{\frac{5}{3}} \left(1+\sqrt{1-\left(\frac{r_h}{r}\right)^{\frac{10}{3}}}\right)\ .
\end{equation}
Recalling the definition of the coordinate $z$ in (\ref{cry_cor}) and using (\ref{alpha_C}), near the horizon one has
\begin{equation}
 u^{\frac{5}{3b}} = 1 + \sqrt{\frac{10}{3}\frac{\pi T}{r_h}} \ z + \ldots
                          = 1 + \frac{5}{3R} \ z +  \ldots  \ \rightarrow \ u = 1 + \frac{b}{R}\, z + \ldots \ .
\end{equation}
Adopting the Cartesian-like coordinates (\ref{P_rho_def}) and recalling (\ref{cry_cor}), in the near-horizon region one has
\begin{equation}
 P \sim u \sim 1 + \frac{b}{R}\, z  \ , \qquad
 \rho \sim \frac{\sqrt{b}}{R}\, y  \ .
\end{equation}
On the critical solution (\ref{y_critical}) the coordinate $P$ becomes
\begin{equation}
 P \sim 1 + \sqrt{\frac{b}{3}}\; \rho\ , \ \frac{dP}{d\rho} \sim \sqrt{\frac{b}{3}}
\end{equation}
which corresponds to the following  in-falling angle in the $(P, \rho)$ plane:
\begin{equation}
\chi_h\,=\,
\arctan\left({2\sqrt{2}\over 3\sqrt{3}}
\right)
\sim 0.498 \ .
\end{equation}

Let us next study the  near-critical black hole solutions, which are described by the following deformation of the critical solution (\ref{y_critical}):
\begin{equation}\label{def}
 y = \sqrt{3}\, z + \xi(z)\ ,
\end{equation}
where $\xi(z)$ is a small function of $z$.  Linearizing (\ref{bh_eom}) in the deformation $\xi$  one gets
\begin{equation}\label{lin_bh}
z^2\ddot \xi + 4(z\dot \xi+\xi) = 0\ .
\end{equation}
Considering a power-law solution $\xi(z) \sim z^\nu$, we get that $\nu$ must satisfy
the quadratic equation $\nu^2+3\nu+4=0$, which has the following two solutions:
\begin{equation}
\nu = -\frac{3}{2} \pm i \frac{\sqrt{7}}{2}\ .
\end{equation}
The general solution to \eqref{lin_bh} is then given by
\begin{equation}\label{bh_cri_gen}
y = \sqrt{3}\, z 
+T^{-5/2} z^{-3/2}\left[A \sin\left(\frac{\sqrt{7}}{2}\ln (T\,z)\right)
+B \cos\left(\frac{\sqrt{7}}{2}\ln (T\, z)\right)\right]\ ,
\end{equation} 
where $A$ and $B$ are numerical coefficients while the power of $T$ has been introduced to comply with the correct physical dimensions. Notice that the perturbed solution
\eqref{bh_cri_gen} does not respect the boundary conditions \eqref{nc_bh_bc} discussed for the near-critical differential problem. The perturbation $\xi(z)$ of the critical solution introduced in \eqref{def} drives us away from criticality: the critical system is dynamically unstable.

The linearized problem \eqref{lin_bh} features a scale invariance, namely if $y(z)=f(z)$ for a certain function $f$ is a solution, then
\begin{equation}\label{res_sol}
\bar y(z) = \frac{f(z\mu)}{\mu}\ ,
\end{equation}
where $\mu$ is a number, is a solution too. The rescaled solution $\bar y$ corresponds to the rescaled boundary conditions 
\begin{equation}
 \bar y(z=0) =\bar y_h = \frac{y_h}{\mu} \ , \qquad \dot {\bar y}(z=0) = 0\ ,
\end{equation}
which implies
\begin{equation}\label{mu_yy}
 \mu = \frac{y_h}{\bar y_h}\ .
\end{equation}

Following Appendix F of \cite{Jokela:2012dw}, one can show that the coefficients $\bar A$ and $\bar B$ of the rescaled solution \eqref{res_sol} are related to the coefficients $A$ and $B$ in (\ref{bh_cri_gen}) as follows:
\begin{equation}
 \left(\begin{array}{c} \bar A\\ \bar B \end{array}\right) = \mu^{-\frac{5}{2}} {\cal M}(\mu) \left(\begin{array}{c} A\\ B \end{array}\right)\ ,
\end{equation}
where the matrix ${\cal M}(\mu)$ is defined as
\begin{equation}\label{mat_M}
 {\cal M}(\mu) = \left(\begin{array}{cc}\cos\left[\frac{\sqrt{7}}{2}\ln (\mu)\right]&-\sin\left[\frac{\sqrt{7}}{2}\ln (\mu)\right]\\
 \sin\left[\frac{\sqrt{7}}{2}\ln (\mu)\right]&\cos\left[\frac{\sqrt{7}}{2}\ln (\mu)\right]\end{array}\right)\ .
\end{equation}
The matrix \eqref{mat_M} satisfies the property 
\begin{equation}
{\cal M}(\mu) = {\cal M}^{-1}(\bar y_h) {\cal M}(y_h) \ .
\end{equation}
where one needs to recall \eqref{mu_yy}. Therefore one has 
\begin{equation}
 \bar y_h^{-\frac{5}{2}} {\cal M}(\bar y_h) \bar A
 = y_h^{-\frac{5}{2}} {\cal M}(y_h) A
 = v\ ,
\end{equation}
where $v$ is a constant vector characterizing the whole family of near-critical black hole embeddings. Since the matrix ${\cal M}(x)$ is periodic when $x$ is shifted by $\frac{\sqrt{7}}{4\pi}\ln(x)$, then 
\begin{equation}
 y_h^{-\frac{5}{2}} A \qquad \text{and}\qquad 
 y_h^{-\frac{5}{2}}  B
\end{equation}
are periodic functions of $\frac{\sqrt{7}}{4\pi}\ln(y_h)$ with period $1$; similarly
\begin{equation}
 \bar y_h^{-\frac{5}{2}}\bar A \qquad \text{and}\qquad 
 \bar y_h^{-\frac{5}{2}}  \bar B
\end{equation}
are periodic functions of $\frac{\sqrt{7}}{4\pi}\ln(\bar y_h)$ with period $1$.

\vskip 1cm
\renewcommand{\theequation}{\rm{D}.\arabic{equation}}
\setcounter{equation}{0}

\section{Thermal screening}
\label{TS}

We wish to study the quark-antiquark potential at non-zero temperature. We consider a string hanging from the boundary and reaching a minimal radial coordinate $r_0$, with its endpoints a distance $\ell$ from each other in the isotropic $x$-direction. We parametrize the string with the radial coordinate, which means that the midpoint will be found when $\frac{1}{x'(r)}=0$. The string metric is given by
\begin{equation}
\mathrm{d}s^2=\mathrm{d}r^2 \left(\frac{\sqrt{h(r)}}{B(r)}+\frac{x_0'(r){}^2}{\sqrt{h(r)}}\right)-\frac{\mathrm{d}t^2
   B(r)}{\sqrt{h(r)}}\,.
\end{equation}
The potential will be given by (minus) the action 
\begin{equation}
    E=\frac{1}{2\pi}\int \mathrm{d}t\mathrm{d}r \; \mathrm{e}^{\phi/2}\sqrt{-|g|}\,,
\end{equation}
regularized by subtracting the energy of strings from the boundary to the horizon.
The equations of motion for $x(r)$ can be written as

\begin{equation}
    x'(r)^2=\frac{r_0^{14/3} R^4 B\left(r_0\right)}{r^{26/3} B(r)^2-r^4 r_0^{14/3} B(r)
   B\left(r_0\right)}\,.
\end{equation}
It follows from this equation that the total separation between the pair can  be written as
\begin{equation}
    \ell=2 \int_{r_0}^\infty\mathrm{d}r\sqrt{\frac{r_0^{14/3} R^4 B\left(r_0\right)}{r^{26/3} B(r)^2-r^4 r_0^{14/3}
   B(r) B\left(r_0\right)}}\,.
\end{equation}
For small temperatures,  $\ell$ be approximated as
\begin{equation}
    \ell = \frac{2 \sqrt{\pi } R^2 \Gamma \left(\frac{5}{7}\right)}{r_0 \Gamma
   \left(\frac{3}{14}\right)}-\frac{\mathcal{J} R^2 r_h^{10/3}}{r_0^{13/3}}+\ldots\,,
   \label{d_r_0}
\end{equation}
where $\mathcal{J}$ is the following integral:
\be
    \mathcal{J}=\int_{1}^{\infty} \text{d$\gamma$}\frac{\left(\gamma ^{4/3}+\gamma ^{2/3}+1\right) \left(\gamma ^6+\gamma ^4-\gamma ^2-1\right)-2 \gamma ^4}{\gamma ^{16/3} \left(\gamma ^4+\left(\gamma ^{4/3}+\gamma ^{2/3}+1\right) \left(\gamma ^2+1\right)\right) \sqrt{\gamma ^{14/3}-1}} \approx 0.0696931 \ .
\ee
Eq. (\ref{d_r_0})  can in turn be reverted for $r_0$,
\begin{equation}
    r_0=2 \sqrt{\pi } \left(\frac{\Gamma
   \left(\frac{3}{14}\right)}{\Gamma \left(\frac{5}{7}\right)}\right)^{-1}\frac{R^2 }{\ell}-
   \frac{ \mathcal{J}
    }{8 \sqrt[3]{2} \pi ^{5/3}
   }\left(\frac{\Gamma
   \left(\frac{3}{14}\right)}{\Gamma \left(\frac{5}{7}\right)}\right)^{10/3}\frac{R^2}{\ell}\left(\frac{r_h \ell}{R^2}\right)^{10/3}+\ldots\,.\label{eq:r0ind}
\end{equation}
Moreover,  the energy of the string can be written as:
\begin{equation}
\begin{split}
    \text{E}_{q \bar{q}}=&\frac{3 \sqrt[3]{3} }{4\ 2^{2/3} \pi  \sqrt[3]{Q_f}} \Bigg[
    -r_0^{4/3}+r_h^{4/3}\\
    &-\frac{4}{3} r_0^{4/3}
   \int_{1}^{\infty}\left(\sqrt[3]{\gamma }-\gamma 
   \sqrt{\frac{r_h^{10/3}-\left(\gamma 
   r_0\right){}^{10/3}}{\left(1-\gamma ^{14/3}\right) r_0^{10/3}+\left(\gamma ^{4/3}-1\right)
   r_h^{10/3}}}\right)\Bigg]\,,
\end{split}
\end{equation}
which for small temperatures can be written
\begin{equation}
    \text{E}_{q \bar{q}}=-\frac{3 \sqrt[3]{3} }{4\ 2^{2/3} \pi  \sqrt[3]{Q_f}}
    \left(\frac{2
   r_h^{10/3} \tilde{\mathcal{J}}}{3
   r_0^2}+\frac{4 \sqrt{\pi } r_0^{4/3} \tan \left(\frac{3 \pi }{14}\right) \Gamma
   \left(\frac{25}{14}\right)}{11 \Gamma
   \left(\frac{9}{7}\right)}-r_h^{4/3}+\cdots\right)\,,
\end{equation}
where $ \tilde{\mathcal{J}}$ is given by:
\begin{equation}
    \tilde{\mathcal{J}}=\int_1^\infty\text{d$\gamma$}\left(\frac{\gamma ^{8/3}+\gamma ^{4/3}+\gamma ^{2/3}+\gamma
   ^2+1}{\gamma ^{2/3} \left(\gamma ^4+\left(\gamma ^{4/3}+\gamma ^{2/3}+1\right) \left(\gamma ^2+1\right)\right) \sqrt{\gamma ^{14/3}-1}}\right)\approx 0.415974\ .
\end{equation}
Finally, inserting \eqref{eq:r0ind} and removing the zero-point thermal energy introduced by the regularization,
\begin{equation}
\begin{split}
    \text{E}_{q \bar{q}}=-\frac{3 \sqrt[3]{3} R^{8/3} }{4\ 2^{2/3} \pi 
   \sqrt[3]{Q_f}}
   \Bigg[2^{4/3}& \pi ^{7/6} \left(\frac{\Gamma \left(\frac{5}{7}\right)}{\Gamma
   \left(\frac{3}{14}\right)}\right)^{7/3} \ell^{-4/3}+\\
   &+\frac{81 \;2^{16/21}\sqrt[3]{\frac{3}{5}} \pi ^{7/3} \Gamma
   \left(\frac{3}{14}\right)^3 }{6125\ \Gamma \left(\frac{5}{7}\right) \Gamma
   \left(\frac{27}{14}\right)}\ell^{-4/3}(T \ell)^{10/3}+\ldots\Bigg]\,,
\end{split}
\end{equation}
where in the last step we used that 
\begin{equation}
    \tilde{\mathcal{J}}-\mathcal{J}=\frac{3 \sqrt{\pi } \Gamma \left(\frac{10}{7}\right)}{14 \Gamma \left(\frac{27}{14}\right)}\,.
\end{equation}
Once again we note that the first temperature correction makes the force less attractive.

Similarly, we can compute the energy with a separation in the anisotropic $z$-direction. This case the metric reads
\begin{equation}
    \mathrm{d}s^2=\mathrm{d}r^2 \left(\frac{\sqrt{h(r)}}{B(r)}+\frac{e^{-2 \phi } z_0'(r){}^2}{\sqrt{h(r)}}\right)-\frac{\mathrm{d}t^2
   B(r)}{\sqrt{h(r)}}\ ,
\end{equation}
with a separation
\begin{equation}
    \ell=\frac{3\ 3^{2/3} \sqrt{\pi } R^2  }{4\sqrt[3]{2 Q_f^2 r_0}}\frac{\Gamma \left(\frac{3}{5}\right)}{\Gamma \left(\frac{1}{10}\right) }
    \left(4-\frac{10}{11}\left(\frac{r_h}{r_0}\right)^{10/3}\right)\ ,
\end{equation}
and an energy
\begin{equation}
\text{E}_{q \bar{q}}=-\frac{9 \pi ^{3/2} R^8 }{4 \ell^4
   Q_f^3 }\frac{\Gamma \left(\frac{3}{5}\right)^5}{\Gamma \left(\frac{1}{10}\right)^5}\Bigg[\frac{729}{4}+\frac{1}{2\, 3^{2/3}}\left(\frac{\Gamma \left(\frac{1}{10}\right)}{\Gamma \left(\frac{3}{5}\right)}\right)^{10}\left(\frac{4}{9\ 5 \sqrt{\pi }}\right)^{10/3}\left(\frac{\ell^3 Q_f^2
   T}{R^4}\right)^{10/3}\Bigg]\,.
\end{equation}

We can also compute the screening corrections to the mass. Indeed, by expanding \eqref{eq:constituentmass} for low temperatures we obtain
\begin{equation}
\begin{split}
    \frac{M_c}{m_q}=1
    -&15^{2/3}\left(\frac{3}{10}\right)^{2}\left(\frac{\pi}{2 Q_f m_q^3}\right)^{1/3}
   \left(R^2 T\right)^{4/3}+\ldots\,\,,
\end{split}
\end{equation}
where we used that, at first order in the correction, $P_0\approx m$ and that $m$ is related to $m_q$ as in  \eqref{mass_dictionary}.

\vskip 1cm
\renewcommand{\theequation}{\rm{E}.\arabic{equation}}
\setcounter{equation}{0}

\section{High temperature black hole embeddings} 
\label{HighT_BH_embed}

Let us consider now black hole embeddings for $d=0$, which we will describe by considering the angle $\chi$ as a function of the radial variable $r$. The equation for $\chi(r)$ written in (\ref{eom_chi_general}) reduces  to:
\bea
&&{\partial \over \partial r}\,\Bigg[ r^5\,\cos^3\chi {\sqrt{1+{1-b\over b}\,\cos^2\chi}\over \sqrt{1+{r^2\over b}\,B\,\chi'^2}}\,B\,\chi{\,'}\Bigg]\rc\rc
&&\qquad\qquad
+\,r^3\,\cos^2\chi \sin\chi\,
{3\,b+4(1-b)\,\cos^2\chi\over 
\sqrt{1+{1-b\over b}\,\cos^2\chi}}\,
\sqrt{1+{r^2\over b}\,B\,\chi'^2}\,=\,0\ . \label{eom_chi}
\eea
We will analyze  the high temperature limit in which $\chi$ is small for all values of the radial variable $r$. At first order in $\chi$ the equation of motion becomes:
\beq
{\partial \over \partial r}\,\big(r^5\,B\,\chi'\big)\,+\,b(4-b)\,r^3\,\chi\,=\,0\,\,.
\label{highT_chi_eq}
\eeq
The horizon radius in (\ref{highT_chi_eq})  can be scaled out by redefining the radial variable and $\chi$, this is equivalent to setting $r_h=1$. The equation for $\chi$  becomes:
\beq
{\partial \over \partial r}\,\Big[r^5\,\Big(1\,-\,{1\over r^{{10\over 3}}}\Big)\chi'\Big]\,+\,b(4-b)\,r^3\,\chi\,=\,0 \ .
\eeq
The general solution of this equation can be written in terms of hypergeometric functions as:
\beq
\chi(r)\,=\,c_1\,r^{-{2\over 3}}\,F\Big({1\over 15}\,,\,{11\over 15}\,; {4\over 5};\,
r^{{10\over 3}}\Big)\,+\,c_2\,F\Big({4\over 15}\,,\,{14\over 15}\,; {6\over 5};\,
r^{{10\over 3}}\Big)\,\,,
\label{chi_r_general_sol}
\eeq
where $c_1$ and $c_2$ are integration constants.  The function on the right-hand side of (\ref{chi_r_general_sol}) is divergent at the horizon $r=1$ for generic values of $c_1$ and $c_2$. Indeed, one has:
\beq
F\Big(\alpha, \beta;\alpha+\beta;z\Big)\approx -{\Gamma(\alpha+\beta)\over \Gamma(\alpha)\Gamma(\beta)}\,\,\log (1-z)\,\,,
\qquad\qquad {\rm as}\,\,z\to 1^{-}\,\,,
\eeq
and, thus, near $r=1$ the angle $\chi$ behaves as:
\beq
\chi(r)\,\approx -\Bigg[
{\Gamma\big({4\over 5}\big)\over 
\Gamma\big({1\over 15}\big)\,\Gamma\big({11\over 15}\big)}\,c_1\,+\,
{\Gamma\big({6\over 5}\big)\over 
\Gamma\big({4\over 15}\big)\,\Gamma\big({14\over 15}\big)}\,c_2
\Bigg]\log\big(1-r^{{10\over 3}}\big)\,\,.
\eeq
To avoid this divergence, the constants $c_1$ and $c_2$ must satisfy
\beq
{c_1\over c_2}\,=\,-\,{\Gamma\big({6\over 5}\big)\,\Gamma\big({1\over 15}\big)\,\Gamma\big({11\over 15}\big)
\over \Gamma\big({4\over 5}\big)\,\Gamma\big({4\over 15}\big)\,\Gamma\big({14\over 15}\big)}\,\,.
\label{c_1_c_2_1}
\eeq

Let us next look at  the UV behavior $r\to\infty$ of $\chi$. In general, for large $z$ the hypergeometric functions behave as
\beq
F(\alpha,\beta;\alpha+\beta; z)\,\approx\,
\Gamma(\alpha+\beta)\,\Bigg[\,
e^{-i\pi\alpha}\,{\Gamma(\beta-\alpha)\over \Gamma^2(\beta)}\,z^{-\alpha}\,+\,
e^{-i\pi\beta}\,{\Gamma(\alpha-\beta)\over \Gamma^2(\alpha)}\,z^{-\beta}
\,\Bigg]\,\,.
\eeq
In particular, we have for $r\to\infty$:
\bear
&&r^{-{2\over 3}}\,F\Big({1\over 15}\,,\,{11\over 15}\,; {4\over 5};\,
r^{{10\over 3}}\Big)\approx\Gamma\big({4\over 5}\big)\Bigg[
e^{-{i\pi\over 15}}\,{\Gamma\big({2\over 3}\big)\over \Gamma^2\big({11\over 15}\big)}\,r^{-{8\over 9}}\,+\,
e^{-{11i\pi\over 15}}\,{\Gamma\big(-{2\over 3}\big)\over \Gamma^2\big({1\over 15}\big)}\,r^{-{28\over 9}}
\Bigg]\rc\rc\rc
&&F\Big({4\over 15}\,,\,{14\over 15}\,; {6\over 5};\,
r^{{10\over 3}}\Big)\approx\Gamma\big({6\over 5}\big)\Bigg[
e^{-{4i\pi\over 15}}\,{\Gamma\big({2\over 3}\big)\over \Gamma^2\big({14\over 15}\big)}\,r^{-{8\over 9}}\,+\,
e^{-{14i\pi\over 15}}\,{\Gamma\big(-{2\over 3}\big)\over \Gamma^2\big({4\over 15}\big)}\,r^{-{28\over 9}}
\Bigg]\,\,.
\label{UV_hyper_chi}
\eear
Thus, at the UV:
\beq
\chi\approx {m_r\over r^{{8\over 9}}}\,+\, {c_r\over r^{{28\over 9}}}\,\,,
\eeq
where $m_r$ and $c_r$ are the mass and condensate parameters in the $r$ variable. Let us prove that, when $c_1$ and $c_2$  fulfill
(\ref{c_1_c_2_1}), $m_r$ and $c_r$ are real.  Using (\ref{UV_hyper_chi}), we get that the coefficient of the leading term of $\chi$ as $r\to\infty$ is:
\beq
e^{-{i\pi\over 15}}\,
{\Gamma\big({4\over 5}\big)\,\Gamma\big({2\over 3}\big)\over\Gamma^2\big({11\over 15}\big)}\,c_1\,+\,
e^{-{4i\pi\over 15}}\,
{\Gamma\big({6\over 5}\big)\,\Gamma\big({2\over 3}\big)\over\Gamma^2\big({14\over 15}\big)}\,c_2\,\,.
\label{UV_leading}
\eeq
Let us now use  the reflection formula for the Gamma function
\beq
\Gamma(z)\,\Gamma(1-z)\,=\,{\pi\over \sin (\pi z)}\,\,,
\label{reflection}
\eeq
to write the relation (\ref{c_1_c_2_1}) between $c_1$ and $c_2$ as:
\beq
{c_1\over c_2}\,=\,-{\Gamma\big({6\over 5}\big)\over \Gamma\big({4\over 5}\big)}\,
{\Gamma^2\big({11\over 15}\big)\over \Gamma^2\big({14\over 15}\big)}\,
{\sin\Big({4\pi\over 15}\Big)\over \sin\Big({\pi\over 15}\Big)}\,\,.
\label{c_1_c_2_2}
\eeq
Using (\ref{c_1_c_2_2}) one can show that the imaginary part of (\ref{UV_leading}) vanishes and that the real part is given by:
\beq
-c_2\,{\Gamma\big({2\over 3}\big)\,\Gamma\big({6\over 5}\big)\,\Gamma\big({1\over 15}\big)\over
\pi\,\Gamma\big({14\over 5}\big)}\,\sin\Big({\pi\over 5}\Big)\,\equiv\,m_r\,\,.
\label{c2_mr}
\eeq
The coefficient of the subleading term is:
\beq
e^{-{11i\pi\over 15}}\,
{\Gamma\big(-{2\over 3}\big)\,\Gamma\big({4\over 5}\big)\over\Gamma^2\big({1\over 15}\big)}\,c_1\,+\,
e^{-{14i\pi\over 15}}\,
{\Gamma\big(-{2\over 3}\big)\,\Gamma\big({6\over 5}\big)\over\Gamma^2\big({4\over 15}\big)}\,c_2\,\,.
\label{UV_subleading}
\eeq
If we now use the relation between $c_1$ and $c_2$ in the form:
\beq
{c_1\over c_2}\,=\,-{\Gamma\big({6\over 5}\big)\over \Gamma\big({4\over 5}\big)}\,
{\Gamma^2\big({1\over 15}\big)\over \Gamma^2\big({4\over 15}\big)}\,
{\sin\Big({14\pi\over 15}\Big)\over \sin\Big({4\pi\over 15}\Big)}\,\,,
\label{c_1_c_2_3}
\eeq
we can easily show that the imaginary part of (\ref{UV_subleading}) vanishes and the real part is given by:
\beq
-c_2\,{\Gamma\big(-{2\over 3}\big)\,\Gamma\big({6\over 5}\big)\,\Gamma\big({11\over 5}\big)\over
\pi\,\Gamma\big({4\over 15}\big)}\,\sin\Big({\pi\over 5}\Big)\,\equiv\,c_r\,\,.
\label{c2_cr}
\eeq
Eliminating $c_2$ between (\ref{c2_mr}) and (\ref{c2_cr}), we get the following relation between the condensate parameter $c_r$ and the mass parameter $m_r$:
\beq
c_r\,=\,-{3\over 2}\,
{\Gamma\big({1\over 3}\big)\over \Gamma\big({2\over 3}\big)}\,
{\Gamma\big({11\over 15}\big)\over \Gamma\big({4\over 15}\big)}\,
{\Gamma\big({14\over 15}\big)\over \Gamma\big({1\over 15}\big)}\,m_r \approx -0.07876 \ .\label{cr_ur}
\eeq
Let us now translate  (\ref{cr_ur}) into a relation between $c$ and $m$, as defined in (\ref{eta_UV}) in terms of the $u$ variable. Actually, the coordinates $u$ and $r$ are proportional to each other in the UV:
\beq
u\approx 2^{{8\over 15}}\,r^{{8\over 9}}\,\,,
\qquad\qquad
(u\,,\, r\to\infty\,\,,\,\,r_h=1)\,\,.
\eeq
Therefore, $(m_r, c_r)$ are related to $(m,c)$ as:
\beq
m_r\,=\,{m\over 2^{{8\over 15}}}\,\,,
\qquad\qquad
c_r\,=\,{c\over 2^{{28\over 15}}}\,\,.
\eeq
It follows that:
\beq
{c\over m}\,=\,2^{{4\over 3}}\,{c_r\over m_r}\,\,.
\eeq
Using this last result we get:
\beq
c\,\approx\,-0.19846\,m\,\,.
\label{c_m_highT}
\eeq
Moreover, we can relate the value of the angle $\chi$ at the horizon to the mass parameter as:
\beq
\chi(r=1)\,=\,{2\pi^2\,(1+\sqrt{5})\over 
\Gamma\big({2\over 3}\big)\,\Gamma\big({1\over 15}\big)\,
\Gamma\big({4\over 15}\big)}\,\,m_r\,=\,
{2^{{7\over 15}}\pi^2\,(1+\sqrt{5})\over 
\Gamma\big({2\over 3}\big)\,\Gamma\big({1\over 15}\big)\,
\Gamma\big({4\over 15}\big)}\,\,m \ .
\eeq

\subsection{On-shell action}
Let us now obtain an approximate expression for the free energy $F$ in the limit of high temperature (or small mass parameter $m$). In this regime our embedding is a black hole embedding with a small value of $\eta(u)$ for all values of the holographic coordinate $u$. In a first, zero-order, approximation we can just take $\eta=0$ in the bulk and boundary actions. We get:
\bear
{\cal I}_{bulk}^{(0)} & = & \int_1^{u_{max}}\,du\,u^{{7\over  2}}\,\tilde f^{{7\over 5}}\,f\rc\rc
 {\cal I}_{bdy}^{(0)} & = & -{2\over 9}\,u^{{9\over 2}}\,\tilde f^{{12\over 5}}\,\Big|_{u=u_{max}}\,=\,-{2\over 9}\,u^{{9\over 2}}_{max}\,-\,{8\over 15}\,u^{{3\over 4}}_{max}\,\,.
\eear
Let us rewrite the  boundary action as an integral, in the form:
\beq
{\cal I}_{bdy}^{(0)}\,=\,-\int_1^{u_{max}}du\,u^{{7\over  2}}\,\big(1+{2\over 5}\,u^{-{15\over 4}}\big)\,-\,
{34\over 45}\,\,.
\eeq
Then, the zero-order free energy $F^{(0)}$ is given by:
\beq
{ F^{(0)}\over {\cal N}}\,=\,
\int_1^{\infty}\,du\,u^{{7\over  2}}\,\Big(
\tilde f^{{7\over 5}}\,f\,-\,1\,-\,{2\over 5}u^{-{15\over 4}}\Big)\,-\,{34\over 45} =-{8\over 9}2^{{2\over 5}} = -2^{{2\over 5}}b \,\,.
\eeq
Therefore, we can approximate $F$ at leading order in $T$ as:
\beq
 F\,\approx\,-{\pi^2\,T_{D7}\over 2\,b^2}\,r_h^4\,\,,
\qquad\qquad
(T\to\infty)\,\,.
\eeq
Notice that this means that $F\propto T^4$ and, therefore it obeys a Stefan-Boltzmann law at leading order  in $T$. To find the corrections to this law, let us consider the terms in the action that are quadratic in $\eta$. For the bulk action these terms are:
\beq
{\cal I}_{bulk}^{(2)}\,=\,\int_1^{u_{max}}\,du\,u^{{7\over 2}}\,
\tilde f^{{7\over 5}}\,f\,\Big[\,{4\over 9}\,u^2\,\dot\eta^2\,-\,{14\over 9}\,\eta^2\Big]\,\,.
\eeq
To evaluate this integral when $\eta(u)$ is a solution to the equations of motion at quadratic order, we apply the method used in appendix C.1 of \cite{Jokela:2012dw}. Suppose that we have an action $S$ which depends quadratically on $\eta(u)$ and $\dot\eta(u)$ as:
\beq
S\,=\,\,\int_1^{u_{max}}\,du\big[\,F_1(u)\,\dot\eta^2\,+\,F_2(u)\,\eta^2\big]\,\,,
\eeq
where $F_1$ and $F_2$ are known functions of $u$. Then, the action $S$ evaluated on a solution to the equation of motion is:
\beq
S^{on-shell}\,=\,F_1\,\eta\,{d\eta\over du}\Big|_{u=1}^{u=u_{max}}\,\,.
\eeq
In our case the function $F_1(u)$ is:
\beq
F_1(u)\,=\,{4\over 9}\,u^{{11\over 2}}\,\tilde f^{{7\over 5}}\,f\,\,,
\eeq
and, since $F_1(u=1)$ vanishes, we get:
\beq
{\cal I}_{bulk}^{(2)}\,=\,
F_1\,\eta\,{d\eta\over du}\Big|_{u=u_{max}\to\infty}\,=\,-{4\over 9}\,m^2\,u_{max}^{{5\over 2}}\,-\,2m\,c\,\,.
\eeq
Moreover, the boundary action at second order is:
\beq
{\cal I}_{bdy}^{(2)}\,=\,-{2\over 9}\,\tilde f\,f^{-1}\,e^{\phi}\,
\sqrt{-\det \gamma}\,\,
 (-2\eta^2)\,\Big|_{u=u_{max}}\,=\,{4\over 9}\,\tilde f^{{12\over 5}}\,u^{{9\over 2}}\,\eta^2\Big|_{u=u_{max}}\,\,.
\eeq
Explicitly:
\beq
{\cal I}_{bdy}^{(2)}\,=\,{4\over 9}\,m^2\,\,u_{max}^{{5\over 2}}\,+\,{8\over 9}\,m\,c\,\,.
\eeq
Then:
\beq
{ F^{(2)}\over {\cal N}}\,=\,
{\cal I}_{bulk}^{(2)}\,+\,
{\cal I}_{bdy}^{(2)}\,=\,-{10\over 9}\,m\,c\,\,.
\eeq
Including this second order correction, we get at  high $T$:
\beq
{ F\over {\cal N}}\,\approx\,-{8\over 9}\,2^{{2\over 5}}-{10\over 9}\,m\,c\,\,,
\qquad\qquad
(T\to\infty). 
\label{high_T_F_subleading}
\eeq
We found in (\ref{c_m_highT}) that $c\sim m$ in this high $T$ regime. Therefore, the last term in (\ref{high_T_F_subleading}) is proportional to $m^2$. As $m\sim T^{-b}$ for fixed quark mass $m_q$, it follows that the second term on the right-hand side of  (\ref{high_T_F_subleading}) gives rise to a subleading contribution that corrects 
the  dominant Stefan-Boltzmann law ($F\propto T^4$) at large $T$. This subdominant contribution grows as $T^{4-2b}=T^{{20\over 9}}$.

\vskip 1cm
\renewcommand{\theequation}{\rm{F}.\arabic{equation}}
\setcounter{equation}{0}

\section{Fluctuations}\label{Fluctuation_appendix}

In this appendix we obtain the equations of motion for the fluctuations around the $\chi=0$ massless embedding of the D7-brane at non-zero temperature and density. Accordingly, let us allow the worldvolume gauge field $A$ to fluctuate as:
\beq
A\,=\,A^{(0)}\,+\,a(r, x^{\mu})\,\,,
\eeq
where $A^{(0)}$ is the one-form written in (\ref{A_t_prime_massless}). Let us  follow the methodology of \cite{Jokela:2015aha} and split the total field strength as
$F\,=\,F^{(0)}+f$ and the DBI matrix as:
\beq
g_8+e^{-{\phi\over 2}} F\,=\,\Big(g_8+e^{-{\phi\over 2}} F^{(0)}\Big)
\Big(1+X\Big)\,\,,
\eeq
where $X$ is defined as:
\beq
X\,\equiv\,\Big(g_8+e^{-{\phi\over 2}} F^{(0)}\Big)^{-1}\,e^{-{\phi\over 2}} \,f\,\,.
\eeq
Then, we expand the DBI determinant in series as:
\beq
\sqrt{-\det (g_8+e^{-{\phi\over 2}}F)}=
\sqrt{-\det (g_8+e^{-{\phi\over 2}}F^{(0)})}\,\Big[1\,+\,{1\over 2}\Tr X\,-\,{1\over 4}\,
\Tr X^2\,+\,{1\over 8}\,\Big(\Tr X\Big)^2\,+{\cal{O}}(X^3)\Big] \ .
\label{DBI_expansion}
\eeq
Let us split the inverse of the matrix $g_8\,+\,F^{(0)}$ as:
\beq
\Big(\,g_8\,+\,e^{-{\phi\over 2}}F^{(0)}\Big)^{-1}\,=\,{\cal G}^{-1}\,+\,{\cal J}\,\,,
\eeq
where ${\cal G}^{-1}$ is the symmetric part and ${\cal J}$ is the antisymmetric part (${\cal G}$ is the so-called open string metric).  It follows that:
\beq
X^{a}_{\,\,\,\,\, b}\,=\, {\cal G}^{ac}\,e^{-{\phi\over 2}}\,f_{cb}\,+\,{\cal J}^{ac}\,e^{-{\phi\over 2}}\,f_{cb}\,\,,
\eeq
where the Latin indexes take values in $a,b,c\in \{t,x,y,r\}$. The traces needed in the expansion (\ref{DBI_expansion}) up to second order in $X$ are:
\bea
\Tr X & = & {\cal J}^{ab}\,e^{-{\phi\over 2}}f_{ba}\rc\rc
\Tr X^2 & = & -{\cal G}^{ac}\,{\cal G}^{bd}\,e^{-\phi}\,f_{cd}\,f_{ab}\,+\,{\cal J}^{ac}\,{\cal J}^{bd}\,e^{-\phi}\,f_{cd}\,f_{ab}\,\,.
\eea
It is easy to prove that the first-order term in $X$ is a total derivative and, therefore, it does not contribute to the equations of motion of the gauge field and can be neglected. Up to second order, the Lagrangian density for the fluctuations takes the form:
\beq
{\cal L}\,\sim\,{H\,e^{{\phi\over 2}}\over \sqrt{d^2+H}}\,
\Big[-\,{1\over 4}\,
\Tr X^2\,+\,{1\over 8}\,\Big(\Tr X\Big)^2\Big]\,\,,
\eeq
where $H$ is the function written in (\ref{H_alpha}). 
If we now define the prefactor ${\cal L}_{*}$ as:
\beq
{\cal L}_{*}\,\equiv\,{H\,e^{-{\phi\over 2}}\over \sqrt{d^2+H}}\,\,,
\eeq
then ${\cal L}$ is given by:
\beq
{\cal L}\,\sim\,{\cal L}_{*}\,
\Big(\,{\cal G}^{ac}\,{\cal G}^{bd}\,-\,
{\cal J}^{ac}\,{\cal J}^{bd}\,+\,{1\over 2}\,{\cal J}^{cd}\,{\cal J}^{ab}
\Big)f_{cd}\,f_{ab}\,\,.
\eeq
The corresponding equation of motion  for  gauge field component $a^d$ is:
\beq
\partial_{c}\,\Bigg[{\cal L}_{*}\,
\Big(\,{\cal G}^{ca}\,{\cal G}^{db}\,-\,
{\cal J}^{ca}\,{\cal J}^{db}\,+\,{1\over 2}\,{\cal J}^{cd}\,{\cal J}^{ab}
\Big)\,f_{ab}\Bigg]\,=\,0\,\,.
\label{eom_general}
\eeq
Before going further, let us write the non-vanishing components of the open string metric for our system:
\bear
&&{\cal G}^{tt}\,=\,-h^{{1\over 2}}\,\,
{H+d^2\over H\,B}\,\,,
\qquad\qquad\qquad\qquad
{\cal G}^{rr}\,=\,h^{-{1\over 2}}\,\,
{H+d^2\over H}\,B\,\,,\rc\rc
&&{\cal G}^{x^1\,x^1}\,=\,{\cal G}^{x^2\,x^2}\,=\,h^{{1\over 2}}\,\equiv\,{\cal G}^{xx}\,\,,
\qquad\qquad
{\cal G}^{x^3\,x^3}\,=\,e^{2\phi}\,h^{{1\over 2}}\,\equiv\,{\cal G}^{zz}\,\,.\qquad
\label{cal_G_values}
\eear
The non-vanishing elements of the antisymmetric tensor are:
\beq
{\cal J}^{tr}\,=\,-{\cal J}^{rt}\,=\,-d\,{\sqrt{d^2+H}\over H}\,\,.
\eeq

Let us write explicitly the equations for the fluctuations. We  choose the gauge in which:
\beq
a_r\,=\,0\,\,.
\eeq
The equation of motion of $a_r$ in this gauge is the following first-order constraint:
\beq
{\cal G}^{tt}\,\partial_t\,f_{tr}\,+\,
{\cal G}^{xx}\,\partial_x\,f_{xr}\,+\,
{\cal G}^{xx}\,\partial_y\,f_{yr}\,+\,
{\cal G}^{zz}\,\partial_z\,f_{zr}\,=\,0\,\,.
\label{Gauss_law}
\eeq
Moreover, the equation for $a_t$ is:
\beq
\partial_r\,\Big[{\cal L}_*\,{\cal G}^{tt}\,{\cal G}^{rr}\,f_{rt}\Big]\,+\,
{\cal L}_*\, {\cal G}^{tt}\,
\Big[{\cal G}^{xx}\,\partial_x\,f_{xt}\,+\,
{\cal G}^{xx}\,\partial_y\,f_{yt}\,+\,
{\cal G}^{zz}\,\partial_z\,f_{zt}\Big]\,=\,0\,\,,
\eeq
and the equations for the components of $a$ along $x\equiv x^1$ and $y\equiv x^2$ are:
\bear
&&\partial_r\,\Big[{\cal L}_*\,{\cal G}^{xx}\,{\cal G}^{rr}\,f_{rx}\Big]\,+\,
{\cal L}_*\, {\cal G}^{xx}\,
\Big[{\cal G}^{tt}\,\partial_t\,f_{tx}\,+\,
{\cal G}^{xx}\,\partial_y\,f_{yx}\,+\,
{\cal G}^{zz}\,\partial_z\,f_{zx}\Big]\,=\,0 \rc\rc
&&\partial_r\,\Big[{\cal L}_*\,{\cal G}^{xx}\,{\cal G}^{rr}\,f_{ry}\Big]\,+\,
{\cal L}_*\, {\cal G}^{xx}\,
\Big[{\cal G}^{tt}\,\partial_t\,f_{ty}\,+\,
{\cal G}^{xx}\,\partial_x\,f_{xy}\,+\,
{\cal G}^{zz}\,\partial_z\,f_{zy}\Big]\,=\,0\,\,.
\qquad\qquad
\eear
Finally, it remains to write the equation for the gauge field along the anisotropic direction $z=x^3$, which is:
\beq
\partial_r\,\Big[{\cal L}_*\,{\cal G}^{zz}\,{\cal G}^{rr}\,f_{rz}\Big]\,+\,
{\cal L}_*\, {\cal G}^{zz}\,
\Big[{\cal G}^{tt}\,\partial_t\,f_{tz}\,+\,
{\cal G}^{xx}\,\partial_x\,f_{xz}\,+\,
{\cal G}^{xx}\,\partial_y\,f_{yz}\Big]\,=\,0\,\,.
\eeq

\subsection{In-plane propagation}

Let us consider a wave propagating in the $x$ direction. This means that all the  gauge field components $a_{\nu}$ only depend on $r$, $t$,  and $x$.  Let us Fourier transform  the gauge field to momentum space as:
\beq
a_\nu(r, t, x)\,=\,\int {d\omega\,dk\over (2\pi)^2}\,
a_\nu(r, \omega, k)\,e^{-i\omega\,t\,+\,i k x}\,\,.
\eeq
From now on  in this section  all the equations are written  in momentum space. The first-order constraint 
(\ref{Gauss_law}) takes the form:
\beq
{\cal G}^{tt}\,\omega\,a_t'\,-\,{\cal G}^{xx}\,k\,a_x'\,=\,0\,\,,
\label{Gauss_in_plane}
\eeq
where the prime denotes derivative with respect to $r$. 
We now define the electric field $E$ as the gauge-invariant combination:
\beq
E\,=\,k\,a_t\,+\,\omega\,a_x\,\,.
\label{E_at_ax}
\eeq
This gauge-invariant combination $E$ appears in the equation for $a_t$ and $a_x$, which are given by:
\bea
\partial_r\,\Big[{\cal L}_*\,{\cal G}^{tt}\,{\cal G}^{rr}\,a_t'\Big]\,-\,k\,{\cal L}_*\,
{\cal G}^{tt}\,{\cal G}^{xx}\,E & = & 0 \rc\rc
\partial_r\,\Big[{\cal L}_*\,{\cal G}^{xx}\,{\cal G}^{rr}\,a_x'\Big]\,-\,\omega\,{\cal L}_*\,
{\cal G}^{xx}\,{\cal G}^{tt}\,E & = & 0\,\,. \label{eom_at_a_x_in_plane}
\eea
Actually, we can use (\ref{Gauss_in_plane}) and the radial derivative of the definition of $E$ to write $a_t'$ and
$a_x'$ in terms of $E'$:
\beq
a_t'\,=\,{{\cal G}^{xx}\,k\over {\cal G}^{xx}\,k^2\,+\,{\cal G}^{tt}\,\omega^2}\,\,E'\,\,,
\qquad\qquad
a_x'\,=\,{{\cal G}^{tt}\,\omega\over {\cal G}^{xx}\,k^2\,+\,{\cal G}^{tt}\,\omega^2}\,\,E'\,\,.
\eeq
Plugging this result in the two equations in  (\ref{eom_at_a_x_in_plane}), we arrive at a unique second-order equation for $E$:
\beq
E''\,+\,\partial_r\log\Bigg[{
{\cal L}_*\,{\cal G}^{tt}\,{\cal G}^{xx}\,{\cal G}^{rr}\over 
 {\cal G}^{xx}\,k^2\,+\,{\cal G}^{tt}\,\omega^2}\Bigg]\,E'\,-\,
 { {\cal G}^{xx}\,k^2\,+\,{\cal G}^{tt}\,\omega^2\over {\cal G}^{rr}}\,\,
 \,E\,=\,0\,\,.
 \label{eom_E_in_plane}
 \eeq
Moreover, the equations for $a_y$ and $a_z$ are  decoupled and given by:
\bea
\partial_r\,\Big[{\cal L}_*\,{\cal G}^{xx}\,{\cal G}^{rr}\,a_y'\Big]\,-\,
{\cal L}_*\,{\cal G}^{xx}\,\Big[ {\cal G}^{xx}\,k^2\,+\,{\cal G}^{tt}\,\omega^2\Big]\,a_y & = & 0\rc\rc
\partial_r\,\Big[{\cal L}_*\,{\cal G}^{zz}\,{\cal G}^{rr}\,a_z'\Big]\,-\,
{\cal L}_*\,{\cal G}^{zz}\,\Big[ {\cal G}^{xx}\,k^2\,+\,{\cal G}^{tt}\,\omega^2\Big]\,a_z & = & 0\,\,.\label{eom_ay_a_z_in_plane}
\eea
Plugging in (\ref{eom_E_in_plane}) the values of the open string metric written in (\ref{cal_G_values}) we get
(\ref{E_eq_in_plane}). 

\subsection{Off-plane propagation}

We now consider waves that propagate in the $z$ direction. In momentum space we write:
\beq
a_\nu(r, t, z)\,=\,\int {d\omega\,dk\over (2\pi)^2}\,
a_\nu(r, \omega, k)\,e^{-i\omega\,t\,+\,i k z}\,\,.
\eeq
The gauge-invariant electric field $E$ is now:
\beq
E\,=\,k\,a_t\,+\,\omega\,a_z\,\,.
\label{E_at_az}
\eeq
The equation of motion for $E$ can be obtained from (\ref{eom_E_in_plane}) by exchanging ${\cal G}^{xx}$ by 
${\cal G}^{zz}$:
\beq
E''\,+\,\partial_r\log\Bigg[{
{\cal L}_*\,{\cal G}^{tt}\,{\cal G}^{zz}\,{\cal G}^{rr}\over 
 {\cal G}^{zz}\,k^2\,+\,{\cal G}^{tt}\,\omega^2}\Bigg]\,E'\,-\,
 { {\cal G}^{zz}\,k^2\,+\,{\cal G}^{tt}\,\omega^2\over {\cal G}^{rr}}\,\,
 \,E\,=\,0\,\,.
 \label{eom_E_off_plane}
 \eeq
This equation reduces to (\ref{E_eq_off_plane}) when the values of ${\cal G}^{tt}$, ${\cal G}^{zz}$, and ${\cal G}^{rr}$ of  (\ref{cal_G_values}) are used.

\vskip 1cm
\renewcommand{\theequation}{\rm{G}.\arabic{equation}}
\setcounter{equation}{0}

\section{Useful integrals}
\label{integrals}

Let us collect some integrals that will be useful in the bulk text and whose expansions will also be needed. First of all, we define the integral $I_{\lambda_1,\lambda_2}(r)$ as:
\bea
I_{\lambda_1,\lambda_2}(r) & \equiv & \int_{r}^{\infty} {\rho^{\lambda_1}\,d\rho\over (\rho^{\lambda_2}+\tilde d^2)^{{1\over 2}}} \\ \label{I_lambda12_def}
 & = & {2\over \lambda_2-2\lambda_1-2} r^{1+\lambda_1-{\lambda_2\over 2}}F\Big({1\over 2}, {1\over 2}-{\lambda_1+1\over\lambda_2};{3\over 2}-{\lambda_1+1\over\lambda_2}
;-{\tilde d^2\over r^{\lambda_2}}\Big) \ . \label{I_lambda12_value}
\eea
For small $r$, assuming that $\lambda_2$ and $\lambda_1+1$ are positive, we have the expansion:
\beq
I_{\lambda_1,\lambda_2}(r) = {1\over \lambda_2}B\Big({\lambda_1+1\over \lambda_2}, {1\over 2}-{\lambda_1+1\over \lambda_2}\Big)\tilde d^{\,2{\lambda_1+1\over \lambda_2}-1}-{r^{\lambda_1+1}\over (\lambda_1+1)\tilde d}+\ldots \ .\label{I_lambda12_expansion}
\eeq
Let us next define $J_{\lambda_1,\lambda_2}(r)$ as follows
\bea
J_{\lambda_1,\lambda_2}(r) & \equiv  & \int_{r}^{\infty}{\rho^{\lambda_1}\,d\rho\over (\rho^{\lambda_2}+\tilde d^{\,2})^{{3\over 2}}}\\ \label{J_integral_definition}
 & = &{2\over 3\lambda_2-2\lambda_1-2}r^{1+\lambda_1-{3\lambda_2\over 2}}F\Big({3\over 2}, {3\over 2}-{\lambda_1+1\over\lambda_2};{5\over 2}-{\lambda_1+1\over\lambda_2}
;-{\tilde d^{\,2}\over r^{\lambda_2}}\Big)\  . \label{J_value}
\eea
For small $r$, when  $\lambda_2$ and $\lambda_1+1$ are both positive, we can expand 
$J_{\lambda_1,\lambda_2}(r)$ as:
\beq
J_{\lambda_1,\lambda_2}(r) = {1\over \lambda_2}B\Big({\lambda_1+1\over \lambda_2}, {3\over 2}-{\lambda_1+1\over \lambda_2}\Big)\,\tilde d^{\,2{\lambda_1+1\over \lambda_2}-3}\,-\,{r^{\lambda_1+1}\over (\lambda_1+1)\tilde d^{\,3}}+\ldots \ . \label{J_lambda12_expansion}
\eeq

\end{document}